\input harvmac
\input ising.defs
\overfullrule=0pt

\newdimen\stringwidth
\setbox0=\hbox{$-$}
\stringwidth=\wd0
\def\p{\hskip\stringwidth}
\def\subsubsec#1{\bigbreak\noindent{\it #1}\nobreak\medskip\nobreak}
\def\({\left(}
\def\){\right)}

\font\ninerm=cmr9
\font\ninei=cmmi9
\font\nineb=cmbx9
\font\nineib=cmmib9
\font\sixi=cmmi6
\def\ninemath{
\textfont0=\ninerm
\textfont1=\ninei \scriptfont1=\sixi \scriptscriptfont1=\scriptfont1}

\def\lesssim{\mathrel
{\hbox{\rlap{\hbox{\lower4pt\hbox{$\sim$}}}\raise1pt\hbox{$<$}}}}
\def\moresim{\mathrel
{\hbox{\rlap{\hbox{\lower4pt\hbox{$\sim$}}}\raise1pt\hbox{$>$}}}}
\def\Xint#1{\mathchoice
   {\XXint\displaystyle\textstyle{#1}}%
   {\XXint\textstyle\scriptstyle{#1}}%
   {\XXint\scriptstyle\scriptscriptstyle{#1}}%
   {\XXint\scriptscriptstyle\scriptscriptstyle{#1}}%
   \!\int}
\def\XXint#1#2#3{{\setbox0=\hbox{$#1{#2#3}{\int}$}
     \vcenter{\hbox{$#2#3$}}\kern-.5\wd0}}

\def\dashint{\Xint-}

\catcode`\@=11 
\def\eqna#1{\DefWarn#1\wrlabeL{#1$\{\}$}%
\xdef #1##1{(\noexpand\relax\noexpand\checkm@de%
{\s@csym\the\meqno\noexpand\f@rst{##1}}{\hbox{$\secsym\the\meqno##1$}})}
\global\advance\meqno by1}
\catcode`\@=12 

\def\table#1#2#3{
\nobreak
\par\begingroup\parindent=0pt\leftskip=1cm\rightskip=1cm\parindent=0pt
\baselineskip=12pt
\midinsert
\centerline{#3}
\vskip 12pt\ninerm\ninemath
{\nineb Table~#1:} #2\par
\endinsert
\endgroup\par
\goodbreak
}

\input epsf
\def\fig#1#2#3{
\nobreak
\par\begingroup\parindent=0pt\leftskip=1cm\rightskip=1cm\parindent=0pt
\baselineskip=12pt \ninerm \ninemath
\midinsert
\centerline{\hbox{#3}}
\vskip 12pt
{\nineb Fig.~#1:}~#2\par
\endinsert
\endgroup\par
\goodbreak
}

\font\cmss=cmss10 \font\cmsss=cmss10 at 7pt
\def\IZ{\relax\ifmmode\mathchoice
{\hbox{\cmss Z\kern-.4em Z}}{\hbox{\cmss Z\kern-.4em Z}}
{\lower.9pt\hbox{\cmsss Z\kern-.4em Z}}
{\lower1.2pt\hbox{\cmsss Z\kern-.4em Z}}\else{\cmss Z\kern-.4em Z}\fi}

\def\pre#1{ (preprint {\tt \hbox{#1}})}

\lref\Ons{L. Onsager, {\sl Crystal Statistics. I. A Two-Dimensional Model 
with an Order-Disorder Transition}, Phys. Rev. 65 (1944) 117--149.}

\lref\Yang{C. N. Yang, {\sl The Spontaneous Magnetization of a 
Two-Dimensional Ising Model}, \hbox{Phys. Rev.} 85 (1952) 808--816.}

\lref\Kaufman{B. Kaufman, {\sl Crystal Statistics. II. Partition Function 
Evaluated by Spinor Analysis}, Phys. Rev. 76 (1949) 1232--1243.}

\lref\Me{A. B. Zamolodchikov, {\sl Integrals of Motion and S-Matrix of the 
(Scaled) $T=T_c$ Ising Model with Magnetic Field}, Int. J. Mod. Phys. A 4 
(1989) 4235--4248.}

\lref\Fateev{V. A. Fateev, {\sl The exact relations between the coupling 
constants and the masses of particles for the integrable perturbed 
Conformal Field Theories}, Phys. Lett. B 324 (1994) 45--51.}

\lref\VicA{M. Caselle, M. Hasenbusch, A. Pelissetto, E. Vicari, {\sl 
High-precision estimate of $g_4$ in the 2D Ising model}, 
J. Phys. A 33 (2000) 8171--8180 \pre{hep-th/0003049}.}

\lref\VicB{M. Caselle, M. Hasenbusch, A. Pelissetto, E. Vicari, {\sl 
The critical equation of state of the 2D Ising Model}, J. Phys. A 34 
(2001) 2923--2948 \pre{cond-mat/0011305}.}

\lref\VicC{M. Caselle, M. Hasenbusch, {\sl Critical amplitudes and 
mass spectrum of the 2D Ising model in a
magnetic field}, Nucl. Phys. B 579 (2000) 667--703 
\pre{hep-th/9911216}.}

\lref\VicD{M. Caselle, M. Hasenbusch, A. Pelissetto, E. Vicari, {\sl 
Irrelevant operators in the two-dimensional Ising Model}, 
DFTT 17/2001, DESY 01-074, IFUP-TH 99/2001, \hbox{Roma1-1963/01}, June 
2001 \pre{cond-mat/0106372}.}

\lref\VicE{M. Caselle, P. Grinza, N. Magnoli, {\sl Correction induced by 
irrelevant operators in the correlators of the 2-d Ising model in a 
magnetic field}, J. Phys. A 34 (2001) 8733-8750 \pre{hep-th/0103263}.}

\lref\McCoyA{T. T. Wu, B. M. McCoy, C. A. Tracy, E. Barouch, {\sl 
Spin-spin correlation functions for the two-dimensional Ising model:
Exact theory in the scaling region}, Phys. Rev. B 13 (1976) 316--374.}

\lref\McCoyB{E. Barouch, B. M. McCoy, T. T. Wu, {\sl Zero-Field 
Susceptibility of the Two-Dimensional Ising Model near
$T_c$}, Phys. Rev. Lett. 31 (1973) 1409--1411.}

\lref\McCoyWuA{B. M. McCoy, T. T. Wu, {\sl Two-dimensional Ising field 
theory in a magnetic field: Breakup of the cut in the two-point function}, 
Phys. Rev. D 18 (1978) 1259--1267.}

\lref\WuMcCoyG{B. M. McCoy, T. T. Wu, { \sl Two-dimensional Ising model 
near $T_c$: Approximation for small magnetic field}, Phys. Rev. B 18 
(1978) 4886--4901.}

\lref\TCSAa{V. P. Yurov, Al. B. Zamolodchikov, {\sl Truncated Conformal 
Space Approach to Scaling Lee-Yang Model}, Int. J. Mod.Phys. A 5 (1990) 
3221--3245.}

\lref\TCSAb{V. P. Yurov, Al. B. Zamolodchikov, {\sl 
Truncated-Fermionic-Space Approach to the \hbox{Critical} 2D Ising Model 
with Magnetic Field}, Int. J. Mod. Phys. A 6 (1991) 4557--4578.}

\lref\TCSAc{G. Mussardo, {\sl Off-critical statistical models: factorized 
scattering theories and bootstrap program}, Phys. Rep. 218 (1992) 
215--379.}

\lref\Book{T. T. Wu, B. M. McCoy, {\sl The Two-Dimensional Ising Model},  
Harvard University Press, 1973.}

\lref\LangerA{J. S. Langer, {\sl Theory of the Condensation Point}, Annals 
Phys. 41 (1967) 108--157, Annals Phys. 281 (2000) 941--990.}

\lref\LangerB{J. S. Langer, {\sl Statistical Theory of the Decay of 
Metastable States}, Annals Phys. 54 (1969) 258--275.}

\lref\LangerC{J. S. Langer, {\sl Metastable States}, Physica 73 (1974) 
61--72.}

\lref\Voloshin{M. B. Voloshin, {\sl Decay of a metastable vacuum in 
(1+1) dimensions}, Yad. Fiz. 42 (1985) 1017--1026, 
Sov. J. Nucl. Phys. 42 (1985) 644--649.}

\lref\Rutkevich{S. B. Rutkevich, {\sl Decay of the metastable phase in 
$d=1$ and $d=2$ Ising models}, Phys. Rev. B 60 (1999) 14525-14528 
\pre{cond-mat/9904059}.}

\lref\Sachdev{S. Sachdev, {\sl Universal, finite temperature, crossover 
functions of the quantum transition in the Ising chain in a 
transverse field}, Nucl. Phys. B 464 (1996) 576--595 \pre{cond-mat/9509147}.}

\lref\Kobzarev{I. Yu. Kobzarev, L. B. Okun, M. B. Voloshin, {\sl Bubbles 
in metastable vacuum}, \hbox{Yad. Fiz. 20} (1974) 1229--1234, Sov. J. 
Nucl. Phys. 
20 (1975) 644--646.}

\lref\Coleman{S. Coleman, {\sl Fate of the false vacuum: Semiclassical 
theory}, Phys. Rev. D 15 (1977) 2929--2936.}

\lref\CC{C. G. Callan, S.Coleman, {\sl Fate of the false vacuum. II. 
First quantum corrections}, \hbox{Phys. Rev.} D 16 (1977) 
1762--1768.}

\lref\Mussardo{G. Delfino, G. Mussardo, P. Simonetti, {\sl Non-integrable 
Quantum Field Theories as Perturbations of Certain Integrable Models}, 
Nucl. Phys. B 473 (1996) 469--508 \pre{hep-th/9603011}.}

\lref\Cardy{J. L. Cardy, {\sl Conformal Invariance and the Yang-Lee Edge 
Singularity in Two Dimensions}, Phys. Rev. Lett. 54 (1985) 1354--1356.} 

\lref\CardyMussardi{J. L. Cardy, G. Mussardo, {\sl $S$-Matrix of the 
Yang-Lee Edge Singularity in Two Dimensions}, Phys. Lett. B 225 (1989) 
275--278.}

\lref\FLZZ{V. Fateev, S. Lukyanov, A. Zamolodchikov, Al. Zamolodchikov,
{\sl Expectation values of local fields in Bullough-Dodd model and 
integrable perturbed conformal field theories}, 
Nucl. Phys. B 516 (1998) 652--674 \pre{hep-th/9709034}.}

\lref\Privman{V. Privman, L. S. Shulman, {\sl Analytic Continuation at 
First-Order Phase Transitions}, J. Stat. Phys. 29 (1982) 205--229.}

\lref\Bugrij{A. I. Bugrij, {\sl The Correlation Function in Two 
Dimensional Ising Model on the Finite Size Latice. I.}, 
\pre{hep-th/0011104}; {\sl Form Factor Representation of the Correlation 
Function of the Two Dimensional Ising Model on a Cylinder}, 
\pre{hep-th/0107117}.}

\lref\YangLee{C. N. Yang, T. D. Lee, {\sl Statistical Theory of Equations 
of State and Phase Transitions. I. Theory of Condensation}, Phys. Rev. 87 
(1952) 404--409.}

\lref\LeeYang{T. D. Lee, C. N. Yang, {\sl Statistical Theory of Equations 
of State and Phase Transitions. II. Lattice Gas and Ising Model},
Phys. Rev. 87 (1952) 410--419.}

\lref\Fisher{M. E. Fisher, {\sl Yang-Lee Edge Singularitry and $\varphi^3$
Field Theory}, Phys. Rev. Lett. 40 (1978) 1610--1613.}

\lref\AlZa{Al. B. Zamolodchikov, {\sl Thermodynamic Bethe Ansatz for RSOS 
Scattering Theories}, Nucl. Phys. B 358 (1991)  
497--523.}

\lref\Gunther{N. J. G\"unther, D. A. Nicole, D. J. Wallace, {\sl Goldstone 
model in vacuum decay and first-order phase transitions}, J. Phys. A 13 
(1980) 1755--1767.}

\lref\LW{M. J. Lowe, D. J. Wallace, {\sl Instantons and the Ising model 
below $T_c$}, J. Phys. A 13 (1980), L381--L385.}

\lref\EH{J. W. Essam, D. L. Hunter, {\sl Classical behaviour of the Ising 
model above and below the critical temperature}, J. Phys. C 1 (1968) 
392--407.}

\lref\Harris{C. K. Harris, {\sl The Ising model below $T_c$: calculation 
of non-universal amplitudes using a primitive droplet model}, J. Phys. A 
17 (1984) L143--L419.}

\lref\Delfino{G. Delfino, {\sl Universal amplitude ratios in the 
two-dimensional Ising model}, Phys. Lett. B 
419 (1998) 291--295 \pre{hep-th/9710019}.}

\lref\Karowski{B. Berg, M. Karowski, P. Weisz, {\sl Construction of 
Green's functions from an exact $S$ matrix}, Phys. Rev. D 19 (1979) 
2477--2479.}

\lref\Lebowitz{J. L. Lebowitz, O. Penrose, {\sl Rigorous Treatment of the 
Van Der Waals-Maxwell Theory of the Liquid-Vapor Transition}, J. 
Math. Phys. 7 (1966) 98--113.}

\lref\MussardoA{G. Delfino, G. Mussardo, {\sl The Spin-Spin Correlation 
Function in the Two-Dimensional Ising Model in a Magnetic Field at 
$T=T_c$}, Nucl. Phys. B 455 (1995) 724--758 \pre{hep-th/9507010}.}

\lref\AlZb{Al. B. Zamolodchikov, {\sl Thermodynamic Bethe Ansatz in 
Relativistic Models: Scaling 3-State Potts and Lee-Yang Models}, 
Nucl. Phys. B 342 (1990) 695--720.}

\lref\AlZc{Al. B. Zamolodchikov, {\sl Mass Scale in the Sine-Gordon Model 
and its Reductions}, \hbox{Int. J. Mod. Phys.} A 10 (1995) 1125--1150.}

\lref\Zinn{S. Zinn, S.-N. Lai, M. E. Fisher, {\sl Renormalized coupling 
constants and related amplitude ratios for Ising systems}, Phys. Rev. E 54 
(1996) 1176--1182.}

\lref\Balog{J. Balog, M. Niedermaier, F. Niedermayer, A. Patrascioiu, 
E. Seiler, P. Weisz, {\sl The Intrinsic Coupling in Integrable Quantum 
Field Theories}, Nucl. Phys. B 583 (2000) 614--670 \pre{hep-th/0001097}.}

\lref\FLZ{P. Fonseca, S. Lukyanov, A. Zamolodchikov, to be published.}

\lref\LeClair{A. LeClair, {\sl Spectrum generating affine Lie algebras 
in massive field theory}, Nucl.~Phys. B 415 (1994) 734--780 
\pre{hep-th/9305110}.}

\lref\KadanoffCeva{L. P. Kadanoff, H. Ceva, {\sl Determination of an 
Operator Algebra for the Two-Dimensional Ising Model}, 
Phys. Rev. B 3 (1971) 3918--3939.}

\lref\YuZ{V. P. Yurov, Al. B. Zamolodchikov, {\sl Correlation Functions of 
Integrable 2D models of Relativistic Field Theory; Ising Model}, 
Int. J. Mod. Phys. A 6 (1991) 3419--3440.}

\lref\ZinnJustin{J. Zinn-Justin, {\sl Quantum Field Theory and Critical 
Phenomena}, Oxford University Press, 1989.}

\lref\Andreev{A. F. Andreev, {\sl Singularity of Thermodynamic 
Quantities at a First Order Phase Transition Point}, Sov. Phys. JETP 18 
(1964) 1415--1416.}

\lref\FisherB{M. E. Fisher, {\sl University of Colorado, Boulder, Summer 
School Lectures, 1964}, University of Colorado Press, Boulder, 1965; 
{\sl Proc. Centennial Conf. Phase Transformations}, Univ. Kentucky, 1965.}

\Title{\vbox{\baselineskip12pt\hbox{RUNHETC-2001-37}}}
{
\vbox{
\centerline{Ising field theory in a magnetic field:}
\vskip2pt
\centerline{analytic properties of the free energy}
}}
\medskip
\centerline{P. Fonseca
and A. Zamolodchikov} \medskip
\centerline{\sl Department of Physics and Astronomy, Rutgers University}
\centerline{\sl Piscataway, NJ 08855-0849, USA}
\vskip 0.6in
\noindent
We study the analytic properties of the scaling function
associated with the $2D$ Ising model free energy in the critical domain
$T\to T_c\,$, $H\to 0$.  The analysis is based on numerical data 
obtained through the Truncated Free Fermion Space Approach. We determine 
the discontinuities across the Yang-Lee and Langer branch
cuts. We confirm the standard analyticity assumptions and propose
``extended analyticity''; roughly speaking, the latter states that 
the
Yang-Lee branching point is the nearest singularity under Langer's
branch cut. We support the extended analyticity by evaluating numerically
the associated ``extended dispersion relation''. 

\Date{12/2001}

\newsec{Introduction}

The $2D$ Ising model is one of the best studied systems in 
statistical mechanics. Nonetheless, some questions concerning its 
criticality, notably in the presence of an external field $H$, remain open.
The Ising model free energy exhibits a singularity at the critical point 
$H=0$, $T=T_c$. The singularity is described in terms of the Euclidean 
quantum field theory known as the Ising Field Theory (IFT). It can be 
defined as a perturbed conformal field theory through the action
\eqn\IFT
{
{\cal A}_{\rm IFT} = {\cal A}_{(c=1/2)} + \tau\,\int\,\epsilon(x)\,d^2 x +
h\,\int \,\sigma(x)\,d^2 x\,,
}
where ${\cal A}_{(c=1/2)}$ stands for the action of $c=1/2$ conformal
field theory of free massless Majorana fermions, $\sigma(x)$ and
$\epsilon(x)$ are primary fields of conformal dimensions $1/16$ 
and~$1/2$. To be precise, we assume that the normalizations 
of these fields are fixed by the usual CFT convention,
\eqn\Norm
{
|x|^2 \,\langle \epsilon(x)\epsilon(0)\rangle \to 1\,; \qquad 
|x|^{1/4}\,\langle \sigma(x)\sigma(0)\rangle \to 1\, \quad {\rm as}
\quad |x| \to 0\,.
}
Under this normalization, the parameters $\tau$ and $h$ in \IFT\ have 
mass dimensions $1$ and~$15/8$, respectively. These parameters represent 
a deviation from the Ising model critical point, 
\eqna\Latt
$$\eqalignno{
\tau &= C_{\tau}\,\Delta T \,\(1 + O\({\Delta T}, H^2\)\)\,,&\Latt a
\cr
h &= C_h \, H \,\(1 + O\(\Delta T, H^2\)\)\,,&\Latt b
}$$
where $\Delta T = 1-T/T_c$, and real positive constants 
$C_{\tau}, C_h$ (as well
as the higher-order terms in the above relations) depend on the
details of the microscopic (lattice) interaction\foot{For the
Ising models with nearest-neighbour interactions on square and triangular
lattices, specific values of these constants can be found in
\refs{\VicC,\VicD}; for higher order terms in Eqs.~(1.3) see~\refs{\VicE, 
\VicD}.}. 
The leading singular
part ${\cal F}_{sing}(T,H)$ of the Ising model specific free energy is 
universal, and it coincides with the vacuum energy density of the IFT 
\IFT; 
in what follows we use the notation $F(2\pi\tau, h)$ for this quantity. 
It can be expressed through a universal scaling function $\Phi(\eta)$ of 
a single variable - the scaling parameter
\eqn\etatau
{
\eta = 2\pi\tau/h^{8/15}
}
(see Eq.~\Phifunction\ below for the precise definition of $\Phi(\eta)$). 
This scaling function is of much interest as it controls 
all thermodynamic properties of the Ising model in the critical
domain. Although there are many exact 
results (obtained through exact solutions of \IFT\ at $h=0$ and all $\tau$ 
\refs{\Ons, \Kaufman, \Yang, \McCoyB, \McCoyA}, and at $\tau=0$ 
and all $h$ \refs{\Me, \Fateev}; these data are collected in \Delfino) as 
well as much numerical data \refs{\EH,\Zinn,\VicC,\VicA,\VicB}
about this function, its complete analytic characterization is still lacking. 

In this work we report preliminary results of a numerical study
of the analytic \hbox{properties} of this scaling function. We use certain 
modification of the well-known Truncated Conformal Space Approach 
(TCSA) \refs{\TCSAa,\TCSAb}, which we call Truncated Free-Fermion Space 
Approach (TFFSA). Although this modification is 
designed specificaly to treat the case of IFT (while TCSA is applicable to
a wide class of perturbed CFT), in our case it produces better accuracy by 
taking full advantage of the fact that at $h=0$ the IFT \IFT\ is a
free-fermion theory \refs{\Ons, \Kaufman}. Using this approach, we compute
numerically the scaling function $\Phi(\eta)$ for real as well as for some
complex values of $\eta$, to the accuracy sufficient to make exploration
of its analytic structure in the complex $\eta$-plane. We locate the
Yang-Lee edge singularity \refs{\YangLee,\LeeYang}, and estimate some of
its characteristics. Also, 
we study the free energy $F(2\pi\tau, h)$ at real $\tau > 0$ 
(low-temperature regime) and complex $h$; in particular, we
determine, with reasonable accuracy, the imaginary part of the metastable
branch of $F$ for small as well as large values of $h$. For small $h$ our
result is in good agreement with the prediction from the critical
droplet calculations \refs{\LangerA, \Voloshin}; moreover, we find a 
leading correction to the droplet model asymptotic. We formulate an
``extended analyticity conjecture'' which states, roughly speaking,
that the continuation of the free energy under Langer's branch cut
is analytic all the way down to the Yang-Lee singularity. We then use
our numerical data to support this conjecture by verifying the associated
``extended dispersion relation''.

The paper is arranged as follows: In Sect.~2 some details of
the TFFSA are presented. In Sect.~3 we
discuss what is known and what is expected on the analytic properties
of the free energy in the critical domain. Basic notations used
throughout the paper are introduced there. The extended analyticity
conjecture is formulated in Sect.~4, where the associated dispersion
relation is also derived. In Sect.~5 we briefly discuss the excited
(``meson'') states of IFT, and the role of the ``false vacuum''
resonance in the formation of the finite-size energy spectrum in this
theory. The relation between the width of this resonance and the
separation between the meson energy levels at the
``near-intersection'' points is presented there. Qualitative pictures
of finite-size energy levels for real and pure imaginary $h$ obtained
through the TFFSA are described in Sect.~6, where we also explain in some
detail how accurate numerical data for the scaling function are
extracted from the finite-size spectra. The data is described in
Sect.~7, and in Sect.~8 we present our analytical interpretation of it. 
That includes approximations of the discontinuities across the
\hbox{Yang-Lee} and Langer branch cuts, and numerical evaluation of the
corresponding dispersion relations. Numerical support of the extended
analyticity is also presented in this Section. In Sect.~9 we discuss
possible physical significance of the extended analyticity.

\newsec{The Truncated Free-Fermion Space Approach}

As is well known (see e.g. \Book), at zero external field the Ising model 
is equivalent to a free-fermion theory. Correspondingly, \IFT\ can be 
written as
\eqn\AIFT
{
{\cal A}_{\rm IFT} = {\cal A}_{\rm FF} + 
h\,\int\,\sigma(x)\,d^2 x \,,
}
\eqn\AFF
{ 
{\cal A}_{\rm FF} = {1\over {2\pi}}\,\int\,\left[\psi{\bar\partial}\psi +
{\bar\psi}\partial{\bar\psi} + im\,{\bar\psi}\psi\right]\,d^2 x \,, 
\qquad m = 2\pi\tau \,.
}
Here $\partial={1\over 2}\,(\partial_{\rm x}-i\partial_{\rm y})$, 
${\bar\partial}={1\over 2}\,(\partial_{\rm x}+i\partial_{\rm y})$,
where $x = ({\rm x}, {\rm y})$ are Cartesian coordinates, $\psi, 
{\bar\psi}$ are chiral components of the Majorana fermi field, and
$\sigma(x)$ is the ``spin field'' associated with this fermion.

As in standard TCSA (see e.g. Ref.~\TCSAc\ and references therein), we 
start with the IFT~\IFT\ in
finite-size geometry, with one of the two Euclidean coordinates 
compactified on a circle of circumference $R$, ${\rm x}+R \sim {\rm
x}$. If ${\rm y}$ is treated as (Euclidean) time, the
finite-size Hamiltonian associated with \AIFT\ can be written as
\eqn\Hamiltonian
{
H_{\rm IFT} = H_{\rm FF} + h\,V\,, \qquad 
V = \int_{0}^{R}\,\sigma({\rm x})\,d{\rm x}\,,
}
where $H_{\rm FF}$ is the Hamiltonian of the free-fermion theory \AFF.
We are interested in the eigenvalues of $H_{\rm IFT}$, particularly in its
ground-state energy $E_0 (R,m,h)$, because for large~$R$ one expects
to have 
\eqn\Eass
{
E_0 (R,m,h) = R\,F(m,h) + O \(\exp\(-M_1\,R\)\)\,,
}
with $M_1$ being the gap in the spectrum of 
$H_{\rm IFT}$ at $R=\infty$, i.e. the mass of the lightest particle of the
field theory \IFT. In what follows, in refering to the eigenvalues of
the Hamiltonian \Hamiltonian\ we typicaly use the notation $E(R)$, with 
the arguments $m,h$ suppressed; in particular, $E_0 (R)$ will stand for
the ground-state eigenvalue of \Hamiltonian.

The free part $H_{\rm FF}$ of the Hamiltonian \Hamiltonian\ is
diagonal in the basis of $N$-particle states of free fermions of mass 
$|m|$. 
At finite $R$ the space of states of \AFF\ splits into two sectors, the 
Neveu-Schwartz (NS) sector and Ramond (R) sector (with $\psi, {\bar\psi}$ 
antiperiodic or periodic as ${\rm x} \to {\rm x}+R$, respectively). 
In each sector the particle momenta are quantized as $p_n = 2\pi\,n/R$ 
where  $n \in \IZ+1/2$ in NS sector, and $n \in \IZ$ in R sector. 
In what follows we typicaly use the notation $n_i$ for integers, and $k_i$ 
for 
half-integers. The $N$-particle states can be obtained from the NS and R 
vacua $|0\rangle_{\rm NS}$ and $|0\rangle_{\rm R}$ by applying 
the corresponding canonical fermionic creation operators,
\eqna\Space
$$\eqalignno{
{\rm NS\ sector}&: \quad |k_1, \cdots , k_N \rangle_{\rm NS} 
= a_{k_1}^{\dagger} 
\cdots a_{k_N}^{\dagger}\,|0\rangle_{\rm NS}\, \quad k_1, \cdots, k_N 
\in \IZ+1/2\,, &\Space a
\cr
{\rm R\ sector}&: \quad |n_1, \cdots , n_N \rangle_{\rm R} 
\,= a_{n_1}^{\dagger} 
\cdots a_{n_N}^{\dagger}\,|0\rangle_{\rm R}\, \quad ~\, n_1, \cdots, n_N 
\in \IZ\,. &\Space b
}$$
The normalizations of these states are fixed (up to phases) by 
conventional anticommutators,
\eqn\ac
{
\{a_k , a^{\dagger}_{k'}\} = \delta_{k,k'}\,, \qquad 
\{a_n , a^{\dagger}_{n'}\} = \delta_{n,n'}\,. 
}
In all cases the energies associated with the $N$-particle states, 
$E_N(R)$, have 
standard form
\eqna\Freelev
$$\eqalignno{
E_{N\,({\rm NS})}(R) &= E_{0\,({\rm NS})}(R) + 
\sum_{i=1}^{N}\,\omega_{k_i}(R)\,,&\Freelev a
\cr
E_{N\,({\rm R})}(R) &= E_{0\,({\rm R})}(R) + 
\sum_{i=1}^{N}\,\omega_{n_i}(R)\,,&\Freelev b
}$$
where
\eqn\omegas
{
\omega_k (R) = \sqrt{m^2 +(2\pi\,k/R)^2}\,; \qquad 
\omega_n (R) = \sqrt{m^2 +(2\pi\,n/R)^2}\,
}
(with positive branch of the square root taken, in particular,
$\omega_0 (R) = |m|$), and
\eqna\Fgren
$$\eqalignno{
E_{0\,({\rm NS})}(R) 
&= R\,F(m,0) - |m|\,\int_{-\infty}^{\infty}\,{{d\theta}\over
{2\pi}}\,\cosh\theta\,\log\(1+e^{-|m|R\cosh\theta}\)\,,&\Fgren a
\cr
E_{0\,({\rm R})}(R) 
&= R\,F(m,0) - |m|\,\int_{-\infty}^{\infty}\,{{d\theta}\over
{2\pi}}\,\cosh\theta\,\log\(1-e^{-|m|R\cosh\theta}\)\,,&\Fgren b
}$$
are the eigenvalues associated with $|0\rangle_{\rm NS}$ and 
$|0\rangle_{\rm R}$, respectively. The term  
\eqn\ons
{
F(m,0) = {{m^2}\over{8\pi}}\,\log m^2
}
in Eqs.~(2.9) accounts for the famous Onsager's singularity of the 
Ising free energy at \hbox{zero $h$ \Ons}. 

In order to treat the IFT with nonzero $h$ we can admit only the states
which \hbox{respect} the periodicity condition for the spin field, 
$\sigma({\rm x}+R,{\rm y})=+\sigma({\rm x}, {\rm y})$. This condition brings 
a distinction between
the cases $m>0$ (the ``high-T regime'') and $m<0$ (the 
\hbox{``low-T regime''}). The admissible states are
\eqna\GSO
$$\eqalignno{
m>0:& \qquad {\rm NS}-states \  with\ N \ even\ , \quad 
{\rm and}\quad {\rm R}-states \ with \ N \  even\ ,&\GSO a
\cr
m<0:& \qquad {\rm NS}-states \  with\ N \ even\ , \quad 
{\rm and}\quad {\rm R}-states \ with \ N \  odd\ .&\GSO b
}$$
(at $m=0$ the odd-$N$ states in the ${\rm R}$ sector can be viewed as 
even-$N$ states, with one $a_0$ particle added). All the above statements are 
well-known (see e.g. \Book).

The operator $h\,V$ in the full Hamiltonian \Hamiltonian\ generates 
transitions 
between the states in NS and R sectors. Fortunately, all its matrix
elements between the above states are known exactly. They are related
in a simple way to the finite-size formfactors of the field~$\sigma(x)$, 
for which an explicit expression exists,
\eqn\fs
{
\eqalign{
_{\rm NS} \langle k_1, k_2, \cdots, k_K |&\sigma(0,0)| n_1, n_2, \cdots,
n_N \rangle_{\rm R}
\cr
&=S(R)\,\prod_{j=1}^{K}\,
{\tilde g}(\theta_{k_j})\,\prod_{i=1}^{N}\,g(\theta_{n_i})\,
F_{K,N}(\theta_{k_1}, \cdots \theta_{k_K}| \theta_{n_1}, \cdots, 
\theta_{n_N})\,,
}}
where $\theta_n$ ($\theta_k$) stand for the finite-size rapidities 
related to the integers $n$ (half-integers $k$) by the equations
\eqn\thetaka
{
|m|R\,\sinh\theta_k = 2\pi\,k\,, \qquad |m|R\,\sinh\theta_n = 2\pi\,n\,.
}
In \fs\ $F_{K,N}$ is the well-known spin-field 
formfactor in infinite-space \Karowski,
\eqn\ffac
{\eqalign{
&F_{K,N}(\theta_1, \cdots, \theta_K | \theta_1', \cdots, \theta_N')
\cr
&=i^{\left[{K+N\over2}\right]}~
{\bar\sigma}\prod_{0<i<j\leq K}\!\!\tanh\({{\theta_i - \theta_j}\over 
{2}}\)\ \prod_{0<p<q\leq N}\!\!
\tanh\({{\theta_p' - \theta_q'}\over {2}} \)\ 
\prod_{{0<s\leq K}\atop { 0<t\leq N}}\!\!
\coth\({{\theta_s - \theta_t'}\over {2}}\)\,,
}}
where
\eqn\sigmabar
{
{\bar\sigma} = {\bar s}\,|m|^{1/8}\,, \qquad {\bar s} = 2^{1/12}\,e^{-1/8}\,
A^{3/2} = 1.35783834...
}
(with $A$ standing for the Glaisher's constant), 
and the rest of the factors represent \hbox{finite-size} effects. 
The overall factor $S(R)$ is essentially the vacuum-vacuum 
matrix element
\eqn\onepoint
{
{\bar\sigma}\,S(R)=
\left\{\eqalign{
_{\rm NS} \langle 0|\sigma(0,0)|0\rangle_{\rm R}
&\qquad\hbox{for}~m>0\,,
\cr
_{\rm NS} \langle 0|\mu(0,0)|0\rangle_{\rm R}
&\qquad\hbox{for}~m<0\,,
}\right.
}
where $\mu(x)$ is the usual dual spin field \KadanoffCeva, and $S(R)$ is 
given by
\eqn\onepointS
{
S(R) = 
\exp\left\{{{(mR)^2}\over 2}
\!\!\!\int\!\!\!\!\int_{-\infty}^\infty
{{d\theta_1 d\theta_2}\over{(2\pi)^2}}
{{\sinh\theta_1\,\sinh\theta_2}\over{\sinh(mR\cosh\theta_1)\,
\sinh(mR\cosh\theta_2)}}\log\left|
\coth{{\theta_1 - \theta_2}\over 2}\right|\right\},
}
which was obtained in \Sachdev. The momentum-dependent leg factors $g$ and 
${\tilde g}$ are
\eqn\gnka
{
g(\theta)=e^{\kappa(\theta)}/\sqrt{|m|R\cosh\theta}\,, \qquad 
{\tilde g}(\theta)=e^{-\kappa(\theta)}/\sqrt{|m|R\cosh\theta}\,,
}
where 
\eqn\legs
{
\kappa(\theta) = \int_{-\infty}^{\infty}\,{{d\theta'}\over{2\pi}}\,
{1\over{\cosh(\theta-\theta')}}\,\log\({{1-e^{-|m|R\cosh\theta'}}\over 
{1+e^{-|m|R\cosh\theta'}}}\)\,.
}
The phase factors $i^{\left[{K+N\over2}\right]}$ (where $[...]$ denotes 
the integer part of the number) appearing in~\ffac\ can be removed by 
an appropriate phase rotation of the states, and thus play no role in the 
TFFSA computations.

The above expression \fs\ can be extracted from the result of recent 
papers \Bugrij. 
In fact, we have obtained it independently, before 
\Bugrij\ appeared, by
a different approach. As our derivation seems to be simpler, we outline
it in Appendix A. This expression is the~$m\neq 0$ generalization of 
corresponding massless matrix elements used in \TCSAb\ in the TCSA study 
of IFT with $m=0$.

It is useful to note that the Hamiltonian \Hamiltonian\ can be rewritten 
to make its scaling form explicit,
\eqn\Hscal
{
H_{\rm IFT} = E_{0\,({\rm NS})}(R) + |m|\,H_0 (r) + |m|\,\xi\,H_{\sigma} 
(r)\,,
}
where
\eqn\xiparam
{
\xi = h/|m|^{15/8} 
}
and the operators $H_0 (r)$ and $H_{\sigma}(r)$ 
(corresponding to the terms $H_{\rm FF}$ and $hV$ in \Hamiltonian) 
depend on a dimensionless parameter $r = |m|R$ only.

The bulk energy density $F(m,h)$ can be extracted from the
large-$R$ asymptotic behaviour \Eass\ of the ground-state eigenvalue of 
\Hamiltonian. As follows from \Hscal, it has the form
\eqn\G
{
F(m,h) = {{m^2}\over{8\pi}}\,\log m^2 + m^2\,G(\xi),
}
where the first term is inherited from \ons; the scaling function $G(\xi)$ 
(and related function~$\Phi(\eta)$ defined in \Phifunction\ below) is the 
main object of our interest in this paper.

It is hardly possible to diagonalize the Hamiltonian \Hamiltonian\ 
exactly. To render it tractable by numerical methods we follow the idea of 
TCSA, i.e. we use finite-dimensional approxi-mations (``truncations'') of 
the infinite dimensional space of the states (2.5), which include only
the states of sufficiently low energy. The truncated Hamiltonian 
can be diagonalized numericaly, yielding an approximation to 
the ground-state energy function $E_0 (R)$; the free energy is then
extracted from its behaviour at sufficiently large $R$, according 
\hbox{to \Eass}. More details about this analysis are presented in 
Sect.~6.

\newsec{Analytic properties of the scaling function}

In this section we discuss known results and expected analytic 
properties of the scaling function in \G.

In fact, to describe the free energy $F(m,h)$ for both positive and 
negative $m$ one needs two scaling functions $G(\xi)$ in \G. Although 
these
two functions are analytically related (see Sect.~3.3 below) we use 
separate notations $G_{low}(\xi)$ for $m>0$ and $G_{high}(\xi)$ for $m<0$; 
these functions describe the free energy in low-T and high-T regimes, 
respectively. Both functions are defined so that $G_{low}(0) = 
G_{high}(0) =0$.

\subsec{The function $G_{high}(\xi)$}

This function is even, i.e. $G_{high}(\xi)=G_{high}(-\xi)$. Around  
$\xi=0$ it can be represented by a convergent  power series in $\xi^2$,
\eqn\highexp
{
G_{high}(\xi) = G_2\,\xi^2 + G_4\,\xi^4 + G_6\,\xi^6 + \cdots\,.
}

\table{1}{Numerical values of the coefficients $G_{2n}$ in \highexp. The
first column contains the data from Refs.~\McCoyB\ and \VicB. The results
obtained through high-T dispersion relation, Eq.~\gIIn, with 
the use of our approximation \empirical, are presented in the second 
column.}
{\vbox{
\offinterlineskip
\halign{
\strut
\hfil$#$\hfil&
\enskip\vrule#\enskip&
$#$\hfil&
$~#$\hfil&
\enskip\vrule#\enskip&
\hskip 12pt $#$\hfil
\crcr
&&\hfil\qquad\hbox{\rm From}&&&\hfil\hskip -12pt\hbox{\rm From high-T}
\cr
&&\hfil\qquad\hbox{\rm References}&&&\hfil\hskip -12pt\hbox{\rm dispersion 
relation}
\cr
\omit&height2pt&&&height2pt&
\cr
\noalign{\hrule}\omit&height2pt&&&height2pt&
\cr
G_2 && -1.8452280...&^{\rm \McCoyB} && -1.8452283\cr
G_4 && \p 8.33370(1) & ^{\rm \VicB} && \p 8.33410\cr
G_6 && -95.1689(4) & ^{\rm \VicB} && -95.1884\cr
G_8 && \p 1457.55(11) & ^{\rm \VicB} && \p 1458.21\cr
G_{10} && -25884(13) & ^{\rm \VicB} && -25889 \cr
G_{12} && \p 5.03(1)\times10^5 & ^{\rm \VicB} && \p 5.02\times 10^5\cr
G_{14}&& \hfil\hbox{\rm ---} & && -1.04\times10^7\cr
}}}

This series coincides with the perturbative expansion in $h$ in the 
field theory \IFT. The first coefficient $G_2$ is known through this 
perturbation theory exactly \McCoyB, and the coefficient $G_4$ was 
obtained with
high accuracy by using formfactor expansion of the \hbox{four-spin}
correlation function \Balog. There is a substantial amount of numerical 
data on the further coefficients. What appears to be rather accurate 
estimates of $G_{2k}$ up to $G_{12}$ are presented in a recent paper 
\VicB. The data on these coefficients are collected in Table~1. 

The function $G_{high}(\xi)$ can
be analyticaly continued to complex values of $\xi$. The \hbox{Yang-Lee}
theory \refs{\YangLee,\LeeYang}
guarantees analyticity of $G_{high}(\xi)$ in the whole complex $\xi$ plane
with possible exception of the imaginary axis. At the imaginary axis one 
expects to observe branch cuts resulting from condensation  of the Yang-Lee
zeroes of the partition function in the thermodynamic limit. These branch cuts 
extend from $i\xi_0$ to $i\infty$ and from \hbox{$-i\xi_0$ to~$-i\infty$}, 
as
shown in Fig.~1, where $\xi_0$ is some positive constant whose numerical 
value will be estimated in the Sect.~6,
\eqn\xizero
{
\xi_0 = 0.18930(5)\,. 
}
\vskip -12pt
\nobreak
\midinsert
\centerline{
\hbox to \hsize{ 
\hfil
\vbox{
\hbox{\hskip 0.5cm \epsfysize=5.5cm\epsfbox{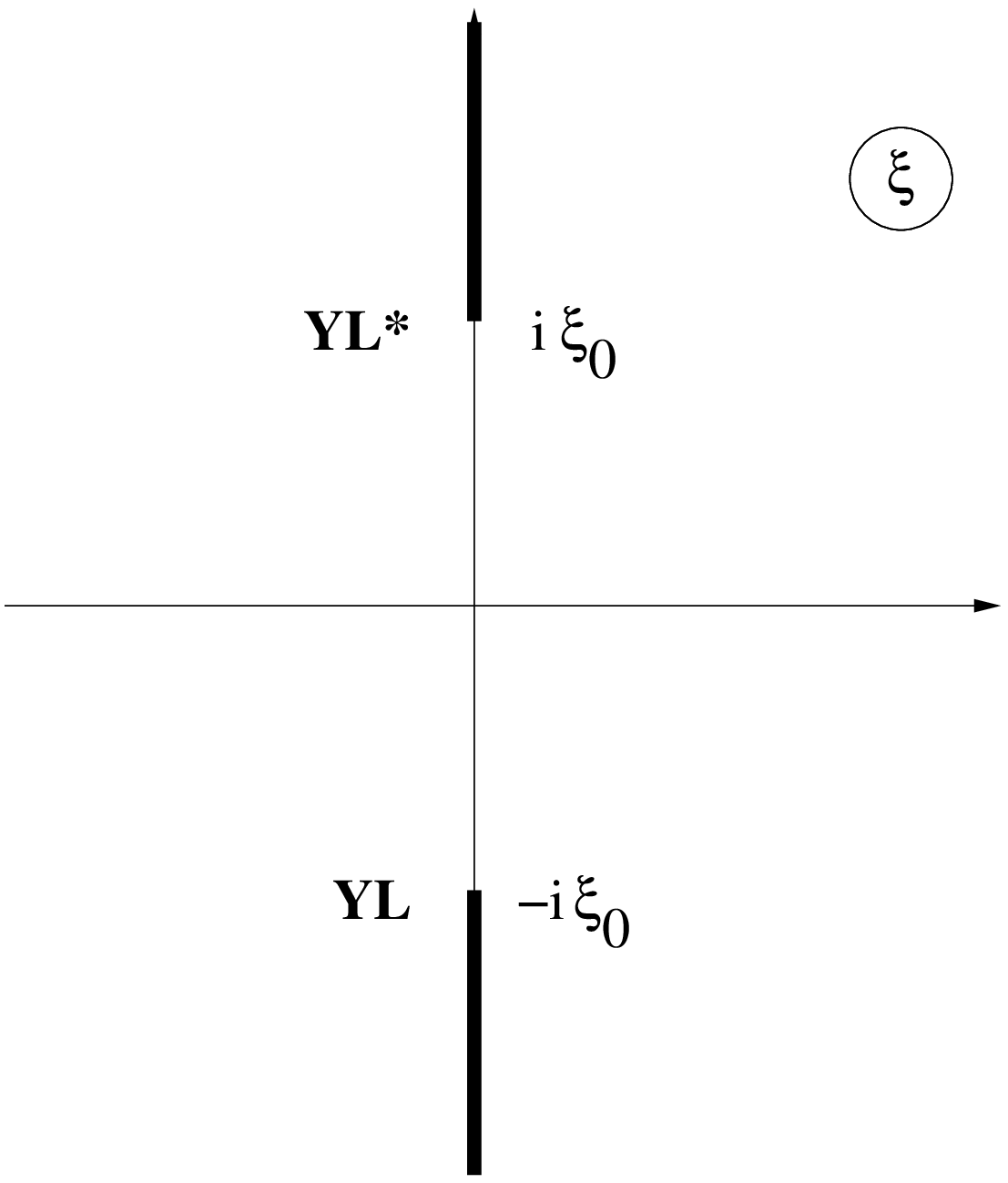}}
\vskip 10pt
\vbox to 54.75pt{\hsize=6cm \baselineskip=12pt \ninerm\ninemath
\noindent
{\nineb Fig.~1:} Analytic structure of $G_{high}(\xi)$ 
in the complex $\xi$ plane. YL and YL$^*$ denote the Yang-Lee edge 
singularities; associated branch cuts are shown as solid lines.
\vfil}}
\hskip 0.5cm
\vbox{
\hbox{\hskip 0.5cm\epsfysize=5.5cm\epsfbox{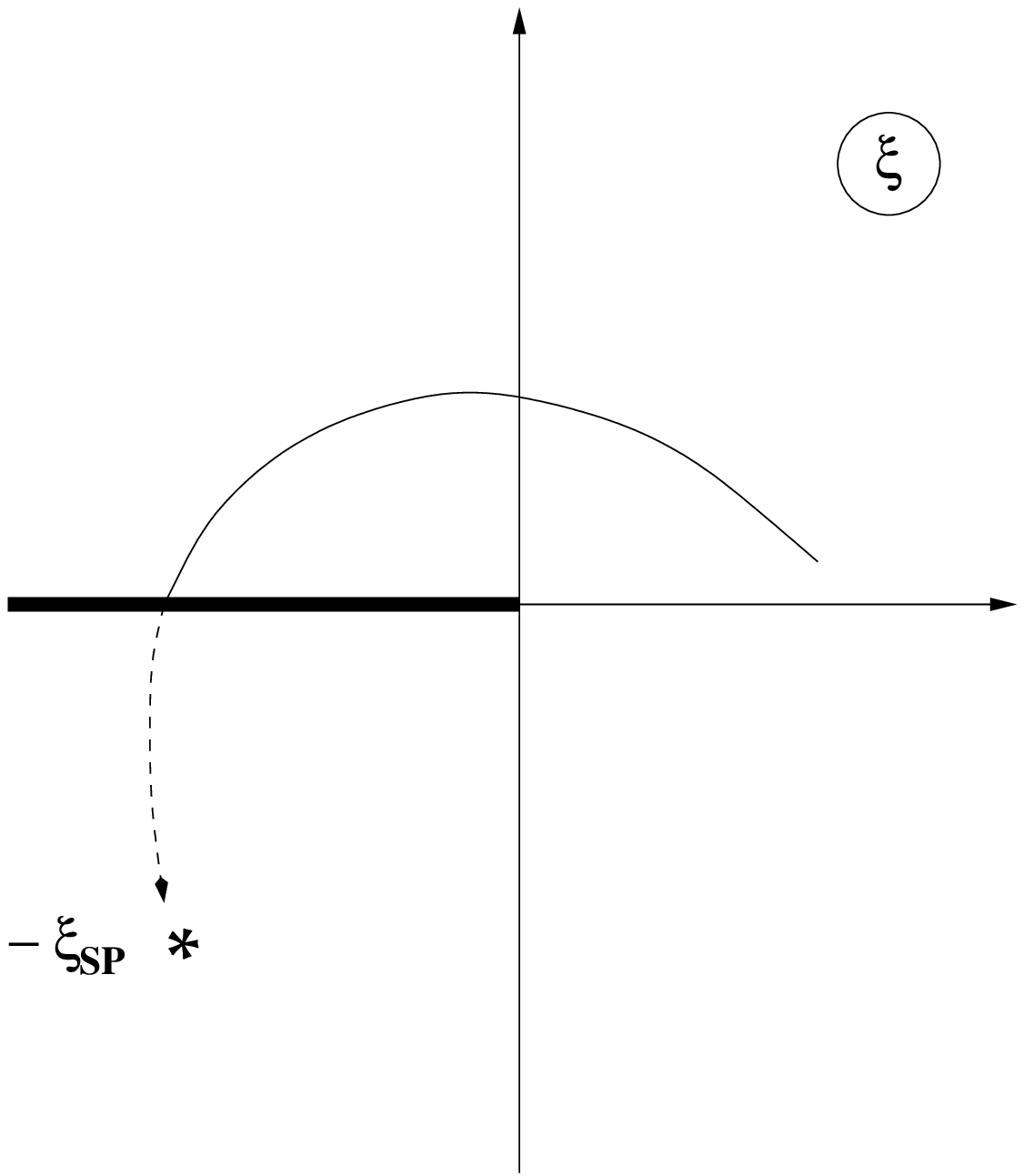}}
\vskip 10pt
\vbox to 54.75pt{\hsize=6cm \baselineskip=12pt \ninerm\ninemath
\noindent
{\nineb Fig.~2:} Analytic structure of {$G_{low}(\xi)$}. The solid 
line is the Langer's branch cut. The star denotes the nearest singularity 
under this branch cut; its significance is discussed in Sect.~9.\vfil}}
\hfil
}
}\endinsert
\goodbreak

The branching points $\pm i\xi_0$ represent
the Yang-Lee edge singularity \YangLee. Combining these analytic 
properties with
the asymptotic behaviour $G_{high}(\xi) \simeq \xi^{16/15}$ (see 
Eq.~\ghighass\ below) one can derive the dispersion relation \LeeYang
\eqn\exphigh
{
G_{high}(\xi) = - \xi^2\,\int_{\xi_{0}}^{\infty}\,
{{2\,\Im m\,G_{imh}(t)}\over{t\,(t^2 + \xi^2)}}\,{{dt}\over {\pi}}\,,
}
where the imaginary part of the function
\eqn\rhohigh
{
G_{imh}(t) \equiv G_{high}(-it+0)
}
relates in the usual way to the discontinuity across either of the
branch cuts in Fig.~1. In particular, the coefficients $G_{2n}$ in 
\highexp\ can be expressed in terms of this function as
\eqn\gIIn
{
G_{2n} = (-)^{n}\,\int_{\xi_{0}}^{\infty}\,
{{2\,\Im m\,G_{imh}(t)}\over{t^{2n+1}}}\,{{dt}\over {\pi}}\,.
}
In what follows we will refer to \exphigh\ as the high-T dispersion 
relation. 

As is known \Fisher, the Yang-Lee edge singularity is a critical
point, and the associated CFT was identified in \Cardy\ as the
(nonunitary) minimal model with central charge~\hbox{$c_{\rm YL} = -
22/5$}. This CFT has only one relevant operator, the primary field
of conformal dimension $\Delta_{\rm YL} = - 1/5$. Therefore, one
expects the singularities of $G_{high}(\xi)$ at~$\xi = \pm\,i\,\xi_0$
to be of the form
\eqn\LYsingular
{
G_{high}(\xi) = {\cal G}_A (\xi) + \(\xi_{0}^2 + \xi^2\)^{5\over
 6}\,{\cal G}_B (\xi) + \(\xi_{0}^2 + \xi^2\)^{5\over
 3}\,{\cal G}_C (\xi)+
{\rm subleading\ singular \ terms}\,,
}
where the functions ${\cal G}_A(\xi), {\cal G}_B (\xi), {\cal G}_C
(\xi), \ldots\,$, are regular at $\xi = \pm\,i\,\xi_0$. To some degree, 
the singular terms in \LYsingular\ can be understood in terms of the 
low-energy {\it effective} action 
\eqn\LYeff
{
\eqalign{
{\cal A}_{\rm eff} =&\, {\cal A}_{\rm LY\ CFT} +
i|m|^{12\over 5}\,\lambda(\xi)\,\int\,\phi_{-1/5}(x)\,d^2 x + 
{{\alpha(\xi)}\over{(2\pi\,m)^2}}\,\int\,
\(T{\bar T}\)(x) \,d^2 x 
\cr
&+ {\rm higher\ irrelevant \ operators}\,,
}}
where $\phi_{-1/5}$ is the relevant primary field mentioned above,
and the rest of the action contains the contributions of irrelevant
operators, the field $(T{\bar T})$ (i.e. the $L_{-2}{\bar L}_{-2}$
descendent of the identity operator) being the lowest of such operators.
The dimensionless coupling constants $\lambda(\xi), \alpha(\xi), 
\ldots\,$, are certain
functions of the scaling parameter $\xi$, universal in the sense that
they are uniquely determined by the original field theory \IFT\
(but taken as ``input'' data in the effective theory \LYeff); 
all these functions are expected to be 
regular at $\xi = \pm\,i\,\xi_0$, and the critical point(s) $\pm i\xi_0$ is
defined by the condition $\lambda(\pm i \xi_0) = 0$. Thus, in the
vicinity of, say, $-i\xi_0$
\eqna\alphazero
$$\eqalignno{
\lambda(\xi) &= \lambda_1\,\(\xi_0^2+\xi^2\) + 
O\(\(\xi_0^2+\xi^2\)^2\)\,;&\alphazero a
\cr
\alpha(\xi) &= \alpha_0 + O\(\xi_0^2+\xi^2\)\,; &\alphazero b
}$$
and similarly for the other couplings.

At $\xi = -i\xi_0$ (as well as at $\xi = i\xi_0$) the field theory
\IFT\ describes the Renormalization Group flow from the Ising model
fixed point with $c=1/2$ down to the Yang-Lee fixed point with
$c_{\rm YL}=-22/5$. In this case the mass gap $M_1$ vanishes, and the
finite-size energy levels are expected to approach the linear
asymptotic \Eass\ with a power-like, not exponential,
accuracy. These power-like corrections to the linear asymptotic \Eass\
can be analysed through the perturbation theory based on the effective
action \LYeff. In particular, for the ground-state level of \Hscal\
with $\xi=\pm i\xi_0$ one obtains
\eqn\Ezeroflow
{
E_0 (R) = F_0 \,R + {{\pi}\over{6\,R}}\(- c_{\rm eff} +
{{\pi\,c_{\rm eff}^2\,\alpha_0}\over{24\,(mR)^2}} -
{{2\pi^2\,c_{\rm eff}^3\,\alpha_0^2}\over{(24)^2\, (mR)^4}} + 
O\left(R^{-{28\over 5}}\right)\)\,,
}
where $c_{\rm eff}=c_{\rm YL}-24\Delta_{\rm YL} = 2/5$ is the 
``effective central charge'' of the Yang-Lee CFT, \hbox{$\alpha_0$ is} the 
leading coefficient in \alphazero b, and $F_0$ stands for the 
free energy $F_0$ associated with this flow; the last quantity is related 
to the value of the scaling function $G_{high}(\xi)$ at the singular 
point,
\eqn\Fzero
{
F_0 = F(m,0) + m^2\,G_{high}(\pm i\,\xi_0) \,.
}
The calculation leading to \Ezeroflow\ is similar to that presented in 
\AlZa; we skip it in this report.

\subsec{The function $G_{low}(\xi)$}

Unlike $G_{high}(\xi)$ above, the analytic continuation of 
$G_{low}(\xi)$ is not an even function of $\xi$. More precisely, for 
$m > 0$ the analytic continuation in $h$ of the
free energy \G\ yields two different analytic functions, depending on
whether one continues from positive or negative parts of the real axis.
In what follows, by $G_{low}(\xi)$ we always understand analytic 
continuation of the free energy from {\it positive} part of the real 
$\xi$ axis. 
Again, according to the Yang-Lee theory, $G_{low}(\xi)$ is analytic in the 
right half-plane $\Re e\, \xi > 0$. 
Langer's \hbox{theory \LangerA} (as well as earlier calculations within 
``droplet models'' \refs{\Andreev, \FisherB}) predicts a weak singularity 
at
$\xi=0$, and it is usually assumed (see e.g.~\refs{\Gunther,\LW,\Harris}) 
that $G_{low}(\xi)$ is analytic in the 
full complex $\xi$-plane with the branch cut from~$0$~to~$-\infty$, 
as is shown in Fig.~2. We will call this the {\it standard 
analyticity assumption}. This assumption will be confirmed by our 
numerical analysis
in Sect.~7. The function $G_{low}(\xi)$ admits an {\it asymptotic}
expansion in powers of $\xi$,
\eqn\lowexp
{
G_{low}(\xi) \simeq {\tilde G}_1 \,\xi + {\tilde G}_2 \,\xi^2 + \cdots\,.
}
Again, the coefficients ${\tilde G}_n$ here in principle can be
computed by means of the perturbation theory in $h$ in \AIFT. Thus, 
the first coefficient is directly related to the spontaneous magnetization
at zero $h$, given by \sigmabar,
\eqn\Gonetilde
{
{\tilde G}_1 = - {\bar s}\,.
}
The coefficient ${\tilde G}_2$ is also known exactly, since its computation
involves integrating the two-spin correlation function, which is 
determined in terms of the Painlev\'e functions \McCoyB. Several further 
coefficients 
were estimated in \WuMcCoyG\ using exact 
large-distance expansions of multi-spin correlation functions at zero
$h$. Numerical estimates from lattice series analysis are also 
available \Zinn. We quote these results in Table 2.

\table{2}{Numerical values of the coefficients ${\tilde G}_n$ in
\lowexp. Exact results for ${\tilde G}_1$ and ${\tilde G}_2$, and
numerical estimates found in literature, are collected in the first
\hbox{column}. The second column shows results obtained through low-T
dispersion relation, Eq.~\dispn, using our approximation \Dmetaapprox.}
{\vbox{
\offinterlineskip
\halign{
\strut
\hfil$#$\hfil&
\enskip\vrule#\enskip&
$#$\hfil&
\enskip\vrule#\enskip&
\hskip 10pt $#$\hfil
\crcr
&&\hfil\hbox{\rm From}&&\hfil\hskip -10pt\hbox{\rm From low-T}
\cr
&&\hfil\hbox{\rm References}&&\hfil\hskip -10pt\hbox{\rm dispersion 
relation}
\cr
\omit&height2pt& &height2pt&
\cr
\noalign{\hrule}\omit&height2pt& &height2pt&
\cr
\tilde G_1 && \hfil-1.35783834... ^{\rm \Book}  && -1.35783835\cr
\tilde G_2 && \hfil-0.0489532...\hskip 1pt ^{\rm \McCoyB} && -0.0489589\cr
\tilde G_3 && \p 0.0387529\,^{\rm \WuMcCoyG}\,;~\p 0.039(1)\hskip 
5pt 
^{\rm \Zinn} && \p 0.0388954\cr
\tilde G_4 && -0.0685535\,^{\rm \WuMcCoyG}\,;~ 
\hbox{$-0.0685(2)$}  
^{\rm \Zinn} && -0.0685060\cr
\tilde G_5 && \hfil\hbox{\rm ---} && \p 0.18453\cr
\tilde G_6 && \hfil\hbox{\rm ---} && -0.66215\cr
\tilde G_7 && \hfil\hbox{\rm ---} && \p 2.952\cr
\tilde G_8 && \hfil\hbox{\rm ---} && -15.69\cr
\tilde G_9 && \hfil\hbox{\rm ---} && \p 96.76\cr
\tilde G_{10} && \hfil\hbox{\rm ---} && -6.79\times10^2\cr
\tilde G_{11} && \hfil\hbox{\rm ---} && \p 5.34\times 10^3\cr
\tilde G_{12} && \hfil\hbox{\rm ---} && -4.66\times 10^4\cr
\tilde G_{13} && \hfil\hbox{\rm ---} && \p 4.46\times 10^5\cr
\tilde G_{14} && \hfil\hbox{\rm ---} && -4.66\times 10^6\cr
}}
}

If one makes the above standard analyticity assumption (as in Fig.~2), the
coefficients~${\tilde G}_n$ can be represented as the integrals 
\LW\
\eqn\dispn
{
{\tilde G}_n = 
(-)^{n+1}\,\int_{0}^{\infty}\,{{\Im m\, G_{meta}(t)}\over{t^{n+1}}}\,
{{dt}\over \pi}\,, \quad n = 2, 3, 4, \ldots,
}
where the function
\eqn\meta
{
G_{meta}(\xi) \equiv G_{low}(-\xi + i0)
}
at positive $\xi$ describes the values of $G_{low}(\xi)$ at real
negative $\xi$, at the upper edge of the branch cut in
Fig.~2. Correspondingly, the discontinuity of $G_{low}(\xi)$ across
this branch cut equals $2i\,\Im m\,G_{meta}(-\xi)$.  
The Eq.~\dispn\ follows directly from the dispersion 
relation~\refs{\Gunther,\LW}
\eqn\lowdisp
{
G_{low} (\xi) = {\tilde G}_1\,\xi - \xi^2\,\int_{0}^{\infty}\,
{{\Im m\,G_{meta}(t)}\over{t^2\,(t + \xi)}}\,{{dt}\over
\pi}\,,
}
where the integral converges at any finite $\xi$ because $G_{low}(\xi)
\sim \xi^{16/15}$ as $\xi \to \infty$ (as follows from \aran\ and
\phiseries\ below).

The function $G_{meta}(\xi)$ deservingly attracts much attention, for
at least two reasons. One is the commonly accepted interpretation of
the function
\eqn\Gmeta
{
F_{meta}(m, h) = m^2\,\({1\over {8\pi}}\log m^2 + G_{meta}(\xi)\)\,, 
\qquad \xi = h/|m|^{15/8}\,,
}
which coincides with the analytic continuation of the free energy \G\ to 
negative values of $h$,  
as the free energy associated with the metastable state at $T<T_c$. In 
fact, to the best of our knowledge, it is the only
mathematically precise definition of the metastable free energy
available today. The function \Gmeta\ takes complex values, and its
imaginary part
\eqn\Gammam
{
\Gamma(m,h) \equiv \Im m\,F_{meta}(m,h) = m^2\,\Im
m\,G_{meta}(\xi)
}
is interpreted as the rate of decay of the
metastable state. More precisely, according to Langer's theory
\refs{\LangerA, \LangerB}, the actual rate of decay is asymptotically 
proportional (with
the constant factor absorbing the time scale) to this imaginary part
in the limit  $h \to 0$ (or $\xi \to 0$). It is tempting to assume 
(as is often
done in the literature on nucleation theory) that similar relation
extends to some finite domain of $h$, although it is understood that
the proportionality coefficient, being sensitive to at least some
details of the kinetic model, may very well depend on $h$, and its
degree of universality is not clear. Another, and much
better understood, interpretation of \Gmeta\ is in terms of the
quantum field theory \IFT\ in $1+1$ Minkowski space-time. Namely, at
$\tau >0$ and $h \neq 0$ this field theory exhibits a global
resonance state commonly refered to as the ``false vacuum'', and the 
quantity \Gmeta\ coincides
with the associated (complex) energy density; in particular, the
imaginary part of \Gmeta\ gives precisely the decay probability (per unit
volume and per unit time) of this resonance state \refs{\Kobzarev, 
\Coleman,\CC}. If $h$ is small, both of the above
interpretations allow one to justify the 
validity of the instanton saddle-point calculation, which in the $D=2$ 
case
yields  the following $h\to 0$ 
asymptotic behaviour of the imaginary part of \Gmeta\ \Voloshin\ 
(see~\refs{\LangerA, \Kobzarev, \Coleman, \CC, \Gunther} for such 
calculations in more general context),
\eqn\voloshin
{
\Gamma(m,h) \to
{{{\bar\sigma}\,h}\over{2\pi}}\,\exp\(-{{\pi\,m^2}\over
{2{\bar\sigma}h}}\) \quad {\rm as} \quad h\to 0\,,
}
where ${\bar\sigma} = {\bar s}\,|m|^{1/8}$ is the magnetization at
zero $h$ (see Eq.~\sigmabar). It is fair to say that there is still some 
controversy about whether \voloshin\ gives the correct numerical 
coefficient in the asymptotic behaviour of this imaginary part 
\Rutkevich. Our analysis in Sect.~7 is completely consistent with the 
coefficient in \voloshin, and actually offers the leading correction to 
this asymptotic (see Eq.~\discass).

\subsec{The function $\Phi(\eta)$}

Although the scaling functions $G_{high}$ and $G_{low}$ above are
defined independently, in fact they can be analytically
related one to another. To do this it is useful to introduce another
scaling variable,
\eqn\etaparam
{
\eta = m/|h|^{8/15}\,,
}
and rewrite the free energy \G\ as
\eqn\Phifunction
{
F(m,h) = {{m^2}\over{8\pi}}\,\log m^2 + |h|^{16/15}\,\Phi(\eta)\,.
}
Of course, the scaling function $\Phi(\eta)$ here is related to
the function(s) $G(\xi)$ in \G, and vice versa. If both $m$ and $h$ are
real and positive we have 
\eqn\bv
{
\eta = 1/\xi^{8/15}\,,
}
and therefore
\eqn\br
{
\Phi(\eta) = \eta^2\,G_{low}\(1/\eta^{15/8}\) \quad {\rm for\ real}
\quad \eta > 0\,.
}
Similarly, if $h > 0$ but $m < 0$ the variables \xiparam\ and \etaparam\ 
are
related as 
\eqn\av
{
\eta = - 1/\xi^{8/15}\,,
}
so that
\eqn\ar
{
\Phi(\eta) = \eta^2\,G_{high}\(1/(-\eta)^{15/8}\) \quad {\rm for\ real}
\quad \eta < 0\,.
}
On the other hand, for fixed $h \neq 0$ the free energy $F(m,h)$, as  
function of $m$, is expected to be analytic at all finite real $m$, including
the point $m=0$, for if $h\neq 0$ the correlation length remains finite even 
at $m=0$. Therefore the scaling function $\Phi(\eta)$ can be written as
\eqn\Phitilde
{
\Phi(\eta) = - {{\eta^2}\over{8\pi}}\,\log\eta^2 + {\tilde\Phi}(\eta)\,,
}
where the function ${\tilde\Phi}(\eta)$ is analytic at all finite real
$\eta$. 

\nobreak
\midinsert
\centerline{
\hbox to \hsize{
\hfil
\vbox{
\hbox{\hskip 12pt \epsfysize=6cm\epsfbox{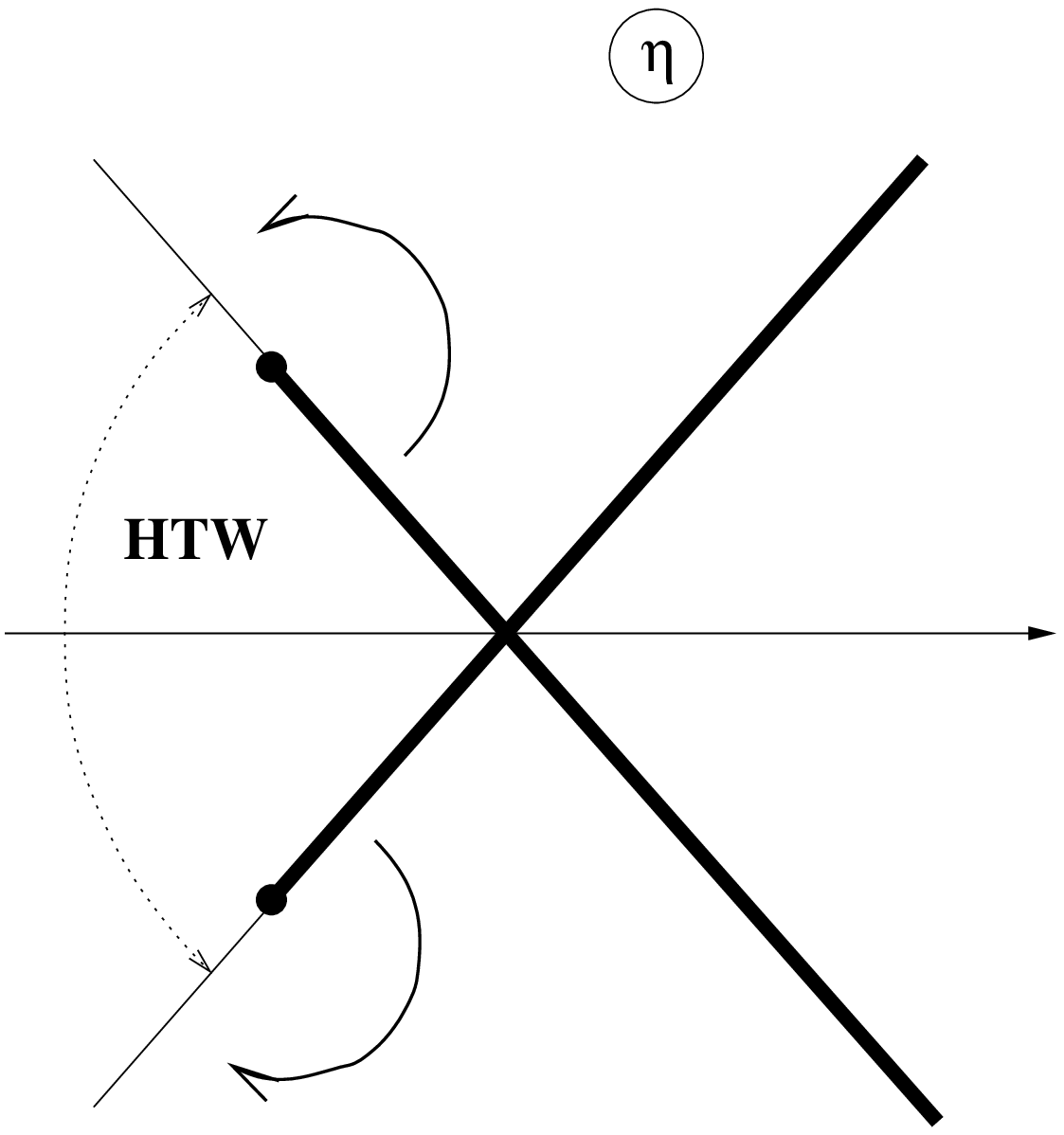}}
\vskip 6pt
\vskip 12pt
\vbox to 80pt{\hsize=6.5cm \baselineskip=12pt \ninerm\ninemath
\noindent
{\nineb Fig.~3:}
Mapping of Fig.~1 under the variable transformation \av; the wedge
HTW \hbox{$\(-{4\pi\over15} < {\rm arg}(-\eta) < {4\pi\over15}\)$}
is the image of the right half-plane in Fig.~1. Rotating the branch cuts
as suggested by the arrows opens the principal sheet of the $\eta$-plane
shown in Fig.~4.
\vfil}}
\hskip 0.5cm
\vbox{
\hbox{\hskip 12pt\epsfysize=6cm\epsfbox{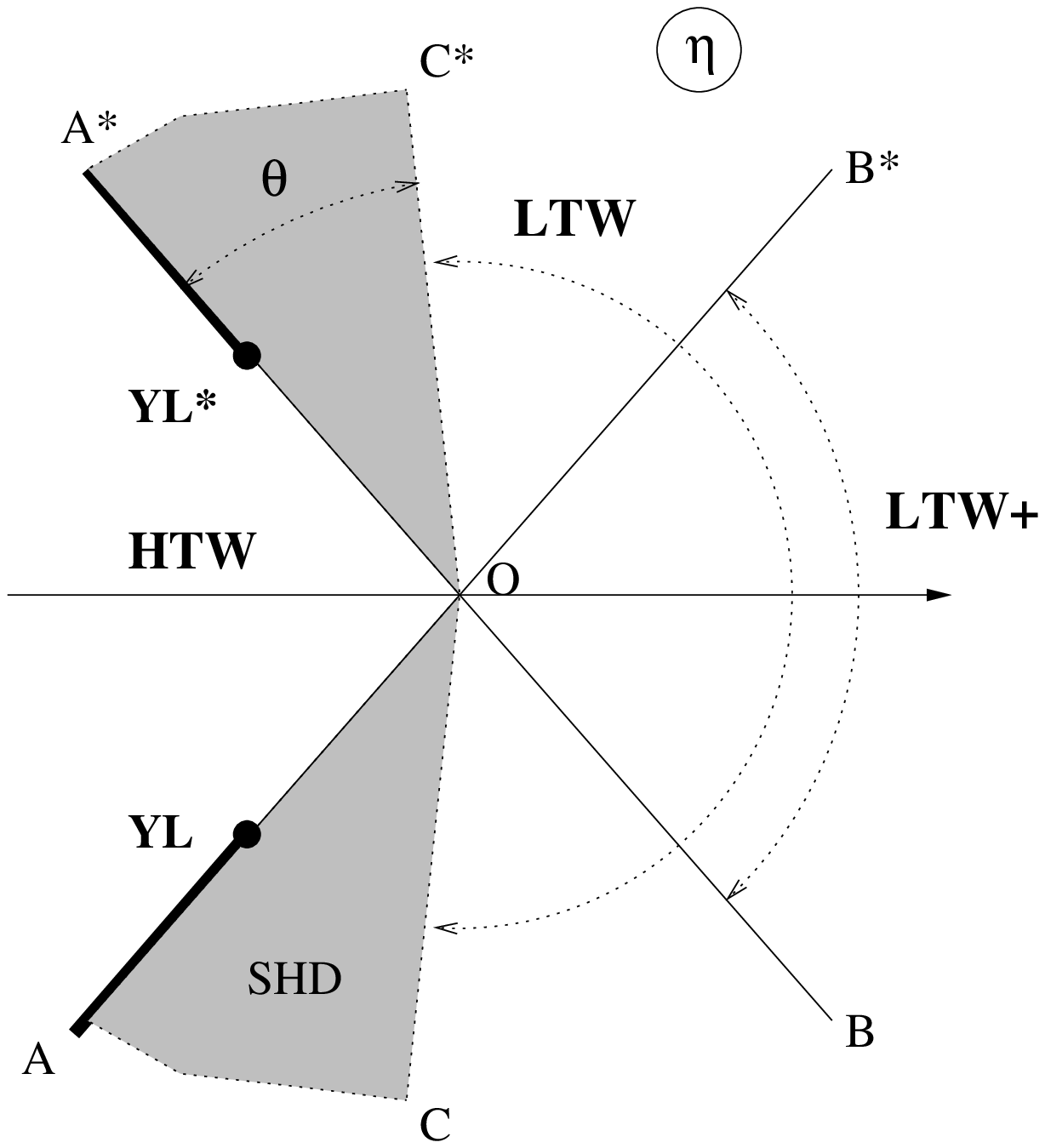}}
\vskip 12pt
\vbox to 80pt{\hsize=6.5cm \baselineskip=12pt \ninerm\ninemath
\noindent
{\nineb Fig.~4:} Principal sheet of the $\eta$-plane for
the scaling function $\tilde\Phi(\eta)$. The wedge LTW
\hbox{\ninemath $\(-{8\pi\over15} <{\rm arg}(\eta)
<{8\pi\over15}\)$} is the image of the principal sheet in Fig.~2 under the
map \bv.
Analyticity~ of $\,\tilde\Phi(\eta)\,$ in the wedge~ SHD
\hbox{$\(-{11\pi\over15}< \arg(\eta) < -{8\pi\over15}\)$} is the
subject
of the "extended analyticity" conjecture in Sect.~4.
\vfil}}
\hfil
}
}
\endinsert\goodbreak\

Eqs. \br\ and \ar\ can be promoted to analytic relations
valid for complex values of $\eta$. One notes that the variable
transformation \av\ maps the right half-plane in Fig.~1 onto the wedge
$-4\pi/15 < {\rm arg}(-\eta) < 4\pi/15$; we call it the ``High-T Wedge'', 
see~Fig.~3, where it is marked as the region HTW. Therefore the
following analytic relation holds in this wedge,
\eqn\bran
{
- {{\eta^2}\over{4\pi}}\,\log(-\eta) + {\tilde\Phi}(\eta) =
\eta^2\,G_{high} \(1/(-\eta)^{15/8}\) \quad {\rm for} \quad 
-4\pi/15 < {\rm arg}(-\eta) < 4\pi/15\,,
}
where the principle branch of the $\log$ function (i.e. $\log(y)$ is
real for real $y>0$) should be taken.
The Yang-Lee branching points are located at the points 
$\eta = -Y_0 \,e^{\pm i\,4\pi/15}$ in the complex $\eta$-plane,
where
\eqn\etazero
{
Y_0 = 1/\xi_{0}^{8/15}\,.
}

The way the corresponding branch cuts are drawn in Fig.~1 corresponds
to placing them along the rays $\eta = y\,e^{\pm i\,4\pi/15}$, with
$y$ running from $-Y_0$ to $+\infty$, as is done in
Fig.~3. However, it is convenient to rotate these branch cuts $180^o$, 
as is suggested by arrows in~Fig.~3. This way one exposes the full
``principal sheet'' of the complex $\eta$ plane, shown in Fig.~4,
where the branch cuts extend along the rays 
$\eta = y\,e^{\pm i\,4\pi/15}, -\infty < y < -Y_0$. In this picture
the way to reach the left half plane in Fig.~1 is to go from the
region HTW under either of the cuts. This complication is not
important, as in view of the symmetry $G_{high}(\xi) = G_{high}(-\xi)$
one expects to see there merely another copy of the principal sheet. 
More importantly, there is now a direct way (through some vicinity of
the real axis, where~${\tilde \Phi}(\eta)$ is analytic) into the
right-hand  part of the principal sheet. Here, in the wedge 
$-4\pi/15 < {\rm arg}(\eta) < 4\pi/15$ (the domain ${\rm LTW}_{+}$), 
which is the image of the
right half plane in Fig.~2 under the variable transformation \bv, the
function ${\tilde\Phi}(\eta)$ is related to $G_{low}(\xi)$ as
\eqn\aran
{
- {{\eta^2}\over{4\pi}}\,\log(\eta) + {\tilde\Phi}(\eta) =
\eta^2\,G_{low} \(1/\eta^{15/8}\) \quad {\rm for} \quad 
-4\pi/15 < {\rm arg}(\eta) < 4\pi/15\,,
}
where again the principle branch of the logarithm is understood. 
Thus, ${\tilde\Phi}(\eta)$ is analytic 
in the wedges HTW and ${\rm LTW}_{+}$ in Fig.~4,
where it matches the functions $G_{high}$~and~$G_{low}$ through \bran\
and \aran, plus it is analytic 
in some finite neighbourhood of the \hbox{point $\eta=0$}. Moreover, the 
transformation \bv\ maps the full principal sheet of the complex
$\xi$-plane in Fig.~2 onto twice wider ``Low-T Wedge'' 
$-8\pi/15 < {\rm arg}(\eta) < 8\pi/15$ (shown as LTW in Fig.~4),
with the upper and lower edges of the branch cut in Fig.~2 
\hbox{represented} by the rays ${\rm arg}(\eta) = \mp\,8\pi/15$. The 
standard analyticity assumption (see Sect.~3.2) implies analyticity of 
${\tilde\Phi}(\eta)$ in this wider wedge. Not unexpectedly, our numerical 
analysis in Sect.~8 will confirm this to an high degree of precision. The 
question interesting indeed is about analytic properties of this scaling 
function in the remaining part of the $\eta$-plane, the ``Shadow Domain'' 
$-11\pi/15 < {\rm arg}(\eta) < -8\pi/15$ (SHD in Fig.~4). Significance of 
this 
problem
will be discussed in Sect.~9. In Sect.~4 below we will formulate what we 
believe to be the most natural conjecture about the shadow domain 
analyticity.

In any case, ${\tilde\Phi}(\eta)$ is analytic at $\eta =0$, and in some 
finite domain around this point it can be represented as a convergent power
series,
\eqn\phiseries
{
{\tilde\Phi}(\eta) = \Phi_0 + \Phi_1\,\eta + \Phi_2\,\eta^2 + \cdots\,.
}
In fact, the first two coefficients
in series \phiseries\ are known exactly,  thanks to integrability
of the field theory \IFT\ with $\tau = 0$ \Me. The coefficient
$\Phi_0$ appears as the amplitude in its free energy,
\eqn\Fzerom
{
F(0,h) = \Phi_0\,|h|^{16/15}\,,
}
which was computed in \Fateev,
\eqn\Phizero
{
\Phi_0 = -{{\Gamma (1/3)\,\Gamma(1/5)\,\Gamma(7/15)}
\over{2\pi\,\Gamma(2/3)\,\Gamma(4/5)}\,\Gamma(8/15)}\,
\({{4\pi^2\,\Gamma^2
(13/16)\,\Gamma(3/4)}\over {\Gamma^2
(3/16)\,\Gamma(1/4)}}\)^{8/15}\,.
}
The coefficient $\Phi_1$ is related to the expectation value of the
energy density field $\epsilon(x)$ in this integrable theory,
\eqn\PhioneA
{
\langle \,\epsilon\,\rangle|_{\tau = 0} = - 2\pi\,\Phi_1\,|h|^{8/15}\,,
}
which was determined in \FLZZ,
\eqn\PhioneB
{
\langle \,\epsilon\,\rangle|_{\tau = 0}=(2.00314...)\,|h|^{8/15}\,.
}
Further coefficients can in principle be computed using perturbation
theory in $\tau$; this however is a hard problem because multipoint 
correlation functions of the $\tau=0$ theory are not known exactly (see 
however Ref.~\MussardoA). Our numerical analysis in Sect.~6 yields next 
six coefficients $\Phi_2$ to $\Phi_7$ with reasonable accuracy (see 
Table~3).

\table{3}{First few coefficients of the power series \phiseries:
Exact results from integrable field theory (first column), estimates from
TFFSA data (second column), and results obtained from the extended
dispersion relation, Eq.~(4.10), using approximation \Deltaapprox\
(third column).}
{
\vbox{
\offinterlineskip
\halign{
\strut
$#$\hfil&
\enskip\vrule#\enskip&
$#$\hfil&
\enskip\vrule#\enskip&
$#$\hfil&
\enskip\vrule#\enskip&
\hskip 10pt$#$\hfil
\crcr
&&\hfil\hbox{\rm From} && \hfil\hbox{\rm From}
&&\hfil\hskip -10pt\hbox{\rm From extended}
\cr
&&\hfil\hbox{\rm References} && \hfil\hbox{\rm TFFSA data}
&&\hfil\hskip -10pt\hbox{\rm dispersion relation}
\cr
\omit
& height2pt & & height2pt & & height2pt &
\cr
\noalign{\hrule}\omit
& height2pt & & height2pt & & height2pt &
\cr
\Phi_0&& -1.1977334...\,^{\Fateev} && -1.1977331 && -1.1977320\cr
\Phi_1&& -0.3188096...\,^{\FLZZ} && -0.3188103 && -0.3188192\cr
\Phi_2&& \hfil\hbox{\rm ---} &&  \p 0.1108867 && \p 0.1108915 \cr
\Phi_3&& \hfil\hbox{\rm ---} && \p 0.0164266 && \p 0.0164252 \cr
\Phi_4&& \hfil\hbox{\rm ---} && -2.64\times10^{-4} &&
-2.64\times10^{-4}\cr
\Phi_5&& \hfil\hbox{\rm ---} && -5.14\times10^{-4} &&
- 5.14\times10^{-4}\cr
\Phi_6&& \hfil\hbox{\rm ---} && \p 2.07\times10^{-4} &&
\p 2.09\times10^{-4}\cr
\Phi_7&& \hfil\hbox{\rm ---} && -4.52\times10^{-5} &&
-4.48\times10^{-5}\cr
\Phi_8&& \hfil\hbox{\rm ---} && \hfil\hbox{\rm ---} &&
\p 3.16\times10^{-7}\cr
\Phi_9&& \hfil\hbox{\rm ---} && \hfil\hbox{\rm ---} &&
\p 4.31\times10^{-6}\cr
\Phi_{10}&& \hfil\hbox{\rm ---} && \hfil\hbox{\rm ---} &&
-1.99\times10^{-6}\cr
}}
}
According to \bran\ and
\aran, the coefficients $\Phi_n$ control the asymptotics of the
functions $G_{high}(\xi)$ and $G_{low}(\xi)$ at large $\xi$; for
instance, for $-\pi/2 < {\rm arg}(\xi) < \pi/2$ and sufficiently large
$|\xi|$  
\eqn\ghighass
{
G_{high}(\xi) = \Phi_0\,\xi^{16\over{15}}-\Phi_1\,\xi^{8\over 15} +
{1\over{15\pi}}\,\log\xi^2 + 
\sum_{n=2}^{\infty}\,(-)^{n}\,\Phi_n\,\xi^{{8(2-n)}\over 15} 
\quad {\rm as} \quad 
\xi\to\infty\,.
}
Similar expansion (without the $(-)^n$ factors) holds for 
$G_{low}(\xi)$ in the wider domain \hbox{$-\pi < {\rm arg}(\xi) < \pi$}. 
Conversely, expansions \highexp\ and 
\lowexp\ can be rewritten as large-$\eta$ expansions of the scaling 
function $\Phi(\eta)$ in the corresponding domains of the $\eta$ plane. 
Thus, the expansion
\eqn\phiassminus
{\eqalign{
- {{\eta^2}\over {4\pi}}\,\log(-\eta)+{\tilde\Phi}(\eta) 
= G_2\,(-\eta)^{-{7\over 4}} + G_4\,(-\eta)^{-{22\over 4}}
+ G_6\,(-\eta)^{-{37\over 4}} + \cdots&
\cr
{\rm for}\quad -{{4\pi}\over 15} < 
{\rm arg}(-\eta) < {{4\pi}\over 15}&
}}
converges to $\Phi(\eta)$ for sufficiently large $\eta$ in the domain
HTW in Fig.~4. In the domain LTW the function $\Phi(\eta)$ has asymptotic 
expansion
\eqn\phiassplus
{\eqalign{
- {{\eta^2}\over {4\pi}}\,\log(\eta)+{\tilde\Phi}(\eta) \simeq {\tilde G}_1\,\eta^{{1\over 8}} + {\tilde
G}_2\,\eta^{-{7\over 4}} + {\tilde G}_3\,\eta^{-{29\over 8}} +
\cdots \quad {\rm as}\quad \eta\to \infty\,,&
\cr 
{\rm for}\quad-{{8\pi}\over 15}\!<\!{\rm arg}(\eta)\!<\!{{8\pi}\over 15}.&
}}

To prepare notations for our analysis in Sects.~6--7 let us briefly 
describe the expected behaviour of the scaling function $\Phi(\eta)$
along the axes ${\rm AOB^{*}}$, ${\rm A^{*}O{B}}$ and on 
the rays ${\rm OC}$, ${\rm OC^{*}}$ in Fig.~4; these
domains are of particular interest for our discussion.

The axis ${\rm AOB^{*}}$ corresponds to 
$\eta = e^{{{4\pi}\over 15}\,i}\,y$ with
real (positive as well as negative)~$y$. The values of the scaling
function $\Phi(\eta)$ along this axis are related to the free energy
$F(m,h)$ at pure imaginary $h$ as follows. Let us introduce the
function
\eqn\Phiimh
{
{\Phi}_{imh}(y) = - {{y^2}\over{4\pi}}\,\log |y| - {{i\,y^2}\over{15}} + 
e^{-{{8\pi}\over 15}i}\ {\tilde\Phi}\(y\,e^{{{4\pi}\over 15}i-i0}\)\,.
}
According to \G, \Phifunction\ and \Phitilde, this function relates to 
the analytic continuations of $G_{high}(\xi)$ and $G_{low}(\xi)$ to pure
imaginary $\xi$. Namely,
\eqna\PhiimhG
$$\eqalignno{
{\Phi}_{imh}(y) &= y^2\,G_{low}\(-i/y^{15/8}+0\)
\quad\quad~~\, {\rm for} \quad y>0\,; &\PhiimhG a
\cr
{\Phi}_{imh}(y) &= y^2\,G_{high}\(-i/(-y)^{15/8}+0\) 
\quad {\rm for} \quad y<0\,. 
&\PhiimhG b 
}$$
Therefore it takes real values for $y \leq -Y_0$ (the upper
edge of the branch cut A-YL in Fig.~4), and becomes complex-valued for 
$y > - Y_0$, where its imaginary part relates to imaginary part of the
function \rhohigh,
\eqn\Dimh
{
D_{imh}(y) \equiv \Im m\,\Phi_{imh}(y) =
y^2\,\Im m \,G_{imh}\(1/(-y)^{15/8}\)\,, 
\qquad {\rm for} \quad y < 0\,.
}

The structure of the singularity of $\Phi_{imh}(y)$ at $-Y_0$ can be
described in terms of the expansion \LYsingular. As in the analysis
in Sects.~6--8 we use the variable $\eta$ rather then $\xi$, let us
introduce here more suitable notations. In the vicinity of $-Y_0$ we
expect to have
\eqn\philyang
{
\Phi_{imh}(y) = A(y) + B(y)\,(-Y_0 - y - i0)^{5\over 6} + C(y)\,(-Y_0 -
y - i0)^{5\over 3} + \cdots\,,
}
where the real analytic functions 
$A(y)$, $B(y)$, $C(y), \ldots\,$, are regular at $y=-Y_0$, i.e. they
admit the power-series expansions
\eqna\ABC
$$\eqalignno{
A(y) &= A_0 + A_1\,(-Y_0 -y) + A_2\,(-Y_0 -y)^2 + \cdots\,, &\ABC a
\cr
B(y) &= B_0 + B_1\,(-Y_0 -y) + \cdots\,, &\ABC b
\cr
C(y) &= C_0 + \cdots\,, &\ABC c
}$$
around the point $-Y_0$, with real coefficients. The structure \philyang\
follows from the expected form
\LYeff\ of the effective action in the vicinity of the critical
point. The first subleading singular term written down in \philyang\
is due to the term $T{\bar T}$ in \LYeff. Simple perturbative
analysis (which we skip in this report) allows one to relate some of the 
coefficients in (3.41) to the parameters in (3.8),
\eqn\BCtoalpha
{
B_0 = f_{\rm YL}\,a_{\rm YL}^2\,
\(15\lambda_{1}Y_0^{-{47\over20}}/4\)^{5/6}\,, \qquad 
C_0 = \alpha_0\,B_{0}^2/4 Y_{0}^2\,,
}
where $f_{\rm YL}= -\sqrt{3}/12$ is the free energy amplitude of
the integrable Yang-Lee field \hbox{theory \refs{\CardyMussardi, \AlZb}}, 
and $a_{\rm YL}=2.6429446...$ 
is the corresponding amplitude in its ``mass to coupling
relation'' \refs{\AlZb, \AlZc}.

Let us also write down the $y$-series expansions of \Dimh, 
which follow directly \hbox{from \phiseries}. For
sufficiently small $|y|$ we have
\eqn\Phiimexp
{
D_{imh}(y) = \sum_{n=0}^{\infty}\,{\bar\Phi}_n\,y^n \,, 
}
where 
\eqn\Phibars
{
{\bar\Phi}_n = \Phi_n\,\sin{{4\pi\,(n-2)}\over{15}} \qquad
{\rm for} \quad n\neq 2\,; \qquad {\bar\Phi}_2 = -{1\over 15}\,.
}

The axis OC in Fig.~4 (i.e. $\eta = y\,e^{-i{{8\pi}\over
15}}$ with real $y>0$) is the image of the upper edge of the branch
cut in Fig.~2. The values of $\Phi(\eta)$ along this axis are
related to the complex free energy \Gmeta,
\eqn\Phimeta
{
\Phi_{meta}(y) \equiv -{{y^2}\over{4\pi}}\,\log y + {{2i\,y^2}\over 15}
+ e^{i{{16\pi}\over 15}}\,{\tilde \Phi}\(y\,e^{-i{{8\pi}\over 15}}\) = 
y^2\,G_{meta}\(1/y^{15\over 8}\)\,.
}
In view of \lowdisp\ and \Gammam, the imaginary part of this function
plays particularly important role, so we introduce for it a separate
notation 
\eqn\Dfunction
{
D_{meta}(y) = \Im m\,\Phi_{meta}(y) = {2\over 15}\,y^2 +
{1\over{2i}}\,\bigg( e^{i{{16\pi}\over 15}}\,
{\tilde\Phi}\(y\,e^{-i{{8\pi}\over 15}}\)-e^{-i{{16\pi}\over 15}}\,
{\tilde\Phi}\(y\,e^{i{{8\pi}\over 15}}\)\bigg)\,.
}
As follows from \Phimeta, this function enjoys a power series
expansion
\eqn\DnsA
{
D_{meta}(y) = \sum_{n=0}^{\infty}\,D_n\,y^n\,,
}
with
\eqn\DnsB
{
D_n =
\Phi_n\,\sin{{8\pi\,(2-n)}\over 15}\,\qquad {\rm for}\quad n\neq 2\,;
\qquad D_2 = {2\over15}\,,
}
which converges for sufficiently small $y$. On the other hand,
$D_{meta}(y)$ decays very fast at large positive $y$. According to
\voloshin,
\eqn\dmetaass
{
D_{meta}(y) \to {{\bar s}\over{2\pi}}\,V(y), \qquad {\rm as} 
\quad y\to+\infty\,,
}
where $V(y)$ stands for the function
\eqn\Vdef
{
V(y) = y^{1\over 8}\,\exp\(-{\pi\,y^{15\over 8}}/2{\bar s}\)\,.
}

\newsec{Extended Analyticity}

In the previous Section we have described the domains of analyticity
of the scaling function ${\tilde\Phi}(\eta)$ which follow from previously
known results. An unexplored area lays between the High-T and Low-T Wedges
in Fig.~4, the ``Shadow Domain'' SHD. In more conventional notations (i.e. 
in
terms of the scaling variable $\xi$) this domain is located under the
Langer's branch cut in Fig.~2, imediately below the negative real axis.
Alternatively, it can be reached by going under either of the branch
cuts in Fig.~1. We believe the problem of establishing the analytic 
properties
of free energy in the shadow domain is an important one, and it was one
of the motivations of this work. Some reasons for our interest in this 
problem are explained in Sect.~9.

The simplest conceivable possibility is that the function 
${\tilde\Phi}(\eta)$ is actually analytic in the wedge 
$-11\pi/15 < {\rm arg}(\eta) < 11\pi/15$ which includes the whole of 
the shadow domain. It is also the most elegant possibility (to our
sense of beauty, that is), because it renders the Yang-Lee point an
additional significance of being the closest singularity under the
Langer's branch cut in Fig.~2. We formulate it as the following 
\vskip 0.1in
{\it Extended Analyticity Conjecture}: The scaling function 
${\tilde\Phi}(\eta)$ is analytic in the whole complex $\eta$-plane with 
two branch cuts extending from the points ${\rm YL}$ and ${\rm YL}^{*}$ 
to infinity, along the rays ${\rm YL-A}$ and ${\rm YL}^{*}-{\rm A}^{*}$ 
in Fig.~4, respectively.
\vskip 0.1in

The extended analyticity is equivalent to certain ``extended'' 
dispersion relation, expressing this scaling function in terms of its 
discontinuity across the branch cuts in Fig.~4. To write it down, let
us introduce the function
\eqn\deltatilde
{
{\tilde\Delta}(y) = i\,e^{-i{8\over 15}\,\pi}\,
\bigg({\tilde\Phi}\(-y\,e^{i{4\over 15}\,\pi-i0}\)-{\tilde\Phi}
\(y\,e^{-i{11\over 15}\,\pi+i0}\)\bigg)\,,
}
which is the discontinuity of ${\tilde\Phi}(\eta)$ across the branch cut
${\rm YL-A}$, with additional phase factor introduced for later 
convenience. Here
$y$ is a positive real variable interpreted as the coordinate along the ray 
$O-A$ in Fig.~4. The function \deltatilde\ vanishes for $y < Y_0$, 
while at $y>Y_0$ it takes complex values. In the last domain it coincides
with certain analytic continuation of the function $D_{imh}(y)$,
\eqn\deltadimh
{
{\tilde\Delta}(Y_0 + z) = D_{imh}\(-Y_0 + z\,e^{-i\pi+i0}\)\,, \qquad
(z>0)\,.
}
Unfortunately, this function grows too
fast at large positive $y\ $ 
(\hbox{${\tilde\Delta}(y) \to -y^2/4\,$} as \hbox{$y\to +\infty$}), and 
writing down the dispersion integral directly for
${\tilde\Phi}(\eta)$ would require at least three subtractions. It is more
convenient to use instead the function
\eqn\deltaaa
{
\Delta(y) = \Delta_{log}(y)+{\tilde\Delta}(y)\,,
}
where
\eqn\deltalog
{
\Delta_{log}(y) = {1\over4}\,y^2\,.
}
The function \deltaaa\ is related to the mismatch between the functions
\eqn\Phihigh
{
\Phi_{high}(\eta)=-{\eta^2\over{4\pi}}\,\log(-\eta) +{\tilde\Phi}(\eta)
}
and
\eqn\Philow
{
\Phi_{low}(\eta)=-{\eta^2\over{4\pi}}\,\log(\eta) +{\tilde\Phi}(\eta)
}
at the ray $\rm O-A$. According to the asymptotics \phiassminus\ and 
\phiassplus,
the function \deltaaa\ exhibits a much slower growth rate, $\Delta(y) \to
\Delta_{ass}(y) \quad {\rm as} \quad y\to+\infty$, where
\eqn\deltaass
{
\Delta_{ass}(y) = - e^{-i{\pi\over 8}}\,{\tilde G}_1\,y^{1\over 8}\,,
}
and hence one can write down the dispersion integral with only one 
subtraction,
\eqn\extdisp
{
{\tilde\Phi}(\eta)=\Phi_0 + {\tilde\Phi}_{log}(\eta) -
{{\eta}\over{\pi}}\,\int_{Y_0}^{\infty}\,{{y\,\Re e \(e^{{4\pi
i}\over 15}\,\Delta(y)\)+\eta\,\Re e\(e^{{8\pi i}\over
15}\,\Delta(y)\)}\over{y^2 +2y\eta\,\cos(4\pi/15) +
\eta^2}}\,{{dy}\over y}\,, 
}
where the term
\eqn\philog
{
{\tilde\Phi}_{log}(\eta) = -{{Y_0\,\eta}\over{4\pi}}\,\cos(4\pi/15) + 
{{\eta^2}\over{8\pi}}\,\log\big(Y_{0}^2 + 2\,\eta Y_0\,\cos(4\pi/15) + \eta^2
\big)
}
comes from the domain $0\leq y<Y_0$ in which $\Delta(y)$ coincides with 
\deltalog.

We will refer to Eq.~\extdisp\ as the ``extended dispersion 
relation''. Its validity is \hbox{equivalent} to the validity of the 
extended 
analyticity assumption stated above, and under this assumption
Eq.~\extdisp\ must hold in the whole complex $\eta$-plane, including the 
shadow domain SHD in Fig.~4.  

From \extdisp\ one readily derives the following expressions for the
coefficients of the Taylor expansion \phiseries,
\eqna\extdispcoeffs
$$\eqalignno{
\Phi_0 &= \Re e\,\bigg[{e^{{8\pi i\over15}}\over{\pi}}\,\int_{0}^{\infty}
\,{{\Delta(y)-\Delta_{ass}(y)}\over{y}}\,dy\bigg]\,, &\extdispcoeffs a
\cr
\Phi_1 &= -{{Y_0}\over{4\pi}}\,\cos(4\pi/15) - 
\Re e \,\bigg[{e^{4\pi i\over15}\over\pi}\,\int_{Y_0}^{\infty}
{{\Delta(y)\,dy}\over y^2}\bigg]\,, &\extdispcoeffs b
\cr
\Phi_2 &= {1\over{4\pi}}\,\log Y_0 + \Re e \,\bigg[{1\over\pi}\,
\int_{Y_0}^{\infty}
{{\Delta(y)\,dy}\over y^3}\bigg]\,, &\extdispcoeffs c
\cr
\Phi_n &= (-)^n\,\Re 
e\,\bigg[{{e^{{4\pi i\over15}(2-n)}}\over\pi}\,\int_{Y_0}^{\infty}
{{(\Delta(y)-\Delta_{log}(y))\,dy}\over{y^{n+1}}}\bigg] \qquad {\rm
for} \quad n>2\,. &\extdispcoeffs d
}$$

In Sect.~8 below we will bring up some numerical evidence 
to the validity of this extended dispersion relation, thus furnishing
a support for the above extended analyticity conjecture. 

\newsec{Excited states. False vacuum}

Although in this work we are mostly interested in the free energy
$F(m,h)$, and hence concentrate most of attention on the ground-state
energy \Eass\ of the finite-size Hamiltonian \Hamiltonian, some analysis 
of excited states will prove to be useful in Sect.~8, where we study the
function $\Phi_{meta}(y)$. To prepare for this discussion let us
briefly remind qualitative properties of the quantum field theory
\AIFT, in infinite space-time, interpreted as a particle theory \McCoyWuA.

For $m>0$ and $h=0$ (i.e. for $\eta = +\infty$) we have just a free
fermion  theory \AFF. In this case the vacuum of the bulk theory
is two-fold degenerate (the ground states in NS
and R sectors degenerate as $R\to\infty$), thus manifesting the
spontaneous breakdown of the $\sigma\to -\sigma$ symmetry. The
particles (i.e. the free fermions of \AFF) are identified with the
``domain walls''. Adding the
interaction term in \AIFT\ creates a confining force between these 
fermions
(which henceforth are refered to as ``quarks''), and for nonzero $h$ the
particle spectrum contains only their bound states - ``mesons''. If
$m>0$ and $h$ is small (i.e. $\eta$ is large positive) the
corresponding ``string tension'' is small ($\sim 2\,{\bar\sigma}\,h$,
where ${\bar\sigma}$ is the spontaneous magnetization at zero field
given by \sigmabar), and there is a
large number of stable mesons, their masses $M_i$ densely filling the
interval between $2m$ and $4m$ (the mesons with $M > 2\,M_1$, $\ M_1$
being the lightest meson mass, are
generally unstable). In this region of small $h$ the lower part of the
meson mass spectrum (with $M_i - 2m \ll m$) can be understood in terms of
non-relativistic quarks interacting via a linear confining potential.
This interpretation agrees with the result of \McCoyWuA\ for the
masses of these lightest mesons,
\eqn\Miass
{
M_i - 2\,m \to {{(2\,{\bar\sigma}\,h)^{2\over 3}\,z_i}\over{m^{1\over
3}}}
\qquad {\rm as} \quad h \to 0\,,
}
where $- z_i$, $i=1, 2, 3, \ldots\,$, are zeroes of the Airy function, 
${\rm
Ai}(-z_i) = 0$.
In fact, a few first relativistic corrections to these masses can be
computed (see Appendix B),
\eqn\Miexp
{
M_i = m\,\left\{ 2 + {{(2{\bar s})^{2\over 3}\,z_i}\over{\eta^{5\over 
4}}} -
{{(2{\bar s})^{4\over 3}\,z_{i}^2}\over{20\,\eta^{5\over 2}}} + 
\bigg({{11\,z_{i}^3}\over 1400}-{57\over 280}+{q_{2}\over 2}\bigg)\,
{{(2{\bar s})^2}\over{\eta^{15\over 4}}} + O\(\eta^{-5}\)\right\}\,,
}
where the constant $q_2$ is the same as in Eq.~\Mquark\ below. 

As $\eta$ decreases, the heavier of the mesons gradualy disappear
from the spectrum of stable particles; when their masses exceed the
stability threshold $2\,M_1$ they (in general) become resonance
states. In particular, when $\eta$ reaches zero (i.e. $m=0$), only three
stable mesons remain below the threshold; in this case however there
are additional five stable particles above the threshold, which owe
their stability to the fact that the field theory~\IFT\ with $\tau =0$
is integrable \Me. This process continues when $\eta$ becomes
negative, until finally at $\eta < \eta_2$ ($\eta_2 \approx -2.09$, see
Sect.~6 below) only one particle remains stable. 
As $\eta \to -\infty$ its mass $M_1$ approaches 
$|m|$ \foot{The numerical value of the constant $a$ in Eq.~\Monehigh\
comes from our estimate
$$
a \approx {\bar s}^2\,\big(247/9\sqrt{3} -23/2 + 14/3\pi\big)
$$
of leading ($\sim h^2$) perturbative mass 
correction. The approximation used in this estimate is similar to that
proposed in \WuMcCoyG. We will present this calculation elsewhere.},
\eqn\Monehigh
{
M_1 = |m| \,\bigg(1 + {{a}/{(-\eta)^{15\over 4}}} +
O\((-\eta)^{-{15\over 2}}\)\bigg)\,; \qquad a \approx 10.75\,.
}
Although a detailed discussion of the mass spectrum is
outside the scope of this paper (we intend to present it separately), 
we show in Fig.~5 the $\eta$ dependence of the first few meson masses 
obtained using the TFFSA.

\fig{5}{Masses of the three lighest particles in the Ising field theory 
\IFT. The solid lines are the plots of the dimensionless ratios 
$M_i/|h|^{8/15}$ ($i=1,2,3$) versus the parameter \etaparam. The dashed 
lines represent the corresponding large-$|\eta|$ expansions, Eqs.~\Miexp\ 
and~\Monehigh\ (with all terms explicitely written in these Eqs. 
included).} 
{\epsfxsize=9cm\epsfbox{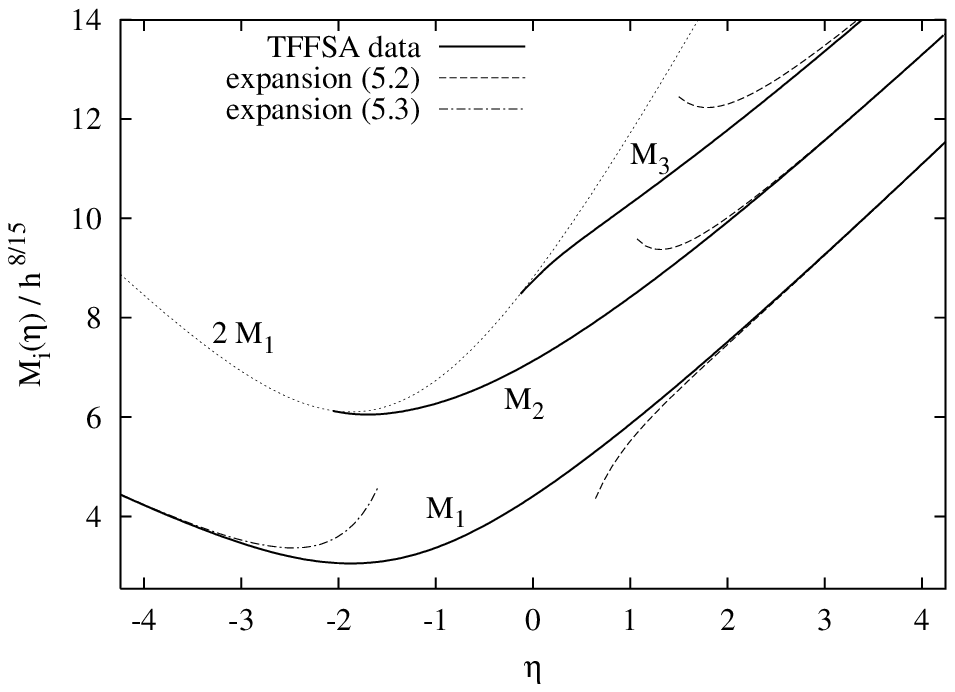}}

The mesons described above are excitations over the stable vacuum of the
system (which is unique for $h\neq 0$). 
If $m>0$ and $h$ is sufficiently small, the system exhibits also an unstable 
``false vacuum'', a global resonance state whose (complex) energy is
an intensive quantity, i.e.
\eqn\Emeta
{
E_{meta} = R\,F_{meta}(m,h)\,,
}
where $R \to \infty$ is the spatial size of the system. The 
corresponding energy density $F_{meta}$ is a complex-valued quantity,
and its imaginary part is
interpreted as the decay probability density (the probability 
per unit volume and unit time)
of the false vacuum. According to standard arguments (see 
\refs{\Kobzarev,\Coleman,\CC}) 
the resonance energy density $F_{meta}(m,h)$ coincides with the analytic
continuation of the vacuum energy density $F(m,h)$ in $h$, from
positive to negative values of $h$, i.e. it can be written as
\eqn\Phimeta
{
F_{meta}(m,h) = {m^2\over{8\pi}}\log m^2 + |h|^{16\over 15}\,
\Phi_{meta}(y)\,, \qquad y = m/|h|^{8\over 15}\,, 
}
where $\Phi_{meta}(y)$ is the function defined in Sect.~3. According to
\voloshin, at small $h$ (i.e. at large $y$) the imaginary part 
\eqn\Gammafunction
{
\Gamma(m,h) = |h|^{16\over 15}\,D_{meta}(y)
}
decays exponentially, see Eq.~\dmetaass. In fact, it was
argued in \Voloshin\ that the asymptotic~\voloshin\ itself holds
with exponential accuracy provided one replaces $2{\bar\sigma}h \to
\Delta F(m,h)$ and \hbox{$m \to m_{q}(m,h)$}, where
\eqn\deltaf
{
\Delta F(m,h) \equiv \Re e\,F_{meta}(m,h) - F(m,h) \approx 2{\bar\sigma}h +
O(h^3)\,,
}
is the ``string tension'', and $m_{q}(m,h)$ is the ``quark
mass''. Although it is not clear how to give a precise definition to the last
quantity, it is natural to assume that at small $h$ it admits an expansion
(perhaps, an asymptotic one) in powers of $h^2$,
\eqn\Mquark
{
m_{q} = m\,\(1 + q_2\,\({{{\bar\sigma}\,h}/{m^2}}\)^2 
+ O\(h^4\)\)\,,
}
where $q_2$ is some constant, whose evaluation constitutes a separate
and somewhat involved problem; at the moment we have only a preliminary
numerical estimate, $q_2 \approx 0.14(1)$ \foot{It results
from estimating the three-quark contribution to the quark self-energy;
we plan to address this problem separately.}.
As the result, the expected asymptotic form of the imaginary part of 
the function $\Phi_{meta}(y)$ is
\eqn\discass
{
D_{meta}(y) \simeq V(y)\, \( V_0 +V_1\,y^{-{15\over
8}}+ \cdots\) \qquad y\to +\infty\,,
}
where $V(y)$ is defined in \Vdef, and
\eqn\Vcoeffs
{
V_0 = {{\bar s}\over{2\pi}} = 0.216106...\,, \qquad 
V_1 = -{{q_2\,{\bar s}^2}\over 2} + 
{{{\tilde G}_3}\over{4\,{\tilde G}_1}} \approx -0.14\,. 
}

In TFFSA we actually study the system in finite-size geometry, with
the spatial compactification length $R$. In this situation the above
stable mesons
$M_i$ correspond to a series of finite-size energy levels $E_i (R)$
which asymptotically (as $R \to \infty$) behave as
\eqn\Eilevels
{
E_i (R) = F\,R + M_i + O\(e^{-{\sqrt{3}\over 2}\,M_1\,R}\),
}
where $F=F(m,h)$ is the vacuum energy density. Of course, there
are no resonance states at finite $R$. If $h$ is sufficiently small, 
the ``false vacuum'' resonance of the $R=\infty$ system is very
narrow, and at finite $R$ it manifest itself as a peculiar
pattern of ``near-intersections'' of the levels $E_i (R)$, see Fig.~$6a$
in Sect.~6.
The near-intersection of the levels $E_i (R)$ and~$E_{i+1}(R)$ 
occurs approximately at $R_{i,i+1} = M_i/\Delta F$, where $\Delta F(m,h)$ 
is the same as in \deltaf, and between the two 
\hbox{near-intersections} the 
level $E_i (R)$ approximately follows the ``resonance'' linear law
\eqn\near
{
E_i (R) \approx \Re e\, F_{meta}(m,h)\,R \qquad {\rm for} \qquad 
R_{i-1,i} < R < R_{i,i+1}\,.
}
This pattern (which was previously observed in \Mussardo, in the the TCSA 
study of IFT \IFT), is clearly visible in Fig.~$6a$ in Sect.~6. 
Calculation presented in Appendix B shows that the separation between 
$E_i(R)$ and $E_{i+1}(R)$ close to their mutual near-intersection follows 
the square-root law
\eqn\opening
{
E_{i+1}(R) - E_{i}(R) \approx m\,t^2\,
\sqrt{(mt)^2\,(R-R_{i,i+1})^2 + 
{{4\,\Gamma\,R}\over{mt^2\,\sigma'_i}}}\,,
}
where $\Gamma = \Gamma(m,h)$ is the resonance ``specific width'', 
$t=(2{\bar s}\xi)^{1/3}$, and the constants ${\sigma'}_i$ are the
derivatives of the function $\sigma(\epsilon)$, Eq.~\Bxv, taken at
its zeroes $\epsilon_i$. According to~\Bxvi\ and \Bxviii, these constants 
have the
small-$\xi$ expansions
\eqn\sigmais
{
\sigma'_i = {{{\rm Ai}'(-z_i)}\over{{\rm Bi}(-z_i)}}\,
\left\{1 + {{z_i}\over 10}\,t^2 - {{19\,z_{i}^2}\over 1400}\,t^4  + 
O\(t^6\)\right\}\,.
}

Eq.~\opening\ will be used in Sect.~7.3 for the numerical determination 
of $\Gamma(m,h)$ at small~$h$.

\newsec{Numerical Analysis}

In the numerical analysis of the Hamiltonian \Hamiltonian\ we used 
the technicaly simplest
(albeit may be not optimal) truncation scheme based on the notion of 
level. We define the level of the state \Space a\ (or \Space b) as 
a half of the sum of absolute values of the 
half-integers~$k_i$~(or the integers~$n_i$).
At $m=0$ this definition coincides with the standard notion of the level
in TCSA. The truncation level $L$ is the maximal level of states admitted 
into the truncated space. For the purposes of this paper we
restricted our attention to the zero-momentum sector, i.e. the states
with $\sum_i n_i$ or $\sum_i k_i$ equal to zero. For a given truncation 
level, the 
Hamiltonian \Hamiltonian\ was numericaly diagonalized in this sector. We 
used different
truncation levels ranging from 10 to 12; the dimensionality of the
truncated spaces then ranges from 487 for $L=10$ to 1186 for $L=12$. This
way some number of lowest energy levels $E_i(R)$ was obtained for various
real values of $h$, corresponding to $\eta$ in the interval $[-5:5]$, as 
well as for pure imaginary $h$, corresponding to $y$ in \Phiimh\ ranging 
in the same interval.

\subsec{Finite-size levels. Qualitative picture}

The qualitative
pattern of the resulting finite-size levels $E_i (R)$ is not very
sensitive to the truncation level (except for some close proximity of
the Yang-Lee point, see below), as long as $R$ is not too large. Of
course, higher truncation levels allow for better accuracy in 
quantitative estimates of its characteristics (energy density and
masses). Also, as is usual with truncated Hamiltonians, the
truncation effects become more prominent for larger values of $R$. 
In quantitative analysis we mostly used the data with 
$R\,|h|^{8\over 15}< 6$.

\nobreak 
\midinsert
\centerline{
\vbox{
\hbox to \hsize{
\hfil
\vbox{
\hbox{\epsfxsize=6.5cm
\epsfbox{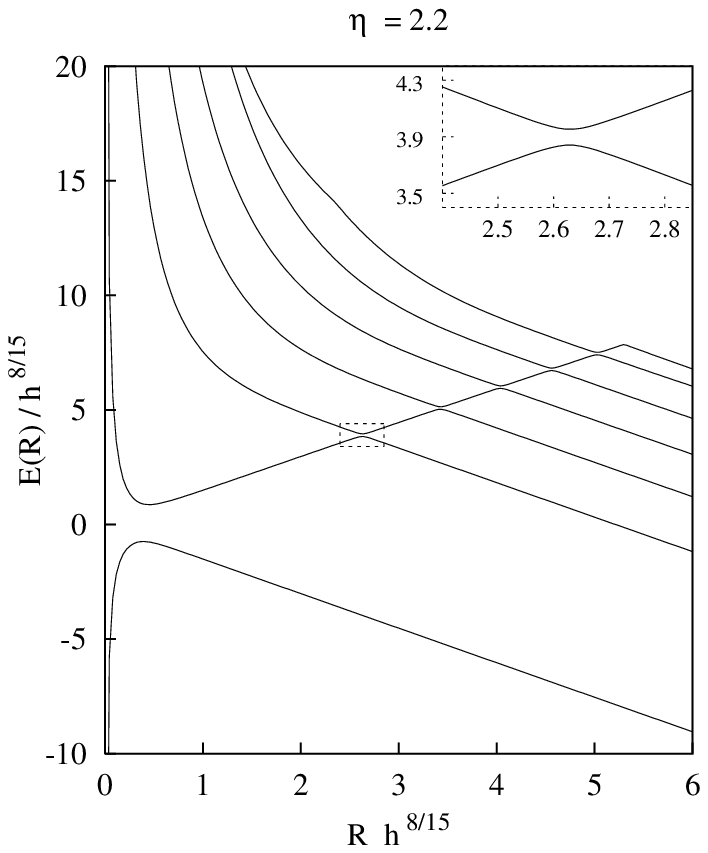}}
\vbox{\hsize=6.5cm \baselineskip=14pt
\centerline{\qquad\nineb Fig.~6\nineib a}
\vfil}}
\hskip 10pt
\vbox{
\hbox{\epsfxsize=6.5cm
\epsfbox{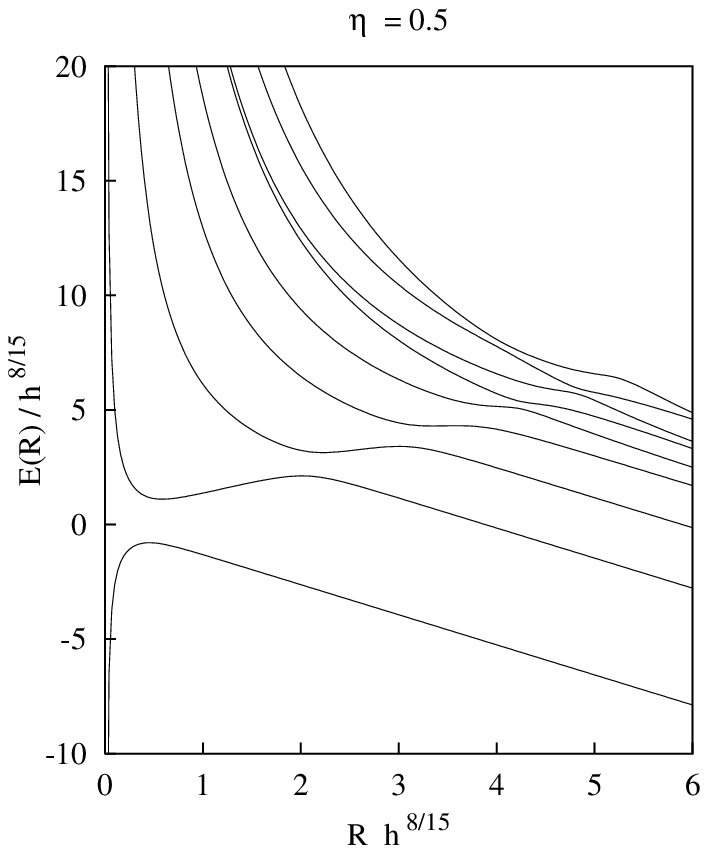}}
\vbox{\hsize=6.5cm \baselineskip=14pt
\centerline{\qquad\nineb Fig.~6\nineib b}
\vfil}}
\hfil}
\vskip 10pt
\hbox to \hsize{
\hfil
\vbox{
\hbox{\epsfxsize=6.5cm
\epsfbox{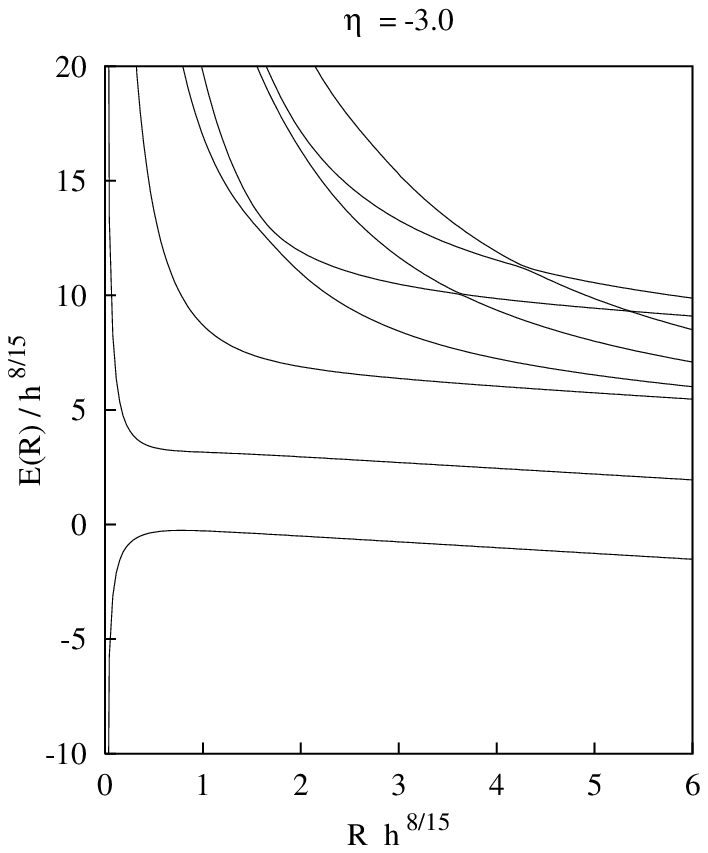}}
\vbox{\hsize=6.5cm \baselineskip=14pt
\centerline{\qquad\nineb Fig.~6\nineib c}
\vfil}}
\hskip 10pt
\vbox to 215pt{\hsize=6.5cm \baselineskip=12pt\ninerm\ninemath
\vbox{\noindent
{\nineb Fig.~6:}~Plots of several lowest energy levels 
$E_i (R)$, $i=0,1,2, \ldots\,$, at different $\eta$.}
\vskip 9pt
\vbox{\noindent
{\nineb {\nineib a})}~\hbox{$\eta=2.2$}. Magnification in the corner 
shows the ``near-intersection'' of the levels $E_1 (R)$ and $E_2 (R)$.}
\vskip 9pt
\vbox{\noindent
{\nineb {\nineib b})}~\hbox{$\eta=0.5$}. Three lowest excited levels 
$E_1(R)$, $E_2(R)$, $E_3 (R)$ are seen to approach straight lines 
\Eilevels\ exponentially; these levels correspond to three stable mesons 
$M_1, M_2, M_3$. The rest of the levels correspond to scattering states.}
\vskip 9pt
\vbox{\noindent
{\nineb {\nineib c})}~\hbox{$\eta=-3.0$}. Only one stable particle 
remains.}
\vfill}
\hfil}
}
}
\endinsert\goodbreak

\subsubsec{6.1.1. Real $h$}

The patterns of the lowest finite-size energy levels $E_i(R)$ for some
real $\eta$ are shown in Figs.~$6a$--$6c$. In the range of $R$
shown in these Figures the difference between the plots for~$L=10$ and 
$L=12$ would not be visible. Qualitatively, the patterns in 
Figs.~$6a$--$6c$ 
are the same as were observed in \Mussardo\ in the TCSA study of \IFT, 
where 
their interpretations in terms of the particle spectra are discussed. 
In this case, we have just a few words to add.

In full accordance with the McCoy and Wu scenario \McCoyWuA, for
sufficiently large positive~$\eta$ (i.e. for $T<T_c$ and small $h$) the
field theory \IFT\ contains a large number of stable particles
(``mesons''). At finite $R$, these particles correspond to the energy
levels $E_i(R)$ which approach, at sufficiently large $R$, the straight
lines $F(m,h)\,R + M_i$, where $M_i$ are the masses of the mesons. This
pattern of levels is clearly visible in the plot of Fig.~$6a$ representing
the case of $\eta = 2.2$. For sufficiently large $\eta$ the meson
masses $M_i$ are in good agreement with \Miexp\ (see Fig.~5).  The plot in
Fig.~$6a$ also shows a characteristic pattern of ``near intersections'' of
the meson levels with the straight line $\Re e \,F_{meta}\,R$, which is
the finite-size manifestation of the ``false vacuum'' already mentioned in
Sect.~5. As is discussed there and in Appendix B, the openings at
these near-intersections are related to the imaginary part \Gammafunction\
of the ``false vacuum'' specific energy, see Eq.~\opening. We will use
this relation in Sect.~8 as a way to estimate the resonance width at
sufficiently large positive $\eta$.

As $\eta$ decreases, the mesons with higher masses gradually disappear
into the \hbox{``continuum''} above the the stability threshold $2\,M_1$,
where $M_1$ is the mass of the lightest particle (see Fig.~$6b$, where
only three stable particle levels are apparent). At the same time the 
resonance ``level'' becomes fussy, in agreement with the expectation
that its width parameter $\Gamma$ grows. The process of depletion
of the ``meson'' spectrum continues as $\eta$ decreases to negative
values. The third lightest particle disapears at $\eta \approx -0.14$, and 
the second one leaves the spectrum of stable particles at $\eta=\eta_2$,
\eqn\etatwo
{
\eta_2 = -2.09(4)\,.
} 
The way these particles disapear is an interesting question which
we hope to address elsewhere. For $\eta$ below $\eta_2$ only one particle
with mass $M_1$ is left, as is manifest in Fig.~$6c$. The 
$\eta$-dependence of the first three masses was shown in Fig.~5. We 
postpone
a detailed quantitative discussion of the particle spectrum to future
publication.

\subsubsec{6.1.2. Pure imaginary $h$}

Much different patterns of the finite-size energy levels appear if
one takes $h$ pure imaginary, which corresponds to $\eta =
y\,e^{{{4\pi i}\over 15}}$ with real $y$. Although in this case the 
Hamiltonian~\Hamiltonian\ is not an Hermitian operator, it has an explicit
symmetry 
\eqn\Rsymm
{
S\,H\,S = H^{\dagger}\,,
}
where the operator $S$ just flips the signs of all states in the R
sector in \Space b. As the result, the eigenvalues of this Hamiltonian 
are either real or come in complex conjugated pairs. 

For sufficiently large negative $y$ the ground state energy $E_0 (R)$
of the truncated Hamiltonian remains real in a wide range of $R$,
including the region where the linear behaviour~\Eass\ is already clearly
visible.  This is also true for the first excited level $E_1 (R)$,
which quickly (i.e. exponentialy) approaches the straight line 
$F\,R + M_1$, i.e. it behaves as a one-particle state with a real mass
$M_1$. The situation 
is exemplified in Fig.~$7a$, where the real parts of the
first few energy levels are plotted. 

\nobreak
\midinsert
\centerline{
\vbox{
\hbox to \hsize{
\hfil
\vbox{
\hbox{\epsfxsize=6.5cm
\epsfbox{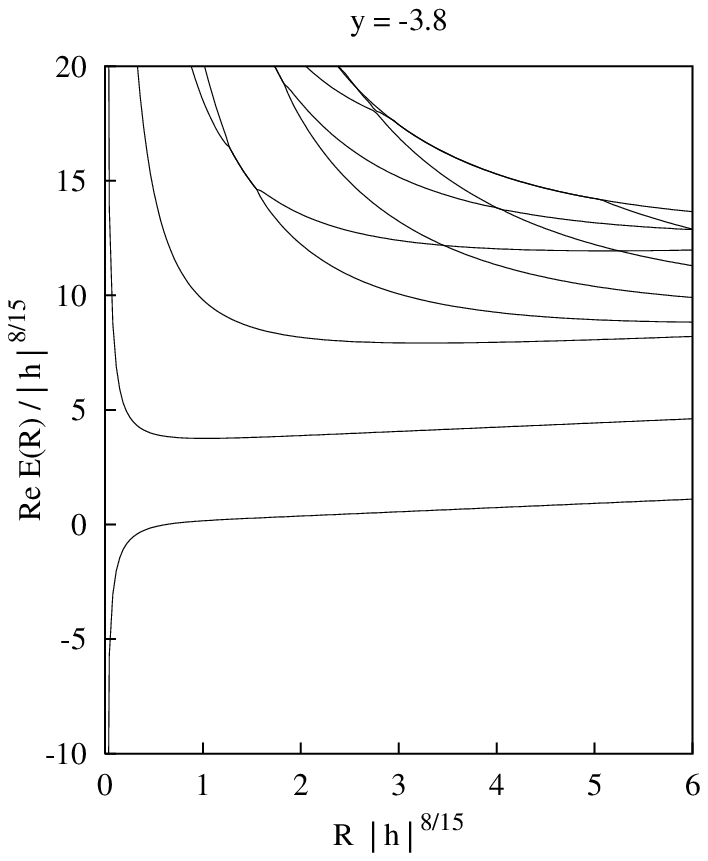}}
\vbox{\hsize=6.5cm \baselineskip=14pt
\centerline{\qquad\nineb Fig.~7\nineib a}
\vfil}}
\hskip 10pt
\vbox{
\hbox{\epsfxsize=6.5cm
\epsfbox{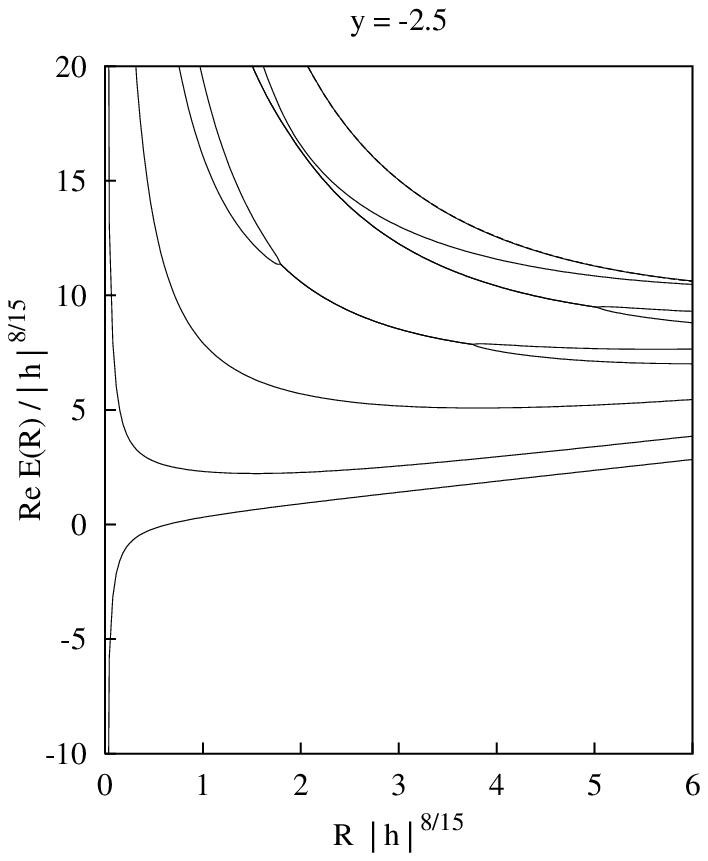}}
\vbox{\hsize=6.5cm \baselineskip=14pt
\centerline{\qquad\nineb Fig.~7\nineib b}
\vfil}}
\hfil}
\vskip 5pt
\hbox to \hsize{
\hfil
\vbox{
\hbox{\epsfxsize=6.5cm
\epsfbox{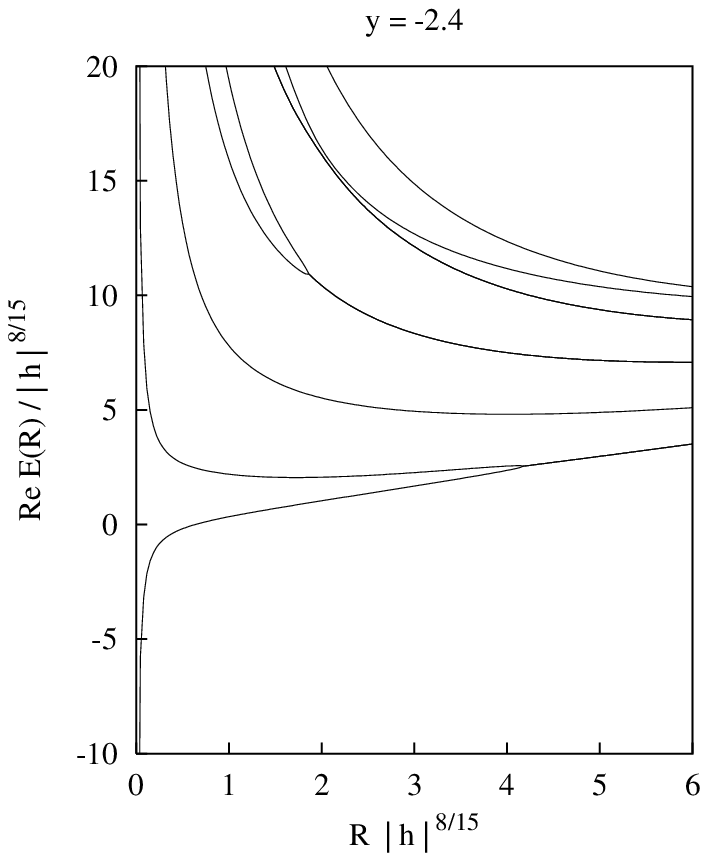}}
\vbox{\hsize=6.5cm \baselineskip=14pt
\centerline{\qquad\nineb Fig.~7\nineib c}
\vfil}}
\hskip 5pt
\vbox to 240.75pt{\hsize=6.5cm \baselineskip=12pt\ninerm\ninemath
\vskip 9pt
\vbox{\noindent
{\nineb Fig.~7:}~Plots of several lowest energy levels $E_i(R)$,
\hbox{$i=0,1,2,\ldots$}, at pure imaginary $h$, i.e. at \hbox{$\eta=
y\,\exp(4\pi i/15)$} with real $y$.~When~a~level becomes complex, 
only its real part is shown.}
\vskip 9pt
\vbox{\noindent
{\nineb {\nineib a})}~\hbox{$y=-3.8$}. The three lowest levels $E_0(R)$, 
$E_1(R)$, $E_2(R)$ are real for all 
\hbox{$R\,|h|^{8\over15} \in [0;6]$}. The level 
$E_1 (R)$ approaches the straight line $F\,R + M_1$ exponentially; it 
corresponds to one-particle state.}
\vskip 9pt
\vbox{\noindent
{\nineb {\nineib b})}~\hbox{$y=-2.5$} (close to the Yang-Lee point 
\hbox{$Y_0 \approx -2.43$}). The three lowest levels remain real for all 
\hbox{$R\,|h|^{8\over15}\in[0:6]$}, but the gap between $E_0 (R)$ and $E_1 
(R)$ 
narrows.
}
\vskip 9pt
\vbox{\noindent
{\nineb {\nineib c})}~\hbox{$y=-2.4$} (on the other side of the YL point). 
The~``nose''~appears:~$E_0 (R)$ and $E_1 (R)$ collide at 
\hbox{$R=R_{nose} \approx 4.3\,|h|^{-{8/15}}$} and become a pair of 
complex-conjugated \hbox{levels} at \hbox{$R > R_{nose}$}.}
\vfil}
\hfil}
}
}
\endinsert\goodbreak

As $y$ increases, the gap $M_1$ becomes smaller (see Fig.~$7b$), 
and for $y$ above $-2.43$ a collision of the first two levels at
some finite $R=R_{nose}(y)$ (the ``nose'') is clearly \hbox{visible}; at
$R > R_{nose}(y)$ the levels $E_0 (R)$ and $E_1 (R)$ form a
complex-conjugated pair, as is seen in Fig.~$7c$. When $y$ furher
increases, the
``nose'' quickly gets shorter, and when it is short enough (as in
Figs.~$8a$--$8b$) it becomes evident that for large \hbox{$R \gg 
R_{nose}$} the 
complex conjugate levels $E_0 (R)$ and $E_1 (R)$ develop into the 
``straight lines'' with complex slopes, \hbox{$E_0 \to F\,R\,$}, 
\hbox{$~E_1 
(R) \to F^{*}\,R$}. We interpret this as the vacuum state acquiring a 
complex energy density $F$ (obviously, in this case there are two 
``complex conjugate'' vacua). Moreover, the next lowest levels $E_2
(R)$ and $E_3 (R)$ eventualy collide and they too become a complex
conjugate pair, both then approaching similar straight lines, shifted
by some complex ``mass'' $M_1$, i.e. \hbox{$E_2 (R) \to F\,R + M_1\,$}, 
\hbox{$~E_3 (R) \to F^{*}\,R + M_{1}^{*}$}; these levels are naturaly
interpreted in terms of the complex-mass particle-like excitations
over the corresponding vacua. 

\nobreak
\midinsert
\centerline{
\vbox{\hsize=13cm
\hbox{
\vbox{
\hbox{\epsfxsize=6.5cm
\epsfbox{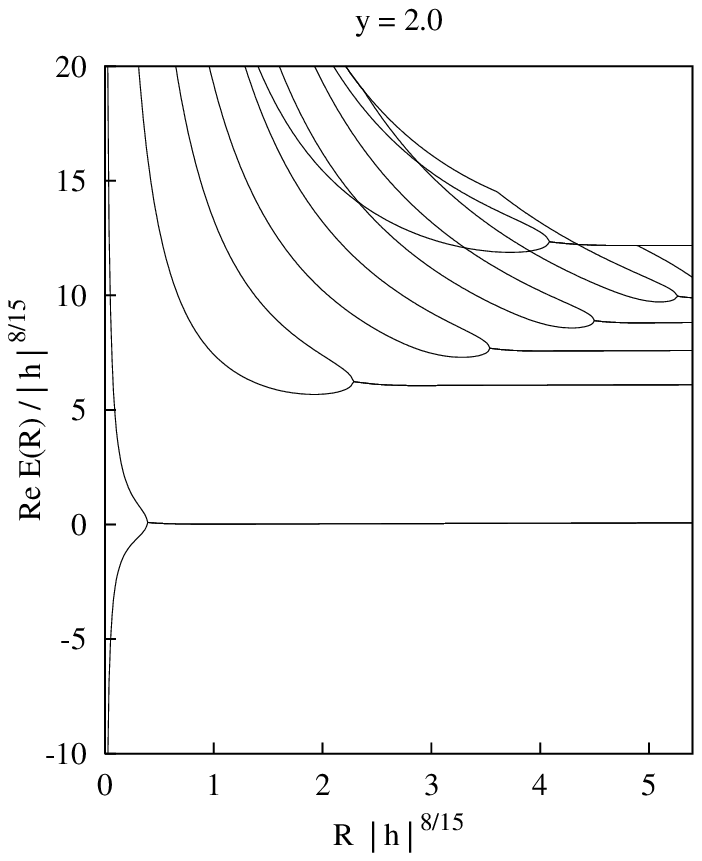}}
\vbox{\hsize=6.5cm
\centerline{\qquad\nineb Fig.~8{\nineib a}}}}
\hskip 10pt
\vbox{
\hbox{\epsfxsize=6.5cm
\epsfbox{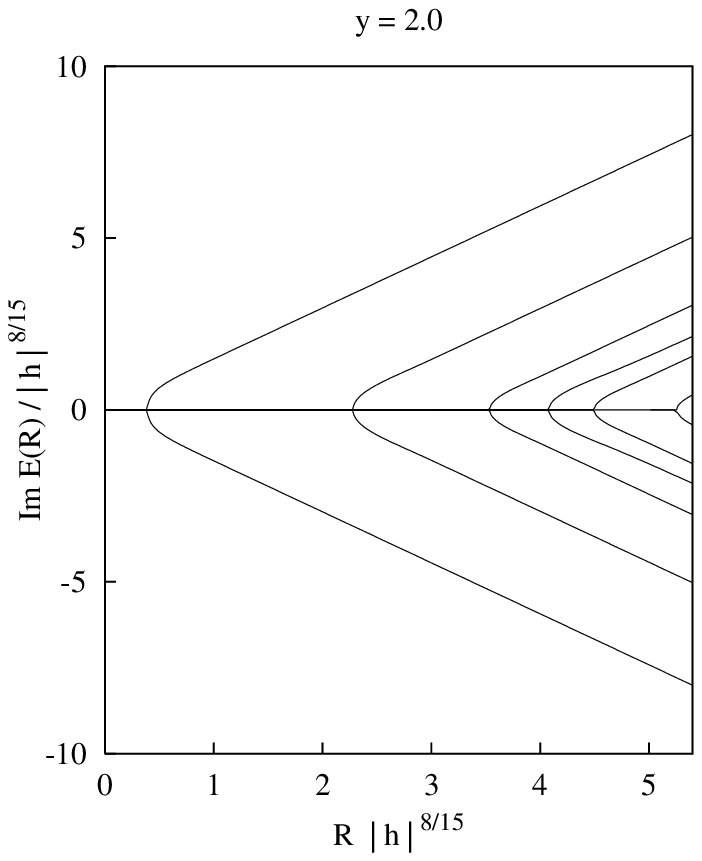}}
\vbox{\hsize=6.5cm
\centerline{\qquad\nineb Fig.~8{\nineib b}}}}
}
\vskip 10pt
\hbox{
\vbox{
\baselineskip=12pt\ninerm\ninemath\noindent
{\nineb Fig.~8:}~The lowest energy levels $E_i(R)$, $i=0,1,2,\ldots\,$,
at $\eta=2.0\times e^{4\pi i/15}$. 
{\nineb {\nineib a})}~Real parts
of~$E_i(R)$; {\nineb {\nineib b})}~Imaginary parts of $E_i(R)$. Two lower
levels, $E_0$, $E_1$, collide at \hbox{$R=R_{nose}\approx 0.389$}
becoming at larger $R$ a complex-conjugated pair. They quickly
approach straight lines, whose slopes are interpreted as $F$ and $F^*$.
Similar collision of the next two levels, $E_2$,~$E_3$, occurs at
$R\approx 2.28$; they also develop into two complex-conjugated  straight
lines, with the same slopes as $F$ and $F^*$. The gap
\hbox{$E_2(R)-E_0(R)$}
approaches a complex constant~$M_1$, interpreted as a  complex mass
(correspondingly, the gap \hbox{$E_3(R)-E_0(R)$} approaches~$M_1^*$).
}\hfill}
}}
\endinsert\goodbreak
\noindent
As $y$ becomes larger the number of such 
complex-mass particles increases. Higher energy levels form a rather 
complicated pattern. Their interpretation in terms of particle scattering 
states (and even legitimacy of such interpretation) in general remains an 
open problem which we hope to address in the future.

Appearence of complex $F$ for $y$ above the Yang-Lee point $-Y_0$ is
well expected in view of the discussion in Sect.~3; $F$ and $F^{*}$ are
just the values of $F(m,h)$ on two edges of the branch cuts in Fig.~1. 
We can say that the patterns in Fig.~$7b$ and in Fig.~$7c$ represent the
situations at the opposite sides of the Yang-Lee edge singularity,
which is located some place in between. Unfortunately (but not
surprisingly), in the close proximity of this critical point 
the results from the truncated Hamiltonians appear to be more
sensitive to the truncation level then elsewhere, especialy for larger
values of $R$. In Fig.~9 we show the plots of the function $C_0 (R)
\equiv {{6R}\over \pi}\,(E_0 (R) - F_0\,R)$, with 
\eqn\Fzero
{
F_0 \approx 0.54743\,|h|^{16/15}\,,
}
computed for different
truncation levels, at $y=-2.4295$ (which we believe to be very close to
the Yang-Lee point $-Y_0$). As the truncation level increases it seems to
develop the expected large-$R$ behaviour \Ezeroflow.

\fig{9}{
The function $C_0(R)=6R[E_0(R)-0.54743|h|^{16\over15}\,R]\,/\,\pi$ 
computed through the TFFSA using different truncation levels $L=10, 11, 
12$, at $\eta=-2.4295\times e^{4\pi i/15}$. The dotted line is the 
expected asymptotic value \hbox{$-c_{\rm eff}=-0.4$}. The dashed line 
represents the two leading corrections written in Eq.~\Ezeroflow\ 
with~\hbox{$\alpha_0=-4.2$}.
} 
{\epsfxsize=9cm\epsfbox{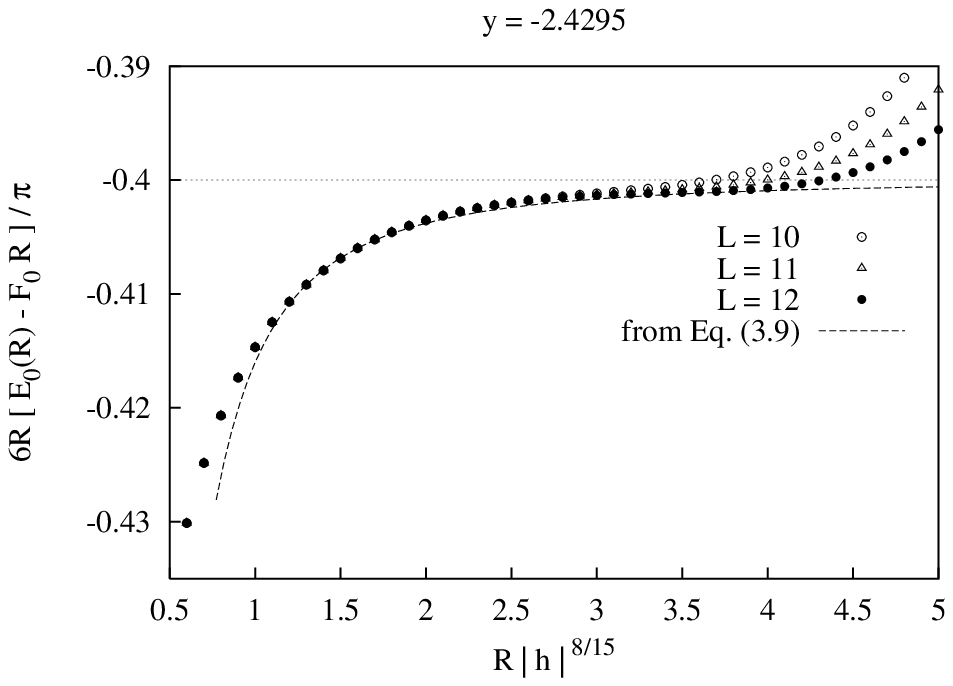}}

\subsec{Quantitative analysis}

The free energy $F(m,h)$ was determined by measuring the slope of 
the $R$ dependence of the ground sate level $E_0 (R)$ at sufficiently
large
$R$. There are two obstacles (well known in TCSA) 
to a precise determination
of this slope. Theoretically, the linear behaviour \Eass\ becomes exact
for large $R \gg M_{1}^{-1}$, while at finite values of $R$ finite-size
corrections may be important. Although the leading finite-size corrections
are related to the masses $M_i$ of the stable particles by well-known
formula (see e.g. \TCSAb)
\eqn\fsize
{
E_0 (R) = R\,F(m,h) - \sum_{i}\,{{M_i}\over\pi}\,K_1 (M_i R) + 
O\(e^{-2\,M_1\,R}\)\,,
}
where $K_1 (x)$ is modified Bessel function, the subleading corrections
depend on the (ge-nerally unknown) scattering amplitudes, and in most 
cases it is not clear how to take them into account. This makes it 
favorable to use larger values of $R$, where these subleading finite-size 
corrections are less significant. 
However, in practical calculations the ground-state energy deviates
from the linear behaviour \Eass\ at large $R$ as the result of the state
space truncation (the ``truncation effects''). Although this deviations
are 
too small to be visible in Figs.~6--8, they substantially reduce 
the precision
of quantitative analysis. The truncation effects bring an upper limit to
the values of $R$ at which precise quantitative analysis can be made.

To reduce the truncation effects, the ground-state energy functions
$E_{0}^{(L)}(R)$ computed with the truncation levels $L= 10, 11, 12$
was extrapolated to $L\to\infty$. The extrapolation was made by fitting
the formula 
\eqn\trunc
{
E_{0}^{(L)}(R) = s_0 (R) + s_1 (R)\,L^{-s_2 (R)}
}
to the 
computed $L$ dependence (at fixed $R$), via $s_0, s_1, s_2$, and
accepting
$s_0 (R)$ as the extrapolated $E_0 (R)$. Although this formula does not 
have 
theoretical justification, the extrapolation procedure appears to work
well for all real as well as imaginary $h$, corresponding to $|\eta|<5$,
substancially improving the linear shape of $E_0(R)$ (e.g. for $R\sim 
20M_1^{-1}$ the deviation from the linear behaviour is reduced by at least 
one order of magnitude).

The extrapolated function $E_0(R)$ was then subjected to a finite-size 
analysis. The masses $M_i$ of the stable particles were estimated from the
higher levels $E_i (R)$, and the expected leading finite-size corrections
(the sum in \fsize) were subtracted from $E_0 (R)$.

For negative $\eta$ in Table 4 we subtracted all one-particle 
corrections and then used
the function $f_0\,R + f_1\,\exp(-2\,f_2\,R)$ to fit (around $R \approx
4\,M_{1}^{-1}$) the remaining part via $f_0, f_1, f_2$, with the best-fit
value of $f_0$ accepted as $F(m,h)$. For positive $\eta$ we subtracted
instead the corrections from the two lightest particles and then used
$g_0\,R + g_1\,K_1 (g_2\,R)$ to fit the resulting curve (around $R \approx
7\,M_{1}^{-1}$). The overall quality of this analysis was controlled by
observing the proximity of the best-fit value of $f_2$ and $g_2$ to the
expected values $2\,M_1$ and $M_3$, respectively (the deviation varied
with $\eta$, but never exceeded $20\%$).

The above procedure of the finite-size analysis is rather labor consuming.
In order to obtain estimates of the coefficients $\Phi_n$ we used similar
procedure in the small interval~\hbox{$[-0.64:0.64]$} around zero but we 
only
subtracted the lighest particle correction before fitting to $g_0\,R +
g_1\,K_1 (g_2\,R)$ around $R \approx 8\,M_{1}^{-1}$.

We believe that the data for $F(m,h)$ obtained through this procedure are 
exact to six significant digits.

Finally, massive data used in Fourier analysis in Sect.~8.2.1 was obtained 
using somewhat simplified procedure, in which only the first term in the 
sum \fsize\ was taken into account, i.e. no subtraction was performed. 
Specifically, the function $g_0\,R + g_1\,K_1 (g_2\,R)$ was fitted to the 
extrapolated data $E_0 (R)$. This procedure produces the estimates which 
agree with the above more elaborate analysis to five significant digits. 

Similar simplified analysis was used in the case of pure imaginary $h$.  
In this case, for~$y>-Y_0$, complex values of the fitting parameters 
$g_0$, $g_1$, $g_2$ were admitted. Exceptionally, for the values on the 
side $y<-Y_0$ presented in Table 5, we subtracted the leading finite size
correction and then fitted to $g_0\,R+g_1\,\exp(-g_2\,R)$. Special was the
vicinity of the critical point $-Y_0$. With $y$ approaching this point the
values of $g_2$ decreased and overall quality of the fit deteriorated. Of
course this is not surprising since near the critical point the
correlation length $M_{1}^{-1}$ is diverging, pushing the domain
of validity of the expansion~\fsize\ to $R\to\infty$, where our analysis
would be severely spoiled by truncation effects. Nonetheless, it was
possible to obtain reasonably accurate data (at least four digits) for
$\Phi_{imh}(y)$ with $y < -2.43$.  On the other side of the critical
point, for $y>-Y_0$, the situation is complicated by the rapid growth of
the ``nose''. In this region we could obtain similar accuracy only for
$-2.1<y$, and our data for $-2.3<y<-2.1$ may be less accurate. In close
vicinity of the critical point a finite-size analysis based on the
effective action \LYeff\ would be more appropriate, and we are planning to
apply it in the future.

\newsec{Numerical results}

\subsec{Scaling function. Real $h$}

Our results for $\Phi(\eta)$ at real values of $\eta$ are shown in
Fig.~10 (see also Table~4 for some numerical values). As expected (see 
Sect.~3), the function appears to be perfectly
smooth everywhere exept for the point $\eta=0$, where the singularity
of the form $-\eta^2\,\log\eta^2/8\pi$ can be observed. The first few 
coefficients of the \hbox{$\eta$-expansion} \phiseries\ can be readily 
extracted
from this data. After subtracting the above singular term, we have fitted
polynomials of various degree to the data in the interval $[-0.64:0.64]$. 
The fit appears stable for the first eight coefficients, $\Phi_0$ to  
$\Phi_7$. The resulting estimates of these coefficients are displayed 
in Table~3. 
The estimates for the coefficients $\Phi_0$ and $\Phi_1$ are
very close to the predictions~\Phizero\ and \PhioneB; this result for
$\Phi_1$ can be viewed as numerical verification of the exact
expectation value $\langle \epsilon \rangle|_{\tau = 0}$ obtained in
\FLZZ. It turns out that the first few terms of the series \phiseries\
provide rather good approximation for $\Phi(\eta)$ with 
$|\eta|\lesssim 2$, 
as shown in Fig.~10. Comparison with the known terms of the
expansions \phiassminus\ and \phiassplus\ is also presented in that
Figure. 

\fig{10}{The scaling function $\Phi(\eta)$ for real $\eta$ (solid line). 
Dashed lines show how this function is approximated by few leading terms 
of the expansions \phiseries, \phiassminus\ and \phiassplus.}
{\epsfxsize=9cm\epsfbox{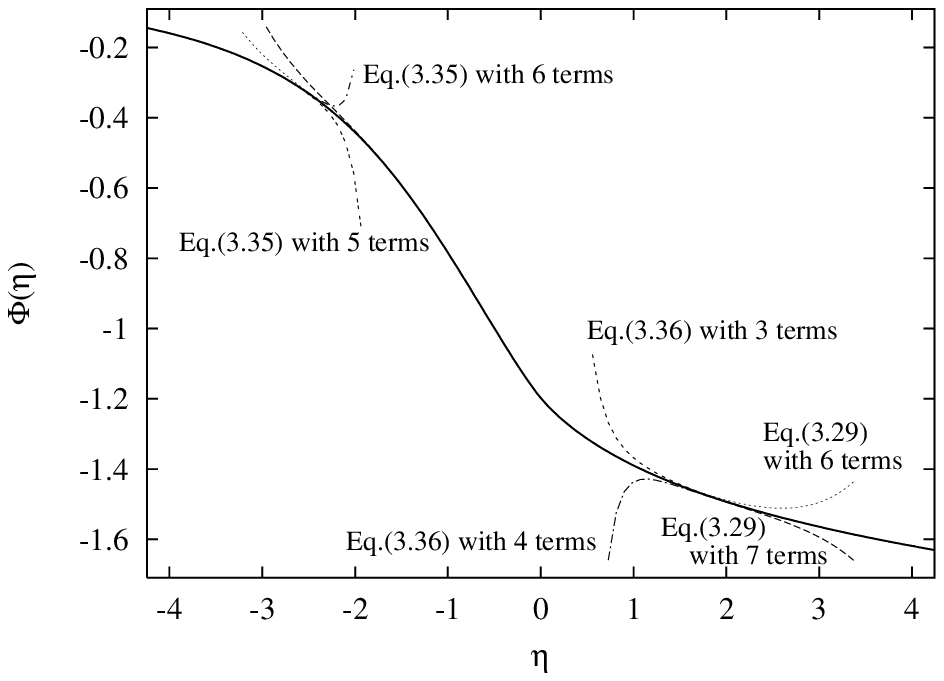}}

\subsec{Scaling function. Imaginary $h$}

\fig{11}{
The TFFSA data for the function $\Phi_{imh}(y)$,
Eq.~\Phiimh\ (empty and full bullets). The imaginary part $\Im m
\Phi_{imh}(y)$  vanishes for $y < -Y_0$. Various dashed lines show how
this function is approximated by its expansions \phiseries, \phiassminus\
and \phiassplus. I6 and I7 are the first seven (including
${\bar\Phi_6}\,y^6$)
and eight (including ${\bar\Phi_7}\,y^7$) terms of
\Phiimexp. Corresponding expansions of the real part are the lines R6 and
R7. One, three, and four terms of the asymptotic expansion \phiassplus\
are
shown as LT1, LT3, and LT4, respectively (obviously, only even (odd)
terms contribute to the real (imaginary) part). Finally, HT12 is the
plot of six leading terms (including $G_{12}$ term) of the low-T
expansion \phiassminus. In all cases the coefficients are taken (or
computed) from Tables 1, 2 and 3.
}
{\epsfxsize=11cm\epsfbox{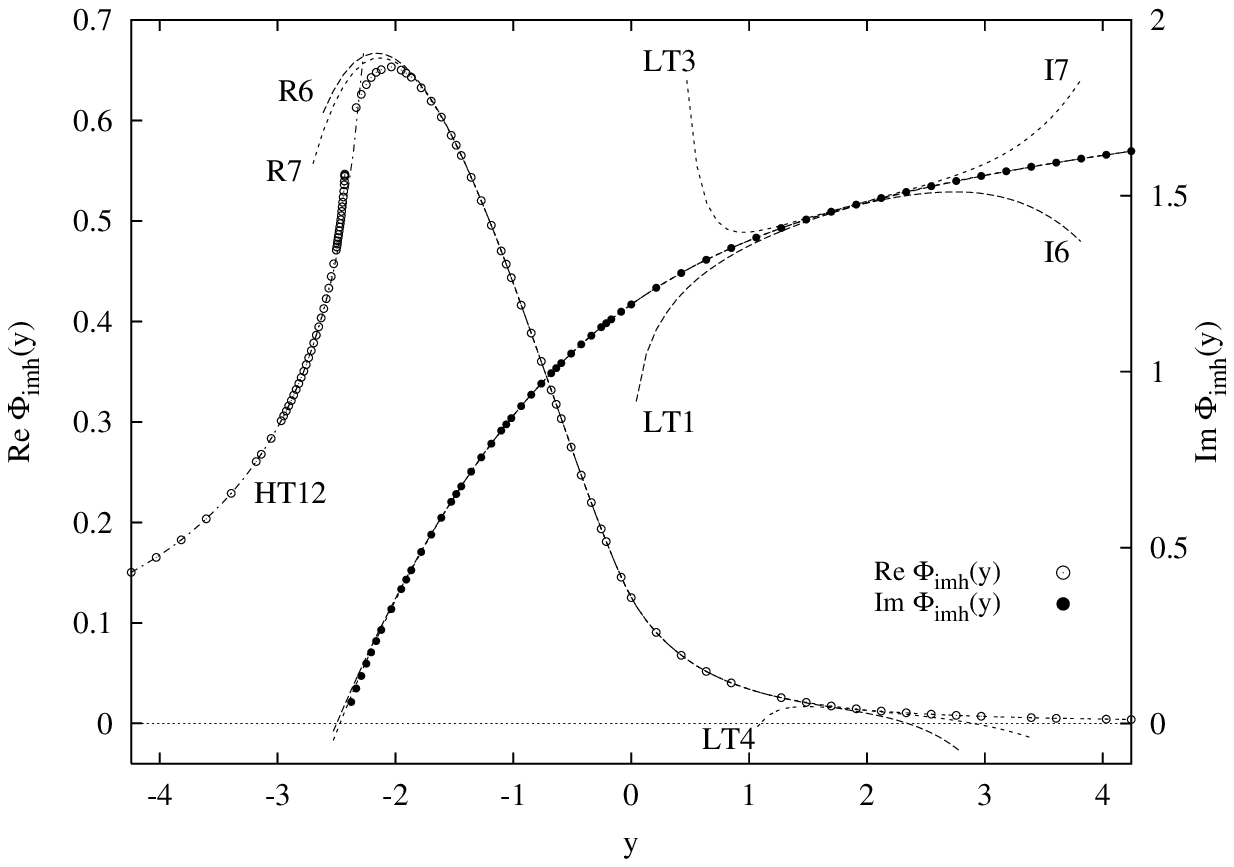}}

Our results for $\Phi_{imh}(y)$ are shown in Fig.~11, where a
singularity at $y = - Y_0$,
\eqn\Yzero
{
Y_0 \approx 2.4295\,,
}
is clearly visible. The function
$\Phi_{imh}(y)$ is real for $y<Y_0$, and it develops an imaginary
part for $y>Y_0$. A comparison with the first few terms of the
expansions \phiassminus\ and \phiassplus\ is also made in the 
Fig.~11. We observe that the first six terms of \phiassminus\
approximate $\Phi_{imh}(y)$
very well for $y<-2.7$. This is not surprising since, according to our
discussion in Sect.~3, the expansion \phiassminus\ is expected to
converge for all $y<Y_0$. In this Figure we also compare the data
with the first eight terms of the expansion \Phiimexp\ (and similar
expansion for the real part of $\Phi_{imh}(y)$). One can notice that
these expansions provide a remarkably good approximation of
$\Phi_{imh}(y)$ for $|y|<2$. We take
this as an indication that it is the Yang-Lee singularity at 
$\eta = - Y_0\,e^{\pm {{4\pi i}\over 15}}$ which determines the
domains of convergence of the small
$\eta$ expansion \phiseries; in other words, it supports our
expectation that the function ${\tilde\Phi}(\eta)$ is analytic in the 
whole disk $|\eta| < Y_0$. 

Our data agree with the expected form \philyang\ of the singularity at
$y= - Y_0$. The imaginary part of $\Phi_{imh}(y)$, Eq.~\Dimh,
receives no contribution from the regular part $A(y)$ in \philyang, 
and therefore it should exhibit the singular behaviour \philyang\ most 
prominently. The log-log plot of this function in Fig.~$12a$ clearly 
confirms the value $5/6$ of the exponent in \philyang\ predicted in 
\Cardy. The ratio $\Im m\,\Phi_{imh}(y)/(y+Y_0)^{5\over 6}$ is shown in 
Fig.~$12b$. 
\vskip12pt
\nobreak
\midinsert
\centerline{
\vbox{\hsize=13cm
\hbox{
\vbox{
\hbox{\epsfxsize=6.5cm
\epsfbox{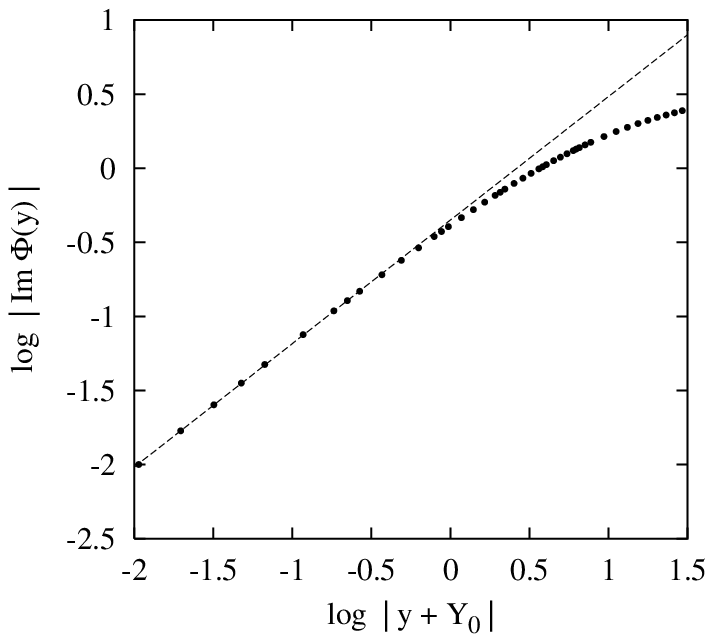}}
\vbox{\hsize=6.5cm
\centerline{\qquad\qquad\nineb Fig.~12{\nineib a}}}}
\vbox{
\hbox{\epsfxsize=6.5cm
\epsfbox{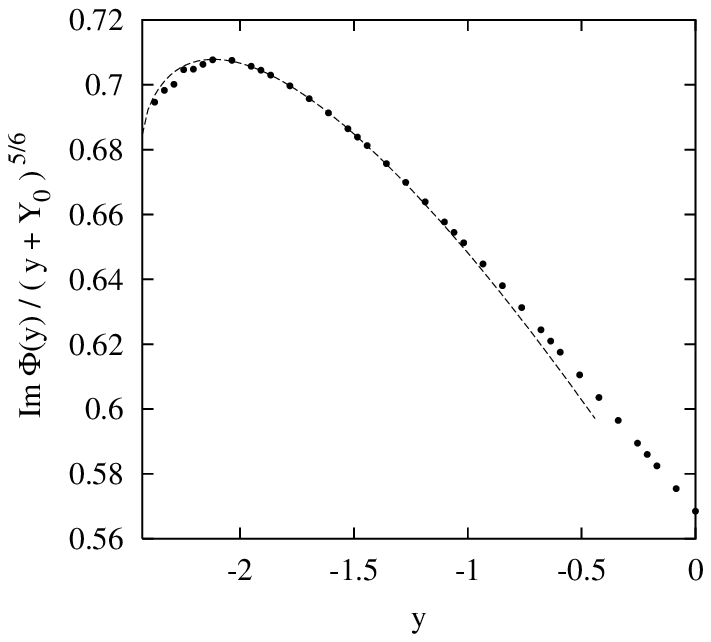}}
\vbox{\hsize=6.5cm
\centerline{\qquad\qquad\nineb Fig.~12{\nineib b}}}}
}
\vskip 10pt
\hbox{
\vbox{
\baselineskip=12pt\ninerm\ninemath\noindent
{\nineb Fig.~12:}~TFFSA data for $D_{imh}(y)$, 
Eq.~\Dimh.~{\nineb {\nineib a})} Log-log plot of
$D_{imh}(y)$ (bullets). The dashed line has slope $5/6$; 
{\nineb {\nineib b})} Ratio 
$D_{imh}(y)/(y+Y_0)^{5/6}$ (bullets), and the graph of
\hbox{$D_{imh}^{\,<}(y)/(y+Y_0)^{5/6}$}, with $D_{imh}^{\,<}(y)$ from 
Eq.~\empminus\ 
(dashed curve).
}\hfil}
}}
\endinsert\goodbreak

This plot reveals the ``fine structure'' of this function near the 
Yang-Lee branching point, which should be attributed to the subleading 
singular terms in \philyang.  Our data is not very precise when taken 
close to the critical point $-Y_0$ (see Sect.~6), and allow for only 
rather rough estimate of the leading coefficients in the expansion 
\philyang, (3.41),
\eqna\Bs
$$\eqalignno{
B_0 =  -1.37(2)\,, \qquad B_1 &= -0.75(5)\,, 
\qquad C_0 = 0.38(7)\,, &\Bs a
\cr
A_0 = 0.5474(3)\,, \qquad A_1 &= 1.06(4)\,, \qquad\quad A_2 = 0.3(1)\,. 
&\Bs b
}$$
These estimates are obtained by fitting a few leading terms of (3.41) to
the data in various intervals of $y$ between $-2.2$ and 
$-1.5$; the
error reflects the dependence on the fitting interval as well as on
the number of the terms used. From \Bs a\ we can
obtain an estimate of the most important parameters in 
\alphazero a, \alphazero b,
\eqn\lavalues
{
\lambda_1 = 3.10(6)\,, \qquad \alpha_0 =  -4.9(9)\,.
}

Independent (and somewhat better) estimate of the parameter $\alpha_0$
can be obtained directly from the finite-size ground-state energy $E_0
(R)$ at the Yang-Lee point $y=-Y_0$. By fitting the expected form
\Ezeroflow\ to the data in Fig.~9 in various regions between
\hbox{$R=1.7\,|h|^{-{8\over 15}}$} and $R=2.8\,|h|^{-{8\over 15}}$ we 
found
\eqn\alphazero
{
\alpha_0 = -4.4(4)\,,
}
(where again the error estimate is from the dependence on the fitting
region). The quality of the fit is also shown in Fig.~9.

\subsec{False vacuum resonance width}

As was discussed in Sect.~5, the finite-size behaviour of the meson levels
close to their ``near-intersection'' points is related through \opening\
to the width $\Gamma$ of the false vacuum resonance, and can be used for
numerical determination of $D_{meta}(y)$. Presently, there are two 
limitations to this approach. First, at small $h$ the ``near-
intersections'' occur at large $R$, where truncation effects are
more significant. In our calculations, a satisfactory stabilization 
of the difference $E_2(R)-E_1(R)$ at $L=12$ was achieved only for 
$\eta < 3.5$. For larger $\eta$ the situation can be somewhat
improved by adding a number of two-quark states of higher levels 
(see below). On the other side of large $h$, a limitation comes from
our unsufficient knowledge about the coefficients $\sigma'_i$ in
\opening; the best we have at the moment is the first three terms
\sigmais\ of their small-$h$ expansions. Nevertheless, one can note
that these three terms were obtained in Appendix B within the same
approximation as the three correction terms in the large-$\eta$
expansion of the meson masses, Eq.~\Miexp. Good agreement of the
last expansion with the data at $\eta > 1.7$ makes us believe that
reasonably accurate estimate of the function $D_{meta}(\eta)$ can be
obtained within this expansion in the same region.

We have performed this analysis in the range $1 < \eta < 4.5$
using the separation \hbox{$E_2 (R)- E_1 (R)$}. The eigenvalues $E_1(R),
E_2(R)$ were computed in ``two-quark extended'' truncated space
(2.5) with $L=12$. Namely, besides all states in (2.5) with 
$L\leq 12$, the two-quark states $a_{k}^{\dagger}a_{-k}^{\dagger}\mid
0\rangle_{\rm NS}$ and $a_{n}^{\dagger}a_{-n}^{\dagger}\mid
0\rangle_{\rm R }$ with the levels $12 < L \leq 100$ were also
admitted. While such extention does not bring any appreciable
difference for $\eta \leq 3.5$, it does substantially improve the
results for $3.5 < \eta < 4.5$. For all values of $\eta$ in the range
$1 < \eta < 4.5$ the separation $E_2 (R)-E_1 (R)$ was in excellent
agreement with the expected square-root behaviour \opening, and it
allowed us to estimate the values of the width parameter $\Gamma$. 
The results for the ratio $D_{meta}(y)/V(y)$ (where $D_{meta}(y)$ is
related to $\Gamma$ through \Gammafunction, and $V(y)$ is the function
\Vdef) are shown in Fig.13. The ratio shows a clear tendency to
approach at large $y$ the expected value $V_0 = 0.2161...\,$. Moreover, 
when
all three terms in \sigmais\ are taken into account, for $y>2.3$
it agrees well with the first correction term in \discass. By direct
fitting the expansion \discass\ to this data we have obtained the
following estimates of the coefficients there,
\eqn\vfits
{
V_0 = 0.2161(4)\,, \qquad V_1 = -0.136(4)\,.
}

\fig{13}{
Data for the ratio $D_{meta}(y)/V(y)$ obtained through the
false vacuum resonance width analysis (Sect.~7.3), using Eq.~\opening\ 
with
three leading terms in Eq.~(5.14). The data obtained using ``two-quark
extended'' truncation space with the truncation level $L=12$ is shown
as empty bullets ($\odot$). The data obtained using two-quark states 
{\it only} is
shown as full bullets ($\bullet$). The solid line is the plot of
Eq.~\Dmetaplus\ and the dotted line is the value $V_0=0.2161$.
}
{\epsfxsize=9cm\epsfbox{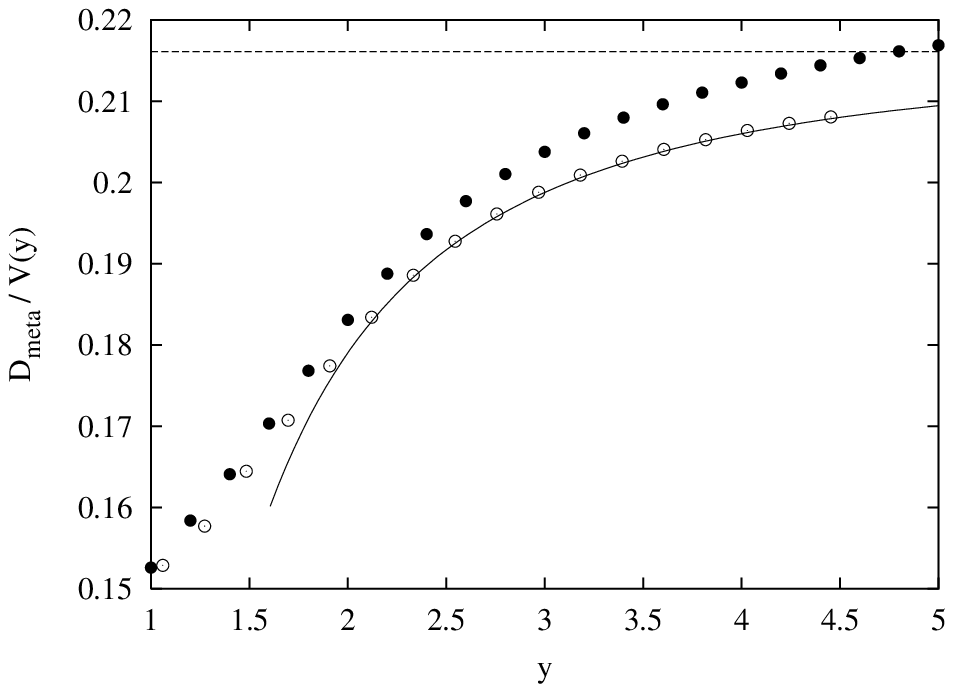}}

In \Rutkevich\ an attempt was made to determine the $h\to 0$
behaviour of $\Gamma$ using two-quark approximation for the resonance
wave function. While correctly reproducing the exponent in \voloshin,
this calculation yields somewhat greater pre-exponential factor 
$\pi {\bar\sigma} h/18$ (which would lead to $V_0 = \pi {\bar s}/18
\approx 0.237$ in our notations). As we explain in Appendix~B, the 
two-quark
approximation breaks down when the quarks are deep inside the
classically unaccessible region, and therefore it is not justified in
the resonance wave function calculation even at $h \to 0$, even though
in this limit the resonance decay goes predominantly through the
two-quark channel\foot{To avoid misunderstanding let us
stress here that in Appendix B we apply the two-quark approximation to
the wave function in the ``meson domain'', not deep under the
barrier. \hbox{Instead} of solving the problem in the last region, we take
a phenomenological approach based on the \hbox{Breit-Wigner} formula
\Bxxiv.}. To verify this being the reason for the above discrepancy,
we have repeated the calculation described in this Section, this time
with the eigenvalues $E_1(R), E_2(R)$ computed purely within the
two-quark approximation. Specifically, in computing these
eigenvalues, {\it only} two-quark states with the levels $\leq 100$
were admitted. As is seen in Fig.~13, the ratio $D_{meta}(y)/V(y)$
obtained this way is hardly consistent with the correct $y\to\infty$ 
asymptotic value $0.2161...\,$.

\newsec{Analyticity}

\subsec{High-T domain (HTW in Fig.~4)}

In principle, our approach allows for direct computation of the
function $\Phi(\eta)$ at complex values of $\eta$ in the wedge HTW in
Fig.~4. However, this data would be redundant because in view of the 
analytic properties of the scaling function discussed in Sect.~3, 
the values of $\Phi(\eta)$ can be computed through the dispersion
relation \exphigh\ once the function $D_{imh}(y)\equiv 
\Im m\, \Phi_{imh}(y)$ is
known. This dispersion relation applies to the function $\Phi(\eta)$
with $\eta$ in the wedge HTW in Fig.~4., where it can be rewritten as
\eqn\etahighdisp
{
\Phi(\eta) = - {{15\,\eta^2}\over{4\,\pi}}\,\int_{0}^{Y_0}\,
{{D_{imh}(-y)\,y^{3\over 4}\,dy}\over{y^{15\over 4}+(-\eta)^{15\over
4}}}\,.
}
Since accurate direct data on $D_{imh}(y)$ in the vicinity of the Yang-Lee
point \hbox{($-Y_0 < y < 2.1$)} is not available at this moment, 
some
extrapolation is needed in order to perform the (numerical)
integration in \etahighdisp. We have tried various extrapolations,
and the results of integration in \etahighdisp\ 
usualy agree with the our direct data on $\Phi(\eta)$ to at least three
significant digits. The best results were obtained with the following
approximation
\eqn\empirical
{
D_{imh}(y) =
\left\{\eqalign{
D_{imh}^{\,<}(y)\qquad &{\rm for}\quad -Y_0 < y < -1.57\,;
\cr
D_{imh}^{\,>}(y)\qquad 
&{\rm for}\quad -1.57 < y < 0\,.
}\right.\,,
}
where
\eqn\empminus
{
D_{imh}^{\,<}(y) =- {1\over 2}\,{\bar B}_0\ (y+Y_0)^{5\over 6} + 
{1\over 2}\,{\bar B}_1\ (y+Y_0)^{11\over 6}+{\sqrt{3}\over 2}\,{\bar C}_0\,
(y+Y_0)^{5\over 3}\,,
}
with
\eqn\BCbars
{
{\bar B}_0 = -1.3693\,, \qquad {\bar B}_1 = -0.74378\,, \qquad {\bar
C}_0 = 0.42446\,,
}
and $D_{imh}^{\,>}(y)$ is just the first eight terms of the $y$-expansion,
\eqn\empplus
{
D_{imh}^{\,>}(y) = {\bar\Phi}_0 + {\bar\Phi}_1\,y + \cdots + 
{\bar\Phi}_7\,y^7\,.
}
with ${\bar\Phi}_{n}$ related to $\Phi_{n}$ from the Table 3 as in
\Phibars. Here we accept \Yzero\ as the value of $Y_0$, while the 
coefficients ${\bar B}_0 , {\bar B}_1, {\bar C}_0$ in 
\BCbars\ are obtained by exact matching \empminus\ with \empplus\ at 
$y = -1.57$ (up to the first derivative), and fine-tuning the remaining
one parameter to achieve better agreement with the data. Note that
these coefficients agree with the estimate \Bs a\ within the stated
accuracy. However, \BCbars\ should not be considered as better
estimates of the coefficients of the expansions ($3.41b$)--($3.41c$), 
since \empminus\ 
ignores higher subleading terms in \philyang.
The quality of the approximation \empirical\ is demonstrated in the
Tables 4 and 5, where the results from \etahighdisp\ are compared to
our data, both for real $\eta < 0$ and $\eta = y\,e^{{4\pi i}\over
15}\,, y < - Y_0$. We also presented in Table 1 the results for 
the first few coefficients $G_{2n}$ in \highexp\ obtained through the 
dispersion relation \gIIn\ with the use of \empirical. The
agreement with the previously known estimates is reasonable, baring in
mind relatively low quality of the approximation \empminus; we
believe it will improve once the structure of singular expansion
\philyang\ is understood in greater quantitative details.

\eject

\table{4}{Numerical values of $\Phi(\eta)$ at selected real points. The
first column shows direct data obtained through TFFSA, as explained in
Sect.~6.2; we believe these numbers are exact to six significant
digits or better. The results from high-T and low-T dispersion
relations, Eqs.~\etahighdisp\ and \lowdispeta, with the use of 
approximations \empirical\ and \Dmetaapprox, are presented in the second 
and the third columns, respectively. The last column contains results from 
extended dispersion relation, Eq.~\extdisp, with approximation 
\Deltaapprox\ used.}
{
\vbox{
\offinterlineskip
\halign{
\strut
$#$\hfil&
\enskip\vrule#\enskip&
\hfil$#$\hfil&
\enskip\vrule#\enskip&
\hfil$#$\hfil&
\enskip\vrule#\enskip&
\hfil$#$\hfil&
\enskip\vrule#\enskip&
\hfil$#$\hfil
\crcr
&
&
\hbox{\rm From}&
&
\hbox{\rm From high-T}&
&
\hbox{\rm From low-T}&
&
\hbox{\rm From extended}   
\cr
\Phi(\eta)&
&  
\hbox{\rm TFFSA data}&
&
\hbox{\rm disp. relation}&
&
\hbox{\rm disp. relation}& 
&
\hbox{\rm disp. relation}   
\cr
\omit&height2pt& &height2pt& &height2pt& &height2pt&
\cr 
\noalign{\hrule}
\omit&height2pt& &height2pt& &height2pt& &height2pt&\cr
\Phi(-5) && -0.1092101 && -0.1092092 && \hbox{\rm ---} && -0.1088626\cr
\Phi(-4) && -0.1592682 && -0.1592643 && \hbox{\rm ---} && -0.1589421\cr
\Phi(-3) && -0.2529928 && -0.2529887 && \hbox{\rm ---} && -0.2527417\cr
\Phi(-2) && -0.4413450 && -0.4413249 && \hbox{\rm ---} && -0.4412136\cr
\Phi(-1) && -0.7839665 && -0.7839668 && \hbox{\rm ---} && -0.7839576\cr
\Phi(0) && -1.1977330 && \hbox{\rm ---} && \hbox{\rm ---} && -1.1977320\cr
\Phi(1) && -1.3898410 && \hbox{\rm ---} && -1.3898417 && -1.3898063\cr
\Phi(2) && -1.4930558 && \hbox{\rm ---} && -1.4930566 && -1.4929849\cr
\Phi(3) && -1.5642732 && \hbox{\rm ---} && -1.5642736 && -1.5641727\cr
\Phi(4) && -1.6188506 && \hbox{\rm ---} && -1.6188510 && -1.6187275\cr
\Phi(5) && -1.6632483 && \hbox{\rm ---} && -1.6632485 && -1.6631076\cr
}}
}

\vskip 16pt

\table{5}{Numerical values of $\Phi_{imh}(y)$, Eq.~\Phiimh, at some real
$y < -Y_0$. Direct data from TFFSA (good to four digits or better,
see Sect.~6.2) are in first column, and results from high-T dispersion
relation with \empirical\ in second column.}
{
\vbox{
\offinterlineskip
\halign{
\strut
$#$\hfil&
\enskip\vrule#\enskip&
\hfil$#$\hfil&
\enskip\vrule#\enskip&
\hfil$#$\hfil
\crcr
&
&
\hbox{\rm From}&
&
\hbox{\rm From high-T}
\cr
\Phi_{imh}(y)&
&
\hbox{\rm TFFSA data}&
&
\hbox{\rm disp. relation}
\cr
\omit & height2pt & & height2pt& 
\cr
\noalign{\hrule}\omit & height2pt & & height2pt
\cr
\Phi_{imh}(-5.0) && 0.1116003 && 0.1115969 \cr
\Phi_{imh}(-4.5) && 0.1349420 && 0.1349380 \cr
\Phi_{imh}(-4.0) && 0.1674520 && 0.1674469 \cr
\Phi_{imh}(-3.5) && 0.2155503 && 0.2155426 \cr
\Phi_{imh}(-3.0) && 0.2946233 && 0.2946092 \cr
\Phi_{imh}(-2.5) && 0.4729294 && 0.4728305 \cr
\Phi_{imh}(-2.4295) && \hbox{\rm ---} && 0.5475373 \cr
}
}
}

\vfil\eject

\subsec{Low-T domain (LTW in Fig.~4)}

\subsubsec{8.2.1 Fourier analysis}

Numerical analytic continuation into this domain was performed using
Fourier \hbox{analysis} in the logarithmic variable $\log \eta$.
According to \phiseries\ and \phiassplus\ the function 
\hbox{$\eta^{13\over16}\,\(\Phi(\eta)-{\tilde G}_1 \,\eta^{{1\over 8}}\)$}
decays sufficiently
fast (exponentially in the above logarithmic variable) both at $\eta\to 0$ 
and at $\eta\to\infty$. Its logarithmic Fourier transform 
\eqn\ft
{
\Psi(\omega) = \int_{0}^{\infty}\,\eta^{{13\over
16}-i\omega}\,\,(\Phi(\eta)-{\tilde G}_1 \,\eta^{1\over 8}) \,{{d\eta}\over\eta}
}
was evaluated numerically from our data for $\Phi(\eta)$. In
principle, using this function the values of $\Phi_{low}(\eta)$,
Eq.~\Philow, along the
rays $\eta = y\,e^{i\alpha}$, with real $y\in (0,+\infty)$, can be 
computed as the inverse Fourier transforms
\eqn\ift
{
\Phi_{low}(y\,e^{i\alpha}) = {\tilde G}_1 \,e^{i{\alpha\over
8}}\,y^{1\over 8} + e^{-i\,{13\over 16}\,\alpha}\,y^{-{13\over
16}}\,\int_{-\infty}^{\infty}\,e^{-\alpha\omega}\,\Psi(\omega)\,
y^{i\omega} {{d\omega}\over {2\pi}}\,.
}
The function $\Phi_{low}(\eta)$ is analytic along the ray provided the
integrand in \ift\ decays faster then any power of $\omega$. The
standard analyticity assumption for the wedge LTW in Fig.~4 is
equivalent to the statement that $|\Psi(\omega)| < \exp\(-{8\over
15}\pi|\omega|\)$ for $|\omega| \to \infty$. Of course it is not
possible to achieve a rigorous conclusion about true large $\omega$
asymptotic of this function using our limited-precision data. For 
$\omega > \omega_{max} \sim 5.5$ the numerical Fourier 
transform $\Psi(\omega)$ is dominated by a 
small ($\sim 10^{-6}$) random-looking function which should be
interpreted as the Fourier transform of the errors associated
with our numerical data for $\Phi$. For this reason the numerical
Fourier transform $\Psi(\omega)$ hardly carries any usefull information 
beyond the window $|\omega| < \omega_{\rm max}$. At the same time 
the function $\Psi(\omega)$ inside this window appears to be
decaying significantly faster then the above exponential, which is
consistent with the standard analyticity assumption. One can actually 
use \ift\ to
compute this function in LTW with reasonable accuracy. Obviously, in 
order to do that one has to introduce some cut-off to eliminate the
high frequency part $|\omega| > \omega_{\rm max}$ of the integral 
\iftdisc\
dominated by the noise; it is expected that only fine details of the 
resulting function would depend on the specifics of the cutoff
procedure. As an example 
(which will turn out to be usefull in our analysis in Sect.~8.3) we show
in Fig.~14 the shape of the function $\Phi_{low}(\eta)$ along the ray 
$\eta = y\,e^{i{\pi\over 3}},\ y>0,$ computed this way. Any changes 
induced by changing the cutoff procedure are by far too small ($<10^{-4}$) 
to be visible in this plot. 

\fig{14}{
The scaling function $\Phi_{low}(\eta)$ along the axis $\eta =
y \,e^{{i\pi}\over 3}$ obtained through numerical analytic
continuation (Sect.~8.2.1). Real and imaginary parts of $e^{-{{2\pi
i}\over 3}}\,\Phi_{low}(y\,e^{{i\pi}\over 3})$ are shown. The $y<0$
part of this axis lays inside the shadow domain SHD in Fig.4, close to
the axis OA. The data here is not very accurate. Nevertheless,
an anomaly around $y \simeq -2.5$ (presumably associated with the 
proximity of the YL singular point) seems to be developing.
}
{\epsfxsize=9cm\epsfbox{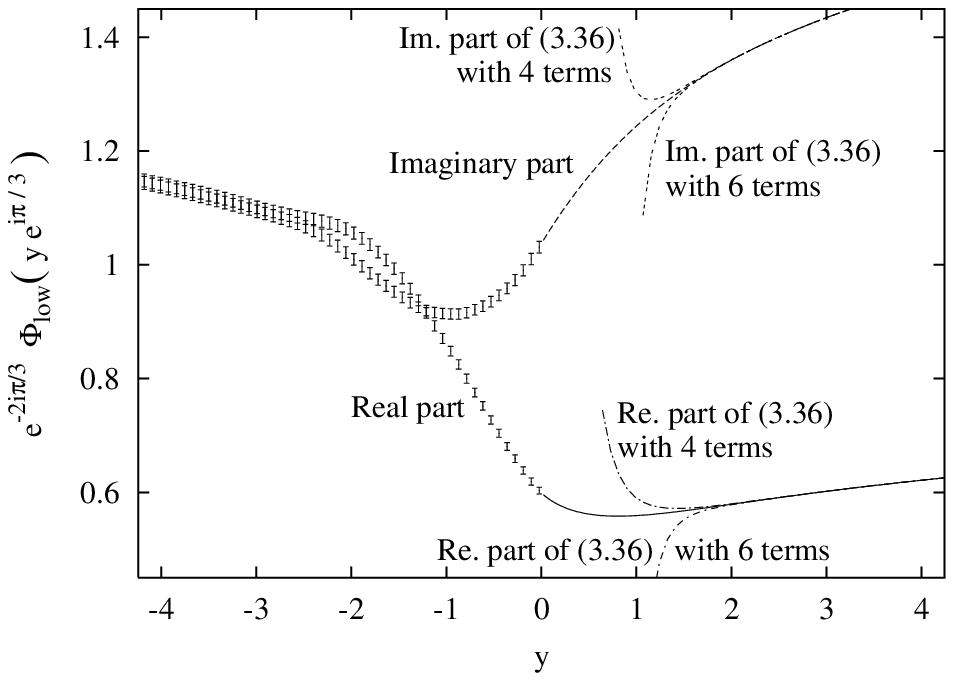}}

\noindent
More interestingly, one can apply the relation
\eqn\iftdisc
{
D_{meta}(y) = - \int_{-\infty}^{\infty}\,y^{-{13\over
16}-i\omega}\,\cosh(8\pi\omega/15)\,\Psi(\omega) \,{{d\omega}\over {2\pi}}
}
(which is equivalent to the dispersion relation \lowdisp) to compute
the function \Dfunction. As expected, the resulting inverse Fourier 
transform \iftdisc\ shows rapid decay at large $y$ (see Fig.~$15a$). 
However, instead of monotonous decay \dmetaass, the cutoff integral \ift\
exhibits small oscillations whose shape is very sensitive to the details
of the cutoff procedure. Therefore, we consider the results for the function
$D_{meta}(y)$ obtained this way only reliable for 
$y < 2.0$. To see how
these results comply with the expected asymptotic form \discass\ we have
plotted in Fig.~$15b$ the ratio $D_{meta}(y)/V(y)$, where $V(y)$ is the 
exponent \Vdef.
The error bars in this plot represent the spread of the data obtained with
different cutoff procedures. From \discass, this ratio is expected to
approach the constant $0.2161...$ as $y\to\infty$. As is seen from the
plot, the ratio reaches a minimum $\approx0.153$ at $y\approx
1.15$, and then shows a tendency to increase towards the 
expected value. 

\nobreak
\midinsert
\centerline{
\vbox{\hsize=13cm
\hbox{
\vbox{
\hbox{\epsfxsize=6.5cm
\epsfbox{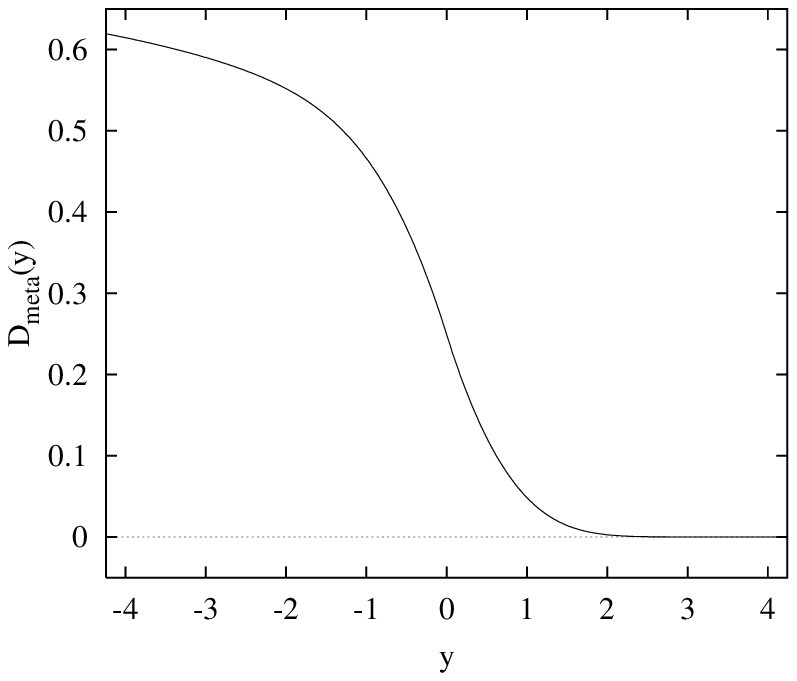}}
\vbox{\hsize=6.5cm
\centerline{\qquad\qquad\nineb Fig.~15{\nineib a}}}}
\hskip 5pt
\vbox{
\hbox{\epsfxsize=6.5cm
\epsfbox{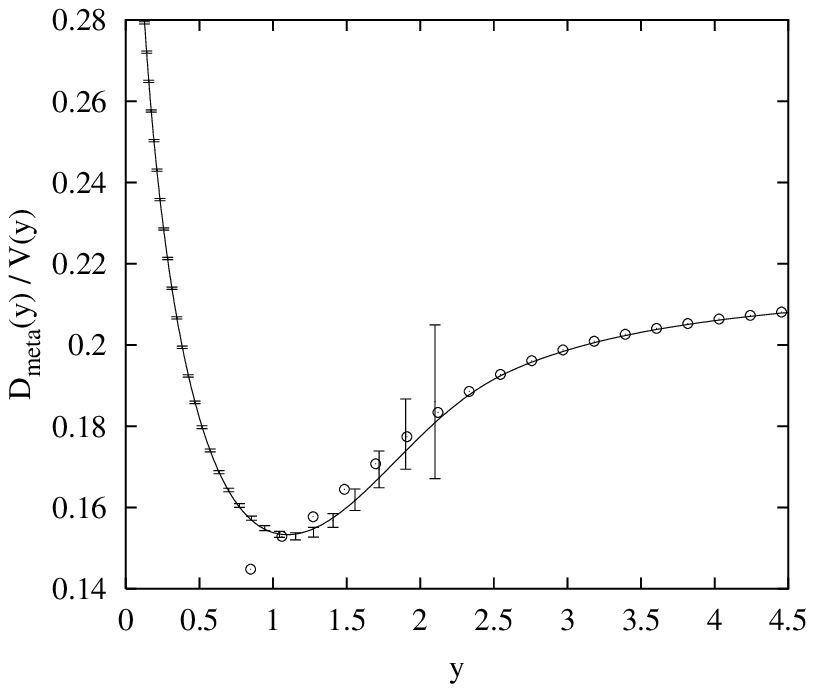}}
\vbox{\hsize=6.5cm
\centerline{\qquad\qquad\nineb Fig.~15{\nineib b}}}}
}
\vskip 10pt
\hbox{
\vbox{
\baselineskip=12pt\ninerm\ninemath\noindent
{\nineb Fig.~15:}~{\nineb{\nineib a})}~The function $D_{meta}(y)$, 
Eq.~\Dfunction, obtained through
numerical analytic continuation in Sect.~8.2.1; {\nineb{\nineib b})} The 
ratio $D_{meta}(y)/V(y)$ (where $V(y)$ is defined in Eq.~\Vdef). The data
from numerical analytic continuation (Sect.~8.2.1) (shown as
error bars) {\it vs.} the data from resonance width analysis (Sect.~7.3) 
(shown as empty bullets). The solid curve is the plot of
Eq.~\Dmetaapprox.
}\hfil}
}}
\endinsert\goodbreak

Unfortunately, as was explained above, for $y > 1.9$ the
data obtained by this method becomes unreliable, and presently neither 
the true large $y$ asymptotic nor the way it is approached can be
established on their basis. But fortunately, the analysis of Sect.~7.3 
apparently provides a complimentary set of data. We replot in Fig.~$15b$ 
the data from the resonance width analysis (identical to those in
Fig.~13) to demonstrate its agreement with the data obtained by the 
Fourier analysis.

In principle, one could try to use this numerical Fourier transform
$\Psi(\omega)$ to continue further, into the shadow domain, by taking
$|\alpha| > 8\pi/15$ in \ift. We found that while for $|\alpha| < 3\pi/5$
the intergal in \ift\ still receives dominating contribution from
$\omega$ well within the window $|\omega| < \omega_{max}$ (and hence
is not too sensitive to the cutoff procedure), for larger $\alpha$ the
domain around $\omega_{max}$ becomes more significant. The result of
such computation with $\alpha=-2\pi/3$ is shown in the left-hand side
of Fig.~14; its accuracy is hardly better then $10^{-2}$, and may be
worse.  For $|\alpha| > 2\pi/3$ the results are too sensitive to the 
cutoff procedure to be
taken seriously. In other words, at present the accuracy of our direct
data about $\Phi_{low}(\eta)$ is not sufficient to support (or unvalidate)
analyticity of this function in the whole of the shadow domain. In
Sect.~8.3 below we use another approach, based on the extended dispersion
relation, Eq.~\extdisp, to verify the extended analyticity.

\subsubsec{8.2.2 Low-T dispersion relation}

According to \lowdisp, in the low-T wedge $-8\pi/15 < {\rm arg}(\eta)
< 8\pi/15$ (the domain LTW in Fig.~4) the scaling function $\Phi(\eta)$ 
is represented in terms of the function $D_{meta}(y)$ as
\eqn\lowdispeta
{
\Phi(\eta)={\tilde G}_1 \,\eta^{1\over 8} - {{15\,\eta^{-{7\over 4}}}\over
{8\,\pi}}\, \int_{0}^{\infty}\,{{D_{meta}(y)\,y^{-{9\over
8}}\,dy}\over {y^{-{15\over 8}}+\eta^{-{15\over 8}}}}\,.
}
It is interesting to check how the above information about $D_{meta}(y)$ 
agrees with our direct data on this scaling function. Inspection of
the plot in Fig.~$15b$ shows that the function $D_{meta}(y)$ is
well approximated by the first eight terms of its power series
expansion \DnsA, i.e. by the polynomial
\eqn\Dmetaminus
{
D_{meta}^{\,<}(y) =D_0 + D_1\,y + \cdots + D_7\,y^7\,,
}
whith the coefficients $D_n$ computed from $\Phi_n$'s in Table 3, all
the way up to \hbox{$y\approx 1.1$}. This plot also shows that 
the above large-$y$ approximation,
\eqn\Dmetaplus
{
D_{meta}^{\,>}(y) = \(0.2161 - 0.136\,y^{-{15\over 8}}\)\,V(y)\,,
}
seems to provide rather accurate approximation of this function for 
$y\moresim 2.5$. The two curves corresponding to \Dmetaminus\ 
and \Dmetaplus\ intersect at $y\approx 1.59$. We have checked that 
even if one just takes \Dmetaminus\ and \Dmetaplus\ as the approximations
for $D_{meta}(y)$ at $y<1.59$ and $y>1.59$ respectively, the
dispersion integral \lowdispeta\ reproduces correctly at least
four significant digits when compared with our direct data at real
positive $\eta$. Of course such accuracy is related to the fact at
$y \sim 1.59$ the function $D_{meta}(y)$ is already rather small. In
fact, much better approximation can be achieved if one uses some
interpolation between \Dmetaminus\ and \Dmetaplus\ in the intermediate 
region. We have obtained very good results with the fifth-order polynomial
interpolation
\eqn\Dmetainterp
{
D_{meta}^{\rm interp}(y) = \bigg[P_0 + P_1\,(y-y_1) + \cdots +
P_5\,(y-y_1)^5 \bigg]\,V(y)\,,
}
with $y_1=1.1$ and the coefficients $P_0, \ldots, P_5$ fixed by the
conditions that \Dmetainterp\ matches smoothly (up to the second
derivative) the approximations \Dmetaminus\ and \Dmetaplus\ at $y=1.1$
and at $y=2.72$, respectively,
\eqn\Interpol{\eqalign{
P_0 = 0.1532863\,, \quad P_1 &= 0.0002611\,, \quad  P_2 = 
0.0499557\,, \quad \cr
P_3 = -0.0209331\,, \quad P_4 &=-0.004408\,, ~~ P_5 = 0.002672\,.
}}
The dispersion integral \lowdispeta\ with
\eqn\Dmetaapprox
{
D_{meta}(y) = 
\left\{\eqalign{
D_{meta}^{\,<}(y) \qquad &{\rm for} \quad 0 < y< 1.1
\cr
D_{meta}^{\rm interp}(y) \qquad &{\rm for} \quad 1.1 < y< 2.72
\cr
D_{meta}^{\,>}(y) \qquad &{\rm for} \quad 2.72 < y< \infty
}\right.
}
reproduces our direct data very accurately (with six significant
digits, i.e. essentially within the estimated accuracy of the direct
data), as is demonstrated in Table 4. In this computation we have
used the exact value \Gonetilde\ of the coefficient ${\tilde G}_1$ in
\lowdispeta, and again the value \Yzero\ of $Y_0$ was assumed. 

The approximation \Dmetaapprox\ can be also used to estimate the
coefficients ${\tilde G}_n$ of the assymptotic expansion \lowexp. It
follows from \dispn\ that
\eqn\tGndisp
{
{\tilde G}_n = (-)^{n+1}\,{{15}\over{8\pi}}\,\int_{0}^{\infty}\,D(y)\,
y^{{(15n-24)}\over 8}\,dy \, \qquad {\rm for}\quad n>1\,.
}
Moreover, the coefficient ${\tilde G}_1$ itself must coinside with the
integral 
\eqn\tGonedisp
{
{\tilde G}_1 = {{15}\over{8\pi}}\,\int_{0}^{\infty}\,(D_{meta}(y)-D_0)\,
y^{-{9\over 8}}\,dy\,,
}
lest analyticity of ${\tilde\Phi}(\eta)$ at $\eta=0$ be violated. 
With our approximation \Dmetaapprox\ the integral \tGonedisp\ reproduces
the coefficient ${\tilde G}_1$ almost exactly (if fact, we have used
this as a condition in choosing the interpolation interval in
\Dmetaapprox), and few higher coefficients obtained through \tGndisp\
are compaired with previous estimates in Table 2. 

\subsec{Extended Dispersion Relation}

Equipped with the results in Sects.~8.1 and 8.2, we made a preliminary 
analysis of analyticity in the shadow domain in order to validate the 
extended  analyticity conjecture formulated in Sect.~4. 
If one assumes the extended analyticity, the estimates in Sect.~8.1 and 
8.2 make it possible to actually reconstruct, with reasonable precision, 
the 
behaviour of the function $\Phi(\eta)$ along the shadow side of the Yang-Lee
branch cut in Fig.~4, thus finding an approximation for the 
discontinuity function 
\eqn\newdelta
{
\Delta(y) = i\,\Phi_{imh}(-y) -i\,e^{-i{8\over 15}\pi}\,\Phi_{low}
\(y\,e^{-i{11\over 15}\pi}\)
}
which enters the extended dispersion relation \extdisp; the expression 
\newdelta\ is a simple consequence of \deltaaa\ and \deltatilde. 
The extended analyticity is then verified by checking this 
dispersion relation against direct data.

Let us note that, on one hand, the results of Sect.~8.2 allow one to 
find the large-$y$ asymptotic 
form of the function $\Phi_{low}(y\,e^{-i{11\over 15}\,\pi})$. Indeed, the 
second (analytic) form of the Eq.~\Dfunction\ which defines $D_{meta}(y)$ 
also defines its analytic continuation to complex values of its argument. The
extended analyticity is equivalent to the statement that $D_{meta}(z)$
is analytic in the wedge $-\pi/5 < {\rm arg}(z) < \pi/5$. Under this 
assumption, it follows from \Dfunction\ that
\eqn\phishadow
{
\Phi_{low}\(y\,e^{-i{11\over 15}\,\pi}\) = 
e^{-i{32\over 15}\pi}\,\Phi_{low}\(y\,e^{{i\pi}\over 3}\) +
2i\,e^{-i{16\over 15}\pi}\,D_{meta}\(y\,e^{-{{i\pi}\over 5}}\)\,.
}
For real positive $y$, this equation expresses the desired values of 
$\Phi_{low}(\eta)$ on the shadow side of the branch cut in terms of its values
well inside the domain LTW in Fig.~4, and also through the analytic 
continuation of the function $D_{meta}(y)$. The first term is already under 
good analytic 
control, see Sect.~8.2.1. In fact, for sufficiently large $y$ the function
$\Phi_{low}(y\,e^{{i\pi}\over 3})$ is approximated very well by first few 
terms
of its asymptotic expansion \phiassplus, as illustrated in Fig.~14. 
Specifically, for $y>3.0$ the first twelve terms of this expansion yield 
an accuracy better than $\sim 10^{-4}$. On the other hand, the large-$y$ 
behaviour of the second term in \phishadow\ follows from the asymptotic 
law \discass. With this, Eq.~\phishadow\ yields approximation
for the second term in Eq.~\newdelta, valid at 
sufficiently large $y$. This readily translates into corresponding
large-$y$ approximation for the function $\Delta(y)$, since the 
\hbox{large-$y$} behaviour of the first term in \newdelta\ follows from 
\PhiimhG b\ 
and \highexp. Finaly, one can check that for $y>3$ the accuracy 
of at least $\sim 10^{-4}$ can be achieved by taking the first 
six terms (i.e. including $G_{12}$ term) of the expansion \highexp. 
Therefore the following formula
\eqn\deltamore
{\eqalign{
\Delta(y) \approx \Delta^{>}(y) =&~\sum_{k=1}^{12}\,
e^{-{{i\pi (k-1)}\over 2}}\,\bigg( G_k - e^{-{{i\pi k}\over 8}}\,{\tilde G}_k 
\bigg)\,y^{{16-15k}\over 8}
\cr
&+2\,\(V_0 + V_1\,e^{{3\pi i\over 8}}\,
y^{-{15\over 8}}\)\,y^{1\over 8}
\,\exp\left\{-\mu\,e^{-{3\pi i\over8}}y^{15\over8}+{3\pi i\over8}\right\}
}}
is expected to approximate the 
discontinuity function well for sufficiently large $y$. In \deltamore\ 
$\mu$ stands for $\pi/2{\bar s}$, and it is understood that $G_{2n+1}=0$.

On the other hand, when $y$ is close to $Y_0$ the relation 
\deltadimh\ is usefull. The approximation \empminus\ for $D_{imh}(y)$
then suggests the following approximation
\eqn\deltaless
{\eqalign{
\Delta(y) &\approx \Delta^{<}(y)  = y^2/4 
\cr
-\theta(y-Y_0)&\(
{1\over 2}\,{\bar B}_0\,
e^{-i{{5\pi}\over 6}}\,(y-Y_0)^{5\over 6}  + {1\over 2}\,{\bar B}_1\,
e^{-i{{5\pi}\over 6}}\,(y-Y_0)^{11\over 6} - {{\sqrt{3}}\over 2}\,
{\bar C}_0 \,e^{-i{5\pi\over3}} \,(y-Y_0)^{5\over3}\)\,,
}}
which is expected to be valid at $y \simeq Y_0$. In \deltaless\
$\theta(x)$ is the step function.

\fig{16}{Plots of real and imaginary parts of $\Delta^{<}(y)$, 
Eq.~\deltaless\ (shown as solid lines), and $\Delta^{>}(y)$, 
Eq.~\deltamore\ (dashed lines).
}
{\epsfxsize=9cm\epsfbox{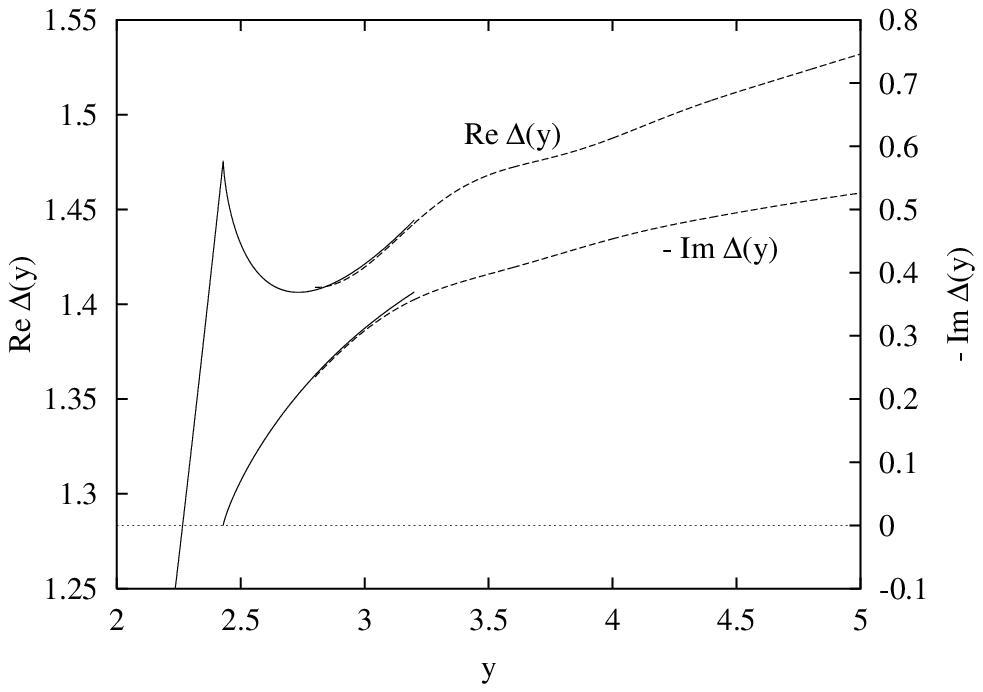}}

The plots of \deltamore\ and \deltaless\ are shown in Fig.~16. In this 
plot the numerical values~\BCbars\ of the coefficients in \deltaless\
are taken. Also, the coefficients $V_0$ and $V_1$ in \deltamore\ are 
identical to those in \Dmetaplus, and the coefficients $G_{k}$ and 
${\tilde G}_k$ are taken from the Tables 1 and 2. In fact, contributions
of the terms with $k>5$ in the sum in \deltamore, as well as the
differences in the choice between the columns in that Tables, are too 
small to be visible in this plots. 
One can see that the approximations \deltamore\ and
\deltaless\ match each other very well in the intermediate region of $y$
around 3. This match itself can be viewed as a support to the extended 
analyticity conjecture. Even more convincing result is obtained by 
evaluating the extended dispersion relation \extdisp\ with the use of 
these approximations. We have evaluated the integral \extdisp\ with
\eqn\Deltaapprox
{
\Delta(y)=
\left\{\eqalign{
\Delta^{<}(y) \qquad &{\rm for}\quad 0 < y < 3
\cr
\Delta^{>}(y)\qquad &{\rm for}\quad 3 < y
}\right.
}
for some real values of $\eta$. In this calculation specific numerical
values of the coefficients $G_{2n}$ from the second column in Table 1
and ${\tilde G}_k$ from the second column in Table 2, as well as the
values \Dmetaplus\ of $V_0, V_1$, were used. The results for the 
function $\Phi(\eta)$ are compared to our direct
data in Table 4. Also, the first coefficients $\Phi_n$ obtained 
through (4.10), within the same approximation, are presented
in Table 3. Although the agreement is not very accurate (the values
of $\Phi(\eta)$ agree with the direct data with only four digits, which
is well below the estimated accuracy of the latter), we take it as a 
strong
support of extended analyticity. The relatively low accuracy should probably
be attributed to low quality of the approximation \deltaless; we believe 
it would improve when better estimates of the coefficients in 
($3.41b$)--($3.41c$) are obtained, and higher subleading singular terms 
in \philyang\ are included.

\newsec{Discussion}

In this work we have used numerical data obtained by TFFSA to study
the scaling function $\Phi(\eta)$ associated with the Ising model
free energy in its critical domain. We estimated some parameters of 
this function, including location of the Yang-Lee singularity~\Yzero\
and leading amplitudes \Bs a, \Bs b\ of associated singular expansion
\philyang. We also examined the analyticity of the scaling function. 
Our analysis confirmed the standard analyticity assumption 
(see Sect.~3), and we have found strong evidence in support of the 
``extended analyticity'' conjectured in Sect.~4. Here we would
like to explain why we regard proving (or invalidating) this extended
analyticity as an important issue. 

As is known, in the classical theory of phase transitions, which does not
incorporate fluctuational effects, the Langer's singularity of
the low-T free  at $H=0$ is not present. Instead, analytic
continuation of the classical low-T free energy shows a branch-cut
singularity at some finite negative value of $H$, $H=-H_{\rm SP}$, known
as the ``spinodal point''. This singularity appears because at
sufficiently large $H$ ($|H|>H_{\rm SP}$) the metastable state becomes
unstable against classical decay. When fluctuations
are taken into account the free energy becomes singular (albeit weakly
singular) at $H=0$, since the metastable phase is always
prone to 
decay through nucleation. But what happens to the spinodal singularity?
There are convincing arguments (see \Privman) to that in
the presence of fluctuations the (analytic continuation of the) free
energy cannot have singularities at real negative $H$. To our opinion, 
the most plausible possibility is that in fluctuational theory the
spinodal singularity does not just disappear but instead gets pushed under
the Langer's branch cut, as is shown schematicaly in Fig.~2. In other
words, the nearest singularity under the Langer's branch cut could be
associated with the ``ideal limit of metastability'' \LangerC, where 
a transition from a decay through nucleation to faster ``classical''  
mechanism of decay takes place. The transition
can be more or less expressed depending on how far under the cut the
singularity is located. The above speculation is made under assumption
that the analytic continuation of the free energy is a quantity
at all relevant to termodynamics of the metastable state, even away from the
domain of vanishingly small $H$. It also ignores all possible kinetic 
aspects of the problem, certainly important in realistic
situations. Nontheless, it was one of the motivations for the
presented study of analyticity under the branch cut in Fig.~2.
The ``extended analyticity'' proposed in Sect.~4 is the statement that 
the nearest singularity under this
branch cut is identical to the Yang-Lee singularity in Fig.~1. In this
terms, the ``true spinodal'' is the Yang-Lee singularity (viewed from
the shadow side, as in Fig.~4). 

Our analysis concerned explicitely with the $D=2$ system. In this case the
\hbox{Yang-Lee} singularity turns out to be located relatively far under 
the branch
cut. The phase of $\xi_{\rm SP}$ in Fig.~2 (which is related to the 
angular size of the ``shadow domain'' in Fig.~4) is~$3\pi/8$. However, on 
a formal level, it is not difficult to extend the analysis of
Sect.~3 to any $1<D<4$. In this generic case the angular size $\theta$
of analogous shadow domain can be expressed in terms of the standard Ising
model critical exponents as 
\eqn\crit
{
\theta = \pi - {{3\pi}\over 2}\,{1\over {2-\alpha-\beta}}\,. 
}

It follows from what we know about the
$D$-dependence of the Ising critical exponents (see e.g. 
Ref.~\ZinnJustin) that when $D$ 
increases from 2 to 4 the angular size of the ``shadow domain'' decreases, 
and goes to zero as the critical exponents approach their classical values
at $D=4$. A simple estimate shows that at $D=3$ this angular size 
is $\approx 0.04\pi$. If similar ``extended analyticity'' assumption
could be validated in this case, one should expect the effect of the 
Yang-Lee singularity being quite
prominent at real negative $H$, at least in the critical domain $T\to
T_c$. Unfortunately, at the moment we do not see any practical way to
check this assumption away from the $D=2$ case.

Even in the $D=2$ case many issues call for refinement. First, although
the extended dispersion relation did check in Sect.~9 rather convincingly,
with present data about the discontinuity $\Delta$ the agreement in 
Table~4 is not
too precise, and hence very weak singularities in the shadow domain are not
excluded. Much better knowledge about the ``fine-structure'' of the 
Yang-Lee 
singularity (in particular, better estimates of the coefficients of the
expansions \ABC b, \ABC c) would be usefull in order to establish the 
extended
analyticity ``beyond any reasonable doubt''. We hope to make progress
in this direction in the future. 
We also want to stress here that even that would
not settle the question of analytic properties of the scaling function
completely. What one should expect when going under the Yang-Lee
branch cut in Fig.~4 from the shadow domain? This and related questions
require at least some understanding of the physics governing the ``shadowy
side'' of the Yang-Lee edge singularity.

Let us also mention some possible further developments, not related
directly to the free energy, which we hope to address in the future:

($i$) Analytic properties of the correlation length (defined as
inverse mass $M_{1}$), can be studied along the same lines. We
expect associated scaling function to enjoy similar extended
analyticity.

($ii$) Detailed quantitative study of the IFT \IFT, viewed as the
particle theory, is possible using TFFSA. This is a beautifull model
of quark confinement, simple but rich in phenomena \McCoyWuA.

($iii$) Physical interpretation of the IFT with pure imaginary $h$
is a particularly intriguing question. Even the qualitative pattern of the
higher finite-size energy levels $E_i (R)$ in this case is yet to be
understood.

\appendix{A}{}

Here we present a simple derivation of the finite-size matrix elements
\fs\ using the symmetries of the ``doubled'' Ising field theory with
$h=0$. This derivation is one of the results of \FLZ, where these
symmetries are exploited to a greater extent (in particular, the
nonlinear differential equations of \McCoyA\ are shown to be a consequence
of these symmetries).

Consider two copies of the free-fermion system \AFF. We will use the 
subscripts $a$ and $b$ to distinguish between the fields corresponding
to these copies. Thus, $(\psi_a , {\bar\psi}_a)$ and $(\psi_b ,
{\bar\psi}_b)$ will stand for the two species of Majorana fermi
fields, $\sigma_a$ and $\sigma_b$ will denote the associated spin
fields, etc. We will also use the notations $a^{\dagger}$, $a$ and 
$b^{\dagger}$, $b$ for the fermion creation and anihilation operators
associated with the two copies; these operators are assumed to satisfy
the canonical anticommutators \ac\ between themselves, and
$a^{\dagger}, a$ anticommute with $b^{\dagger}, b$. Of course, this
doubled Majorana fermion system is equivalent to a single copy of a free
Dirac fermion.

We will study this system in a finite-size geometry, with the spatial
coordinate ${\rm x}$ compactified on a circle of circumference
$R$. Let us call ${\cal H}_{diag}$ the space of states generated by
the creation operators $a^{\dagger}, b^{\dagger}$ from the
``diagonal'' vacua \hbox{$\mid 0 \rangle_{\rm NS, NS} \equiv 
\mid 0\rangle_{{\rm NS},a}\otimes \mid 0 \rangle_{{\rm NS}, b}$} and 
$\mid 0 \rangle_{\rm R, R} \equiv \mid 0\rangle_{{\rm R},a}\otimes \mid 0
\rangle_{{\rm R}, b}$. Consider the operator
\eqn\Ai{
{\bf Y} = {1\over {2\pi}}\,\int_{0}^{R}\,\bigg[\psi_a \partial \psi_b
+ {{im}\over 2}\,{\bar\psi}_a \psi_b \bigg]\,d{\rm x}
}
acting in the space ${\cal H}_{diag}$. It is easy to check that it 
commutes 
with the Hamiltonian of this doubled system, and anihilates both the above
``diagonal'' vacua $\mid 0 \rangle_{\rm NS, NS}$ and $\mid 0
\rangle_{\rm R, R}$. This operator is just one of the generators of
the infinite-dimensional algebra ${\widehat {SL(2)}}$ of the local 
integrals
of motion of the free Dirac field theory \LeClair; in fact, for our
present limited purpose no knowledge about the other generators is
needed. It is straightforward to derive the commutators
\eqna\Aii$$\eqalignno{
[{\bf Y}, \mu_a (x)\mu_b (x) ] &= \sigma_a (x)\,\partial\sigma_b (x)-
\partial\sigma_a (x) \,\sigma_b (x)\,,&\Aii a
\cr
[{\bf Y}, \sigma_a (x)\mu_b (x) ] &= -i\,\big(\mu_a
(x)\,\partial\sigma_b (x) - \partial\mu_a (x) \,\sigma_b (x)\big)\,,
&\Aii b
}$$
where again $\partial$ stands for ${1\over 2}(\partial_{\rm x} -
i\partial_{\rm y})$, and $\mu(x)$ are the dual spin field \KadanoffCeva. 
It is also straightforward to compute the action
of this operator on arbitrary states containing $a$ and $b$
particles. For instance, for the excitations over the vacuum $\mid 0
\rangle_{\rm R, R}$ one obtains
\eqn\Aiii{
{\bf Y} = \sum_{n \in {\IZ}}\,{m\over
2}\,e^{\theta_n}\,\left[a_{n}^{\dagger} b_{n}^{\ } + b_{n}^{\dagger}
a_{n}^{\ } \right]\,,
}
where $\theta_n$ are defined in \thetaka. Similar expression (with 
$n \to k \in {\IZ}+{1\over 2}$) holds for the action of ${\bf Y}$ in
the NS-NS sector. We are going to show that the general structure of
the matrix elements \fs\ (with the exception of the explicit
expressions \legs\ for the ``leg factors'' $g$ and $\tilde g$,
which will be derived in the second part of this section) is a simple
consequence of the relations (A.2) and \Aiii.

To save on writing, we will restrict attention to the case $m>0$, and 
explicitely derive only the particular case
$K=0$ of the expression \fs; the analysis can be extended to a
other cases in a straightforward way. We will use the notations
\eqn\Aiv{
M_{0,N}(n_1, ..., n_N) = \left\{
\eqalign
{ _{\rm NS}\langle 0 \mid 
\sigma(0,0) 
\mid
n_1, ..., n_N \rangle_{\rm R} &\quad {\rm for\  even}~N
\cr
_{\rm 
NS}\langle 0 \mid \mu(0,0) \mid
n_1, ..., n_N \rangle_{\rm R} &\quad {\rm for\  odd}~~N} \right\}  
}
If $N$ is even, take the identity
\eqn\Av{
0 = _{\rm NS, NS}\langle 0 \mid {\bf Y}, \mu_a (x) \mu_b (x)
a_{n_1}^{\dagger} \cdots a_{n_{N}}^{\dagger}\mid 0 \rangle_{\rm R, R}\,.
}
which follows from $_{\rm NS, NS}\langle 0 \mid {\bf Y} = 0$. If $N$
is odd, similar identity, with $\mu_a (x) \mu_b (x)$ replaced by
$\sigma_a (x) \mu_b (x)$, should be used. The
matrix element in the r.h.s. of \Av\ can be written as a sum
of the matrix element of the commutator \Aii a\ and the matrix element of 
$\mu_a(x) \mu_b(x) {\bf Y}$. Using \Aiii\ to compute the latter,
one obtains the relation 
\eqn\Avi{\eqalign{
{M}_{0,0}\,{M}_{0,N}(n_1, \cdots,
n_{N})&\,\sum_{j=1}^{N}\(e^{\theta_{n_j}}\) = 
\cr
I\,\sum_{j=1}^{N}
\,(-1)^{j-1}\,e^{\theta_{n_j}}\,M_{0,1}&({n_j})\,M_{0,N-1}(n_1, \cdots,
n_{j-1}, n_{j+1}, \cdots , n_{N})\,,
}}
where $I=i$ if $N$ is even, and $I=1$ if $N$ is odd. This relation
allows one to obtain recurrently all the matrix elements \Aiv\ in
terms of $M_{0,0}$ and $M_{0,1}(n)$. It is not difficult to show that
the solution of this recurrent relation has the form
\eqn\Avii{
M_{0,N}(n_1, \cdots, n_N) = i^{[N/2]}\,M_{0,0}\,\(\prod_{i=1}^{N}
\,g(\theta_{n_i})\) \,f_{N}(\theta_{n_1}, \cdots,
\theta_{n_N})\,,
}
where $[N/2]$ denotes the integer part of $N/2$,
\eqn\Aviii{
g(\theta_n) \equiv M_{0,1}(n)/M_{0,0}\,,
}
and
\eqn\Aix{
f_{N}(\theta_1, \cdots, \theta_N) = 
\prod_{i<j}^{N}\,\tanh\({{\theta_{n_i}-\theta_{n_j}}\over 2}\)\,.
}
This follows from the well known identities \YuZ\ that the products
\Aix\ obey,
\eqn\Ax{
f_{0,N+1}(\theta, \theta_1, \cdots, \theta_N) =
\sum_{j=1}^{N}\,(-)^{j-1}\,\tanh\bigg({{\theta - \theta_j}\over
2}\bigg)\,f_{0,N-1}(\theta_1, \cdots, \theta_{j-1}, \theta_{j+1},
\cdots, \theta_N )
}
If one expands \Ax\ around $\theta = +\infty$, the term $\sim
e^{-\theta}$ yields exactly \Avi.

It remains to determine the factors $M_{0,0}$ and $g(\theta)$ in
\Avii. The first is known from Ref.~\Sachdev, see 
Eqs.~\onepoint--\onepointS. 
To find the factors
\hbox{$g(\theta_n) = _{\rm NS}\langle 0 | \mu(0) |n \rangle_{\rm 
R}/
_{\rm NS}\langle 0 | \sigma(0) |0 \rangle_{\rm R}$}, let us
return to the single Majorana fermion system \AFF, and consider the
matrix element
\eqn\Axi{
\Psi({\rm x, y}) = _{\rm NS}\langle 0 \mid \mu(0,0) \,\psi ({\rm x, 
y})\mid 0
\rangle_{\rm R}/ _{\rm NS}\langle 0 \mid \sigma(0,0) \mid 0
\rangle_{\rm R}\,,
}
and similar matrix element ${\bar\Psi}({\rm x, y})$ defined as in 
Eq.~\Axi\
with $\psi({\rm x, y})$ replaced by ${\bar\psi}({\rm x, y})$. Here
${\rm x, y}$ are the cartesian coordinates on the euclidean space-time
cilinder, ${\rm x}$ being the periodic direction. The functions $\Psi$
and ${\bar\Psi}$ have the following properties:

\noindent
$i$) They satisfy the Majorana field equations
\eqn\Axii{
{\bar\partial}\Psi({\rm x,y}) = {{im}\over 2}\,{\bar\Psi}({\rm
x,y})\,, \qquad {\partial}{\bar\Psi}({\rm x,y}) = 
- {{im}\over 2}\,{\Psi}({\rm x,y})\,;
}

\noindent
$ii$) $\Psi({\rm x, y})$ is an analytic function for all real ${\rm x,
y}$ except at the points \hbox{$({\rm x, y}) = (0({\rm mod}\,R), 0)$}, and
it obeys the (anti)periodicity conditions
\eqn\Axiii{
\Psi({\rm x}+R, {\rm y}) = \left\{
\eqalign{
\Psi({\rm x,y})&\quad{\rm for}\quad{\rm y}<0
\cr
 -\Psi({\rm x,y})&\quad{\rm for}\quad{\rm y}>0} \right\}\,.
}
The function $\bar\Psi$ obeys the same conditions.

\noindent
$iii$) Their singularities at $({\rm x, y}) = (0,0)$ are such that
\eqn\Axiv{
\lim_{{\rm x,y} \to 0}\,\sqrt{2\,z}\,\Psi({\rm x, y}) = e^{-{{i\pi}\over
4}}\,, \qquad \lim_{{\rm x,y} \to 0}\,\sqrt{2\,{\bar z}}\,
{\bar\Psi}({\rm x, y}) = e^{{i\pi}\over 4}\,,
}
where $z = {\rm x}+i {\rm y}$, ${\bar z} = {\rm x}-i {\rm y}$. Also,
the functions $\Psi({\rm x,y})$ and ${\bar\Psi}({\rm x, y})$ are
bounded when $|{\rm y}| \to \infty$.

It is possible to show that the above conditions $i$)--$iii$) define the
functions uniquely. The solution to these conditions has the form
\eqna\Axv$$\eqalignno{
\Psi({\rm x, y}) &= \quad\quad\sum_{n\in {\IZ}}\,
\sqrt{{\pi}\over{R\,\cosh\theta_n}}\,e^{{\theta_n}\over
2}\,{g}(\theta_n)\,e^{m{\rm y}\,\cosh\theta_n - im{\rm
x}\,\sinh\theta_n }\,,\quad~{\rm for~y}<0\,,\qquad&\Axv a
\cr
\Psi({\rm x, y}) &= -i\,\sum_{k\in {\IZ}+{1\over
2}}\,\sqrt{{\pi}\over{R\,\cosh\theta_k}}\,e^{{\theta_k}\over
2}\,{\tilde g}(\theta_k)\,e^{-m{\rm y}\,\cosh\theta_k + im{\rm
x}\,\sinh\theta_k }\,,~{\rm for~y}>0\,,\qquad&\Axv b
}$$
where $\theta_n$ and $\theta_k$ are the solutions of
the Eqs.~\thetaka, and the
functions $g(\theta)$ and ${\tilde g}(\theta)$ are defined through the
integrals \gnka. The function ${\bar\Psi}({\rm x,y})$ is given by the
same sums \Axv a, \Axv b, but with the factor $e^{\theta\over 2}$ replaced 
by
$e^{-{\theta\over 2}}$. It is clear that the sums \Axv a, \Axv b\ satisfy
\Axii\--\Axiv, and one only needs to check that the analytic continuation 
of
\Axv a\ to ${\rm y} > 0$ coincides with \Axv b, and vice versa. This is
done using standard tricks with contour deformations. One can write
\Axv a, \Axv b\ in terms of the contour integrals
\eqna\Axvi$$\eqalignno{
\Psi({\rm x, y}) &=~\,\sqrt{\pi m}\,\int_{{\cal C}_{-}+{\cal C}_{+}}\,
{{d\theta}\over{2\pi}}\,{{G(\theta)\,e^{\theta\over
2}}\over{1-e^{-imR\sinh\theta}}} \,e^{m{\rm y}\cosh\theta - im{\rm
x}\sinh\theta}  \quad {\rm for~y} < 0\,,\quad&\Axvi a
\cr
\Psi({\rm x, y}) &=  i\,\sqrt{\pi m}\,\int_{{\cal C}_{-}+{\cal C}_{+}}\,
{{d\theta}\over{2\pi}}\,{{G(\theta)^{-1}\,e^{\theta\over
2}}\over{1+e^{imR\sinh\theta}}} \,e^{-m{\rm y}\cosh\theta + im{\rm
x}\sinh\theta} \quad {\rm for~y} > 0\,,\quad&\Axvi b
}$$
where $G(\theta) = \exp(\kappa(\theta))$, with
$\kappa(\theta)$ defined as the integral \legs, $G(\theta)^{-1}$
stands for $1/G(\theta)$ (not for the inverse function), and the 
integration
is performed along the contours ${\cal C}_{-}$, which runs from
$-\infty$ to $+\infty$ just below the real axis, and ${\cal C}_{+}$,
which returns back to $-\infty$ just above the real axis. The function
$G(\theta)$ is meromorphic, and both $G(\theta)$ and $G(\theta)^{-1}$ 
are analytic in the strip $-\pi < \Im m\,\theta < \pi$. It is easy to
verify that
\eqn\Axvii{
G(\theta+i\pi/2) G(\theta-i\pi/2) =
{{e^{mR\cosh\theta}-1}\over{e^{mR\cosh\theta}+1}}\,.
}
If, for
instance, $0 < {\rm x} <R$, then the contour ${\cal C}_{-}$ in \Axvi a\ 
and 
\Axvi b\ can be shifted downward, $\theta\to\theta-i\pi/2$, while the
contour ${\cal C}_{+}$ admits upward shift $\theta\to \theta+i\pi/2$. 
After this deformation the integral \Axvi a assumes the form
\eqn\Axviii{\eqalign{
\Psi({\rm x, y}) = \sqrt{\pi m}\,\int_{-\infty}^{\infty}\,
{{d\theta}\over{2\pi}}\,&\left[{{G(\theta-i\pi/2)\,
e^{{\theta\over 2}-i{\pi\over 4}}}\over{1-e^{-mR\cosh\theta}}}
\,e^{-im{\rm y}\sinh\theta-m{\rm x}\cosh\theta}\right.
\cr
&\left.-\,{{G(\theta+i\pi/2)\,e^{{\theta\over
2}+i{\pi\over 4}}}\over{1-e^{mR\cosh\theta}}}
\,e^{im{\rm y}\sinh\theta + m{\rm x}\cosh\theta}
\right] 
}}
As long as ${\rm x}$ remains within the above interval, this integral
now converges at all ${\rm y}$, both negative and positive; it thus
defines the analytic continuation of \Axvi a\ to all  ${\rm y}$. The same
contour deformations applied to \Axvi b\ yields the integral which
coincides with \Axviii\ in virtue of \Axvii. 

In fact, it is not difficult to find in a similar way a complete basis of
functions $\Psi({\rm x}, {\rm y})$ which satisfy \Axii\---\Axiv, but are 
allowed to grow at $|{\rm y}| \to \infty$; this basis then can be used
to derive the full set of matrix elements \fs\ directly. We will
discuss this topic in greater detail elsewhere.

\appendix{B}{}

In this section we describe some properties of the energy spectrum of
the Hamiltonian~\Hamiltonian\ in the low-T domain $m>0$. In subsection BI 
below we warm up with the derivation of the small $h$ expansion \Miexp\ of
the meson masses, and in BII we discuss the relation of the
near-intersection pattern in Fig.~$6a$ to the characteristics of the
``false vacuum'' resonance.

\vskip 0.1in

{\bf BI.} As is well known, in the low-T regime the interaction term
$h\,\int\, \sigma(x)\,d^2 x$ in \AIFT\ gives rise to a confining
interaction between the fermions of \AFF\ (the ``quarks''). As the
result, the particle spectrum of the IFT \IFT\ with $h \neq 0$ consists
of bound states of the quarks - the ``mesons''. If $h$ is
sufficiently small, the lower part of the meson spectrum can be
studied within the two-quark approximation. One looks for the meson
state (in its rest frame) in the form
\eqn\Bi{
\mid \Psi \rangle = {1\over
2}\,\int_{-\infty}^{\infty}\,{{dp}\over{2\pi}}\,{\tilde\Psi}(p)\,
a_{p}^{\dagger} a_{-p}^{\dagger}\,\mid 0 \rangle\,,
}
and ignores all multiquark components. Here $a_{p}^{\dagger}$ are the
quark creation operators normalized according to the canonical
anticommutators $\{ a_{p}, a_{p'}^{\dagger}\} = 2\pi\,\delta(p-p')$,
and ${\tilde\Psi}(p)$ is the momentum-space wave function, which is
assumed to be antisymmetric, i.e. ${\tilde\Psi}(p) = -
{\tilde\Psi}(-p)$. Within this approximation the eigenvalue problem
for the Hamiltonian~\Hamiltonian\ (with $R=\infty$) leads to the 
Bethe-Salpeter
equation 
\eqn\Bii{
(2\omega(p) - E)\,{\tilde\Psi}(p) = {{t^3}\over
2}\,\dashint_{-\infty}^{\infty}
\,{{m^2}\over{\omega(p)\,\omega(p')}}\,\left[
\({{\omega(p)+\omega(p')}\over{p-p'}}\)^2 + {1\over
2}\,{{p\,p'}\over{\omega(p)\,\omega(p')}} \right]\,{\tilde\Psi}(p')\, 
{{dp'}\over{2\pi}}\,,
}
where the principal value of the integral in the right-hand side is
understood, $E$ stands for the meson's rest energy, and the notations
$\omega(p) = \sqrt{m^2 + p^2}$ and
\eqn\Biii{
t = \bigg({{2{\bar\sigma}h}\over{m^2}}\bigg)^{1\over 3} \equiv (2{\bar
s}\xi)^{1\over 3}
}
are used. The easiest way to derive this equation is to use the
finite-size matrix elements~\fs\ with $N = M = 2$ and take the limit
$R\to\infty$.

The two-particle approximation is justified if $h$ is small and $E$ is
sufficiently close to $2m$. In this case one can consistently treat
the momenta $p, \,p'$ in \Bii\ as small as compaired to $m$. Making a
rescaling 
\eqn\Biv{
p = (mt)\,k\,, \qquad p' = (mt)\,k'\,,
}
and expanding the operators in (B2) in the powers of $t$, one has
\eqn\Bv{\eqalign{
\left[-\epsilon + k^2 - {{t^2}\over 4}\,k^4 + {{t^4}\over 8}\,k^6 +
\cdots\right] \,{\tilde\Psi}((mt)\,k)&
\cr
=\dashint_{-\infty}^{\infty}\,
\left[ {2\over{(k-k')^2}} + {{t^4}\over 8}\,(k+k')^2 +
\cdots \right]&\,{\tilde\Psi}((mt)\,k')\,{{dk'}\over{2\pi}}\,,
}}
where $\epsilon$ is defined as
\eqn\Bvi{
E - 2m = (mt^2)\,\epsilon\,.
}

It is useful to write down the configuration-space form of the
Eq.~\Bv,
\eqn\Bviia{\eqalign{
\left[-\epsilon + |X| - {{d^2}\over{dX^2}} - {{t^2}\over
4}\,{{d^4}\over {dX^4}} - {{t^4}\over 8}\,{{d^6}\over {dX^6}} +
\cdots\right]& \psi(X) = 
\left[-{{t^4}\over 2}\,\delta'(X) +
\cdots\right]\psi(X)\,,
}}
where the configuration-space wave function 
\eqn\Bviib{
\psi(X) =
\int_{-\infty}^{\infty}\,{{dk}\over{2\pi}}\,e^{ikX}\,{\tilde\Psi}((mt)\,k) 
}
is written in terms of the rescaled coordinates,
\eqn\Bviic{
X = (mt)({\rm x}_1 - {\rm x}_2)\,,
}
${\rm x}_1, {\rm x}_2$ being the positions of the quarks. This
function also is antisymmetric,
\eqn\Bviii{
\psi(X) = -\psi(-X)\,.
}
In Eqs.~\Bv\ and \Bviia\ the omitted terms $\cdots$ are $\sim t^6$ or
smaller.

In the leading order in $t^2$ the Eq.~\Bviia\ is just the Schroedinger
equation with the linear potential $|X|$. It leads to the assymptotic
formula \Miass\ for the meson masses (first obtained in \McCoyWuA). One 
can calculate further corrections by solving \Bviia\ perturbatively 
in~$t^2$. Let $A(X)$ be a solution of the Airy equation
\eqn\Bx{
\left[X - {{d^2}\over{dX^2}}\right]\,A(X) = 0\,.
}
It is straightforward to check that the function
\eqn\Bxi{\eqalign{
F(X) =&\, A(X) - {{t^2}\over 20}\,\bigg[4X\,A(X) +X^2\,A'(X)\bigg]
\cr
&+{{t^4}\over
4}\,\left[\(-{{X^2}\over{7}}+{{X^5}\over{200}}\)\,A(X) +
\(-{11\over 35} + {{X^3}\over{35}}\)\,A'(X)\right] + O(t^6)
}}
(where prime denotes the derivative) is the perturbative solution of
the equation
\eqn\Bxii{
\left[X - {{d^2}\over{dX^2}} - {{t^2}\over 4}\,{{d^4}\over{dX^4}} -
{{t^4}\over 8}\,{{d^6}\over {dX^6}} + O(t^6)\right]\,F(X) = 0\,,
}
and that the function
\eqn\Bxiii{
\psi(X) = {\rm sign}(X)\,F(|X|-\epsilon)
}
solves (in the same perturbative sense) the equation \Bviia, provided
\eqna\Bxiv$$\eqalignno{
&F(-\epsilon) = O(t^2)\,;&\Bxiv a
\cr
&F(-\epsilon) + {{t^2}\over 2}\,F''(-\epsilon) = O(t^4)\,;&\Bxiv b
\cr
&F(-\epsilon) + {{t^2}\over 4}\,F''(-\epsilon) + {{t^4}\over
8}\,F''''(-\epsilon) - {{t^4}\over 4}\,F'(-\epsilon) = O(t^6)\,.
&\Bxiv c
}$$
Now, let ${\rm Ai} (X)$ and ${\rm Bi}(X)$ be the two canonical solutions 
of the
Airy equation \Bx, and $F_A (X)$ and $F_B (X)$ - the associated
perturbative solutions \Bxi\ of \Bxii. A straightforward calculation
shows that the linear combination
\eqn\Bxv{
F(X) = F_A (X) + \sigma(\epsilon)\,F_B (X)
}
with
\eqn\Bxvi{\eqalign{
\sigma(\epsilon) =
-{{{\rm Ai}(-\epsilon)}\over{{\rm Bi}(-\epsilon)}}\,&\left[1 +
 {{t^2\,\epsilon^2}\over{20}}\, \({{{\rm Bi}'(-\epsilon)}\over
 {{\rm Bi}(-\epsilon)}} - {{{\rm Ai}'(-\epsilon)}\over {{\rm 
Ai}(-\epsilon)}}\)\right.
\cr
-{{t^4}\over{20}}\,\({{{\rm Bi}'(-\epsilon)}\over
 {{\rm Bi}(-\epsilon)}} - {{{\rm Ai}'(-\epsilon)}\over
 {{\rm Ai}(-\epsilon)}}\)&\({{2\epsilon^3}\over{35}} -
 {{57}\over{14}} -
 {1\over{20}}\,{{{\rm Bi}'(-\epsilon)}\over{{\rm Bi}(-\epsilon)}} \) +
 O(t^6)\bigg]
}}
satisfies the conditions \Bxiv a\---\Bxiv c, i.e. \Bxiii\ with this $F(X)$ 
solves 
\Bviia.

In the bound-state problem one is interested in the normalizable
solutions of \Bviia, $\int_{-\infty}^{\infty}\,|\psi(X)|^2\,dX <
\infty$, i.e.
\eqn\Bxvii{
\psi(X) \to 0 \qquad {\rm as} \qquad |X| \to \infty\,.
}
Threfore, the bound-state eigenvalues of the energy parameter
$\epsilon$ in \Bviia\ coincide with the zeroes of the amplitude
$\sigma(\epsilon)$ in \Bxv. From \Bxvi\ one finds for these zeroes,
\eqn\Bxviii{
\epsilon_i = z_i - {{t^2}\over 20}\,z_i^2 - {{t^4}\over 280}\,
\(57 - {{11}\over 5}\,z_i^3 \) + O(t^6)\,.
}
where $-z_i$, $i=1,2,3,\ldots\,$, are the zeroes of the Airy function, 
${\rm Ai}(-z_i)=0$.

It is straightforward to continue this perturbative expansion to higher
orders in $t^2$. However, without proper modifications this does not
make much sense because, starting with the order $t^4$ in \Bviia, this
expansion exceeds the accuracy of the two-particle approximation. The
exact meson state $\mid \Psi \rangle$ contains four-quark, six-quark,
and higher components which are neglected in \Bi. One can check that
these multi-quark components lead to certain corrections to the r.h.s.
of \Bii\ which start with the terms $\sim t^4$. Therefore, strictly
speaking, even the terms $\sim t^4$ in \Bviia\ and in \Bxviii\ go beyond 
the
accuracy of the two-quark approximation. However, it is possible to
argue that the whole of the leading ($\sim t^6$) correction to \Bii\
due to the multi-quark contributions has the form of the quark mass
renormalization \Mquark, i.e. it ammounts to adding a
momentum-independent term $t^4\,Q_2 /2$ to the l.h.s. of \Bviia. In
turn, this leads to the shift $\epsilon_i \to \epsilon_i + t^4\,q_2
/2$ in \Bxviii. Using \Bvi\ one arrives at the expansion \Miexp.

\vskip 0.1in

{\bf BII.} In the low-T domain $m>0$ the ground state of the IFT with
$R=\infty$ and $h=0$ is two-fold degenerate, with the two vacuum
states $\mid 0 \rangle_{\pm}$ corresponding to the oposite values of
the spontaneous magnetization,
\eqn\Bxix{
_{\pm}\langle 0 \mid \sigma(x) \mid 0 \rangle_{\pm} = \pm \,{\bar
\sigma}\,.
}
Adding the interaction term $h\,\int \sigma(x)\,d^2 x$ with small
positive $h$ lowers the energy density associated with the state $\mid
0 \rangle_{-}$ and rises the energy density of $\mid 0
\rangle_{+}$. That is, the state~$\mid 0 \rangle_{-}$ becomes the true
vacuum of the perturbed system, while the state $\mid 0 \rangle_{+}$
looses its stability and becomes the ``false vacuum'' - the global
resonance state with the complex energy~\Gmeta. 

If $h$ is small, the ``false vacuum'' decay rate can be computed
using the saddle-point analysis in the Euclidean space-time 
\refs{\Kobzarev, \Coleman, \CC, \Voloshin}. This analysis suggests the 
following qualitative
picture of the decay process in real time. The quantum mechanical
tunneling process results in the simultaneous formation of two quarks
separated by the distance $R_d = 2m/\Delta F \approx m/{\bar\sigma}h$, 
which then
speed away from each other, the expanding region between the quarks
being the nucleus of the true vacuum. The distance $R_d$ is the size
of the ``critical droplet'' in the droplet model calculation 
\refs{\LangerA, \Voloshin}. As
long as the separation $R_{1,2}$ between the quarks remains close to
$R_d$, the state can be described in term of non-relativistic quantum
mechanics of the two-quark system with the {\it repulsive} linear
potential~$-2h{\bar\sigma}\,R_{1,2}$. If $|R_{1,2}-R_{d}| \ll R_d$, a
generic state of such two-quark system is described by the wave
function
\eqn\Bxx{
\psi(Y) \sim A_{+}(Y - \epsilon') + S_{res}(E)\,A_{-}(Y - \epsilon')\,,
}
where $Y = (mt)\,(R_d - R_{1,2})$ (the parameter $t$ is defined in
\Biii), $(mt^2)\,\epsilon' = E - E_{res}$ is suitably normalized energy
of the system over the energy $E_{res}$ of the ``false vacuum'',  
and~$A_{\pm}$ are the solutions of the Airy equation \Bx,
\eqn\Bxxi{
A_{\pm}(Y) = {\rm Ai}(Y) \pm i\,{\rm Bi}(Y)\,.
}
These are the ``running wave'' solutions, the wave $A_{-}(Y)$ being
the state of the quarks speeding away from each other. The
``scattering amplitude'' $S_{res}$ in \Bxx\ should be determined (in
principle) by solving the problem in the domain where the quarks are
deep inside the classically unaccesible region $R_{1,2} < R_d$. In
this region one expects the relativistic effects, in particular the
multi-quark components of the state, to play a significant role. The
solution taking into account these effects is not yet 
available\foot{An attempt to apply the two-particle 
approximation in this region was made in an interestiong recent paper 
\Rutkevich. We comment upon the result of this work at the end of 
Sect.~7.2.}, 
and the exact form of~$S_{res}$ is not known. However, even in the absence 
of 
the explicit solution, one can
predict that the amplitude $S_{res}(E)$, as the function of the total
energy $E$ of this state, must exhibit a resonance pole at the complex
energy \Gmeta. If $E$ is close to the resonance energy (given by the
real part of \Gmeta), this amplitude can be approximated by the 
Breit-Wigner formula
\eqn\Bxxiv{
S_{res}(E) \approx {{E/R - \Delta F + i\Gamma}
\over{E/R - \Delta F - i\Gamma}}\,,
}
where $R$ is the spatial size of the system, and $\Delta F = E_{res}/R
\simeq 2\,{\bar\sigma}\,h + O(h^3)$ is the same as in \deltaf. 
With this approximation for the ``resonance scattering amplitude''
$S_{res}(E)$ in~\Bxx, and with the relativistic corrections described
in BI taken into account, the wave function \Bxx\ can be written as
\eqn\Bxxv{
\psi(Y) = {\rm const.}\,\bigg(F_A (Y - \epsilon') + 
\sigma_{res}(\epsilon')\,
F_B (Y - \epsilon')\bigg)\,,
}
where
\eqn\Bxxvi{
\sigma_{res}(\epsilon') = - {{\Gamma(m,h)\,R}\over{mt^2\,\epsilon'}}\,,
}
and $F_A$ and $F_B$ are the same perturbative solutions of \Bxii\ as in
\Bxv. 

Now let us come back to the problem of the meson states, this time
assuming a finite-size geometry of the system, with the spatial
coordinate ${\rm x}$ compactified on a circle of finite circumference
$R$. If $h$ is sufficiently small, almost all the analysis of BI
remains valid, with the exception of the bound-state condition \Bxvii,
which has to be slightly modified. When the meson (i.e. the two-quark
system with the confining interaction generated by the magnetic field
$h$) is put on a finite-size circle, its energy levels should be
determined from some sort of periodicity condition imposed on the
state wave function, rather then by \Bxvii. Again, unfortunately the
two-quark approximation in general breaks down in the classically
unaccessible domain $|X| \gg \epsilon$ (here $|X|$ is the rescaled
quarks separation~\Bviic, and $\epsilon$ relates to the energy $E$ as
in \Bvi), where multi-quark components of the wave function become
significant. Although the solution for the full wave function in this
domain is not known, one can note that this problem is closely related
to the above problem of determining the ``scattering amplitude''
$S_{res}$ in \Bxx, and the form \Bxxvi\ of the wave function still
applies if one makes identifications
\eqn\Bxxvii{
\epsilon' = \epsilon - mt\,(R - R_d )\,, \qquad Y - \epsilon' = X -
\epsilon\,, 
}
which follow from the periodicity of the problem. With this, by
demanding agreement between \Bxv\ and \Bxxv, one arrives at the
equation
\eqn\Bxxviii{
\sigma(\epsilon) = \sigma_{res}(\epsilon - mt\,(R - R_d ))\,,
}
which determines the $R$ dependence of the finite-size energy levels 
$E_{i}(R) - E_{0}(R) = 2m + mt^2\,\epsilon_i (R)$ associated with 
the
meson states; here $\epsilon_i (R)$ are the solutions of the equation     
\Bxxviii. 

The equation \Bxxviii\ is derived within the two-quark
approximation which is justified if $h$ is small. It applies when $R$
is sufficiently greater then $R_d$ (otherwise the domain of validity
of \Bxxv\ does not exist). It is not difficult to see that in this
domain \Bxxviii\ leads to a pattern of the finite-size energies very
similar to what is shown in the Fig.~$6a$. Indeed, for small $h$ the
resonance width $\Gamma$ is expected to be exponentially small (see
Eq.~\voloshin), and hence the function $\sigma_{res}$ in the
r.h.s. of \Bxxviii\ is very small everywhere exept when $\epsilon$ is
close to its resonance pole $\epsilon_{res}(R) \equiv mt\,(R -
R_d)$. Therefore, Eq.~\Bxxviii\ can hold either by virtue of $\epsilon$
being close to one of the zeroes of the function $\sigma(\epsilon)$,
which leads to $E(R)-E_{0}(R)$ close to one of the meson masses $M_i$,
or by virtue of $\epsilon$ being close to $\epsilon_{res}(R)$, leading
to $E(R)-E_0 (R)$ close to $\Delta F \,R$. When the both conditions
are met, which happens when $R$ approaches one of the
``near-intersection'' points $R_{i,i+1} = M_i /\Delta F$, one can use
the linear approximation $\sigma(\epsilon) \simeq (\epsilon -
\epsilon_i)\, \sigma' (\epsilon_i)$, which results in a quadratic
equation for~$\epsilon(R)$. This way one arrives at Eq.~\opening\
for the separations $E_{i+1}(R) - E_{i}(R)$.

\vskip 0.4in

\centerline{\bf Acknowledgments}

\vskip 0.1in

We are grafteful to Al.~Zamolodchikov and S.~Lukyanov for many discussions
and suggestions at various stages of this work. P.~F. would like to thank
R.~Roussev and E.~Vitchev for helpful computational advices and also 
\hbox{P.~Zinn-Justin}, V.~Terras and C.~Bolech for
motivating conversations. A.~Z. acknowledges helpful discussions with
V.~Kazakov, J.~Lebowitz, A.~Larkin and G.~Falkovich. The work of P.~F. was
supported by the schollarship PRAXIS~XXI BD 9102/96, from FCT, Portugal,  
and by a fellowship from {\sl Calouste Gulbenkian Foundation}, 
Portugal. The work of A.~Z. is supported by DOE grant 
\#DE--FG02--96ER10919.

\listrefs
\bye